\documentclass[aps,longbibliography,showpacs,floatfix,superscriptaddress,twocolumn]{revtex4-2}

\usepackage{graphicx,color}
\usepackage{amsfonts}
\usepackage[figuresright]{rotating}  
\usepackage{amssymb}
\usepackage{bm}
\usepackage{amsmath}
\usepackage{mathtools}
\usepackage{psfrag}
\usepackage{multirow}
\usepackage{tabularx}
\usepackage{textcomp}
\usepackage{units}
\usepackage{lipsum}
\usepackage{soul}
\usepackage{titlesec}
\usepackage{times}
\usepackage[colorlinks=true,citecolor=blue,linkcolor=magenta,hypertexnames=false]{hyperref}
\usepackage{enumitem}
\usepackage[dvipsnames]{xcolor}
\usepackage{subfigure}

\DeclareMathAlphabet\mathbfcal{OMS}{cmsy}{b}{n}

\titlespacing*{\section}
{0pt}{1ex plus .25ex}{1ex plus 1ex}
\titlespacing*{\subsection}
{0pt}{1ex plus .25ex}{1ex plus 1ex}
\titlespacing*{\subsubsection}
{0pt}{1ex plus .25ex}{1ex plus 1ex}

\titleformat{\section}
 {\fontsize{10}{15} \bfseries }
  {\thesection}
  {1em}
  {}

\titleformat{\subsection}
  {\bfseries }
  {\thesubsection}
  {2em}
  {}

\titleformat{\subsubsection}
  {\itshape}
  {\thesubsubsection}
  {2em}
  {}

\def\beq{\begin{eqnarray}}
\def\eeq{\end{eqnarray}}

\renewcommand{\v}[1]{\ensuremath{\mathbf{#1}}} 
 
\let\baraccent=\= 
\renewcommand{\=}[1]{\stackrel{#1}{=}} 

\newcommand{\red}[1]{\textcolor{blue}{#1}} 

\makeatletter

\titleclass{\subsubsubsection}{straight}[\subsection]

\newcounter{subsubsubsection}[subsubsection]
\renewcommand\thesubsubsubsection{\thesubsubsection.\arabic{subsubsubsection}}

\titleformat{\subsubsubsection}
   {\normalfont\normalsize\itshape \centering}{\thesubsubsubsection}{1em}{}
\titlespacing*{\subsubsubsection}
{0pt}{1ex plus .3ex}{1ex plus .3ex}

\makeatletter
\renewcommand\paragraph{\@startsection{paragraph}{5}{\z@}%
  {3.25ex \@plus1ex \@minus.2ex}%
  {-1em}%
  {\normalfont\normalsize}}
\renewcommand\subparagraph{\@startsection{subparagraph}{6}{\parindent}%
  {3.25ex \@plus1ex \@minus .2ex}%
  {-1em}%
  {\normalfont\normalsize}}
\def\toclevel@subsubsubsection{4}
\def\toclevel@paragraph{5}
\def\toclevel@paragraph{6}
\def\l@subsubsubsection{\@dottedtocline{4}{7em}{4em}}
\def\l@paragraph{\@dottedtocline{5}{10em}{5em}}
\def\l@subparagraph{\@dottedtocline{6}{14em}{6em}}
\makeatother

\begin{document}

\title{Time-reversal invariant finite-size topology}
\author{R. Flores-Calderon}
\affiliation{Max Planck Institute for the Physics of Complex Systems, Nöthnitzer Strasse 38, 01187 Dresden, Germany}
\affiliation{Max Planck Institute for Chemical Physics of Solids, Nöthnitzer Strasse 40, 01187 Dresden, Germany}
\author{Roderich Moessner}
\affiliation{Max Planck Institute for the Physics of Complex Systems, Nöthnitzer Strasse 38, 01187 Dresden, Germany}

\author{Ashley M. Cook}
\affiliation{Max Planck Institute for the Physics of Complex Systems, Nöthnitzer Strasse 38, 01187 Dresden, Germany}
\affiliation{Max Planck Institute for Chemical Physics of Solids, Nöthnitzer Strasse 40, 01187 Dresden, Germany}

\begin{abstract}

We report finite-size topology in the quintessential time-reversal (TR) invariant systems, the quantum spin Hall insulator (QSHI) and the three-dimensional, strong topological insulator (STI): previously-identified helical or Dirac cone boundary states of these phases hybridize in wire or slab geometries with one open boundary condition for finite system size, and \textit{additional, topologically-protected, lower-dimensional} boundary modes appear for open boundary conditions in \textit{two or more} directions. For the quasi-one-dimensional (q(2-1)D) QSHI, we find topologically-protected, quasi-zero-dimensional (q(2-2)D) boundary states within the hybridization gap of the helical edge states, determined from q(2-1)D bulk topology characterized by topologically non-trivial Wilson loop spectra. We show this finite-size topology furthermore occurs in 1T'-WTe\textsubscript{2} in ribbon geometries with sawtooth edges, based on analysis of a tight-binding model derived from density-functional theory calculations, motivating experimental investigation of our results. In addition, we find quasi-two-dimensional (q(3-1)D) finite-size topological phases occur for the STI, yielding helical boundary modes  distinguished from those of the QSHI by a non-trivial magneto-electric polarizability linked to the original 3D bulk STI. Finite-size topological phases therefore exhibit signatures associated with the non-trivial topological invariant of a higher-dimensional bulk, clearly distinguishing them from previously-known topological phases. Finally, we find the q(3-2)D STI also exhibits finite-size topological phases, finding the first signs of topologically-protected boundary modes of codimension greater than $1$ due to finite-size topology. Finite-size topology of four or higher-dimensional systems is therefore possible in experimental settings without recourse to thermodynamically large synthetic dimensions.


 \end{abstract}
\maketitle
\section{Introduction}



\begin{figure*}
 \includegraphics[width=\textwidth]{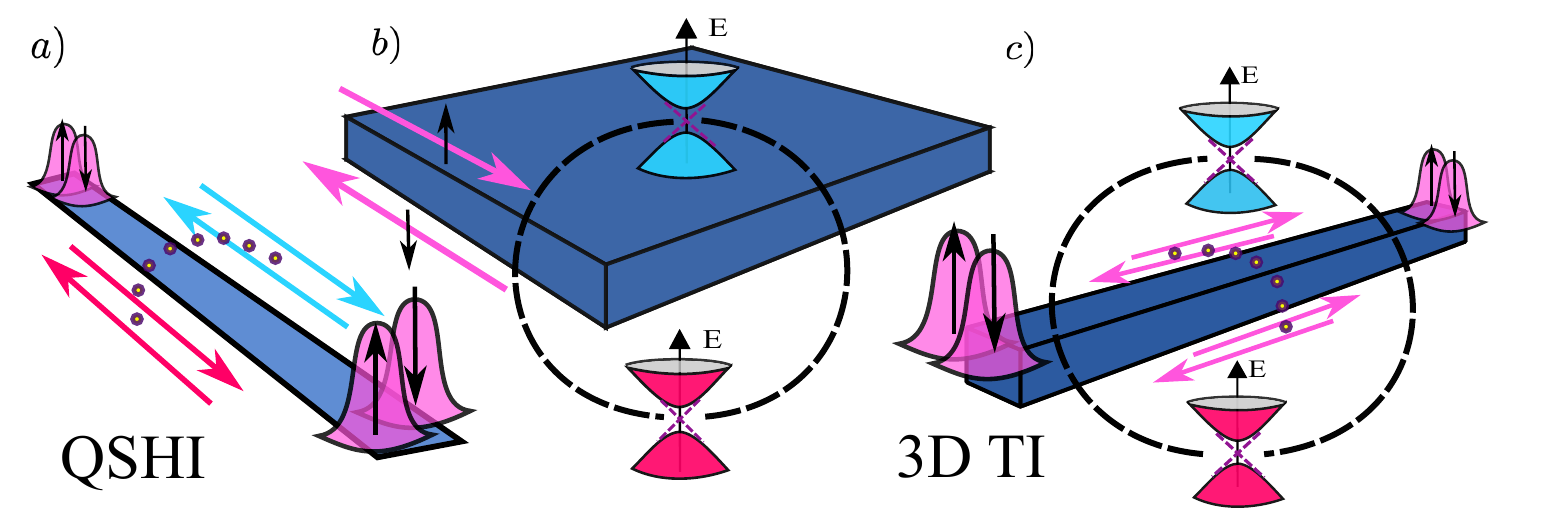}
  \caption{Schematic of the finite-size TRI systems studied. From left to right, a) QSHI wire, b) slab of 3D TI and c) 3D TI wire. Blue and red cones are schematic of the gap openings  of the 3D TI due to the hybridization of the the Dirac cones. Similarly blue and red helical edge states get hybridized (yellow/purple) in the QSHI wire and the finite-size quasi-1D edge states (blue and red) get hybridized (yellow/purple) for the 3D TI wire. Topological edge states (pink) are present as quasi-0D modes or quasi-1D modes polarized in spin, for wire or slab configurations respectively.}
  \label{system-diagram}
\end{figure*}

\begin{figure*}
 \includegraphics[width=0.8\textwidth,height=7cm]{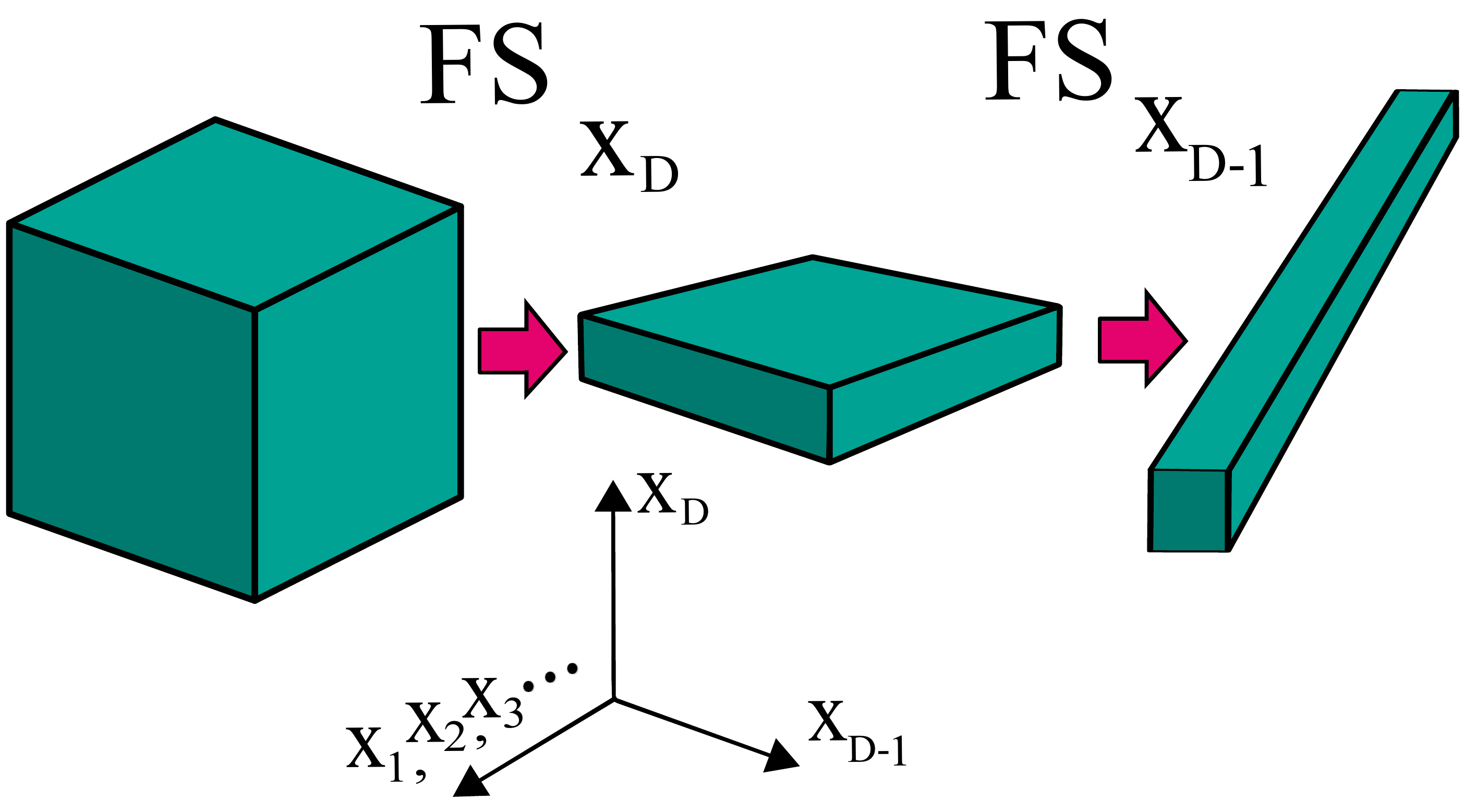}
  \caption{Schematic of the finite-size process for topological insulators. From left to right a $D$ dimensional phase gets shrunk in one direction $x_{D}$ to give rise to a quasi $D-1$ dimensional phase. This phase can be further shrunk in a remaining $x_{D-1}$ direction so that $x_1,x_2,\dots x_{D-2}$ are still periodic directions and now a quasi $D-2$ dimensional phase is realized.}
  \label{quasidim_schematic}
\end{figure*}



The discovery of the first topological insulator (TI), the quantum spin Hall insulator (QSHI) in HgTe quantum wells \cite{QSHI-HgTe-Theory,QSHI-HgTe-Exp} heralded a paradigm shift in condensed matter physics towards broad study of topological phases of matter. Understanding and characterization of topology is now central to the field, with major applications ranging from fault-tolerant quantum computing  \cite{Topo-quantum,KITAEV20032} to unconventional superconductivity \cite{Ersatz-fermi-SPT}. Consequently, searching for novel, experimentally-accessible topological systems is a major theme of the last few decades~\cite{QHE-VonKlitzing,FQH-exper,FQH-Laughlin,QSHI-HgTe-Exp,3DTI-exp2,noh2018topological,peterson2018quantized,imhof2018topolectrical,serra2018observation}. These efforts usually target experimental confirmation of a hallmark of topological phases known as bulk-boundary correspondence: a non-trivial topological invariant of the system bulk is associated with topologically-robust, gapless boundary states. While it has long been understood that a $D$-dimensional bulk topology yields $\left(D-1\right)$-dimensional gapless boundary states for most topological phases \cite{HOTI}, the recent discovery of additional bulk-boundary correspondence even in the canonical phases, known as finite-size topology \cite{fst_one}, shows this foundational aspect of topological physics is richer than previously-thought. If a system is characterized by a topological invariant computed in the $D$-dimensional infinite bulk, but is finite in size and thin in one direction as illustrated in Fig.~\ref{system-diagram} (for the QSHI $D=2$ while for the 3D TI $D=3$ ), such that topologically-protected boundary states interfere with one another, this \textit{quasi-$(D-1)$- or q$(D-1)$-dimensional} bulk is characterized by an additional topological invariant. When this additional invariant takes non-trivial values, open boundary conditions in a second direction yield an additional set of quasi-$(D-2)$-dimensional, topologically-protected boundary states localized on this boundary of the quasi-$(D-1)$-dimensional system. As these quasi-$(D-2)$-dimensional states are localized on the boundary in correspondence with a non-trivial value for a topological invariant of the quasi-$(D-1)$-dimensional bulk, and robust against local perturbations respecting the symmetries protecting the topological phase \textit{in the $D$-dimensional infinite bulk}, they constitute previously-unidentified topological phases of matter. \\

Following the previous thinning process we end up with a q$(D-1)$-dimensional bulk with topological edge states in one less dimension, the situation is then just like at the start of the program, but with $D$ replaced by $D-1$. Thus one may think of applying the thinning process once again, now thinning the $x_{D-1}$ dimension and hybridizing the previous q$(D-2)$ edge states. We then arrive at a q$(D-2)$ dimensional bulk with q$(D-2-1)$ dimensional edge states which again can be subjected to the same procedure. The general process is illustrated in Fig.~\ref{quasidim_schematic}, while Fig.~\ref{system-diagram} c) shows the specific case of the 3D TI q$(3-2)$ bulk. We note that this procedure could in principle be applied until there are no more number of dimensions to thin down.

Although theoretical discovery of the Chern insulator \cite{Haldane-model} preceded theoretical prediction of the TR-invariant QSHI \textit{derived from it}~\cite{kane2005quantum,kane2005z2}, experimental confirmation of the QSHI \cite{QSHI-HgTe-Exp} occurred within one year of the prediction, while more than two decades passed for the Chern insulator \cite{chern-discovery}. This reflects a broader trend in the field, of TR-invariant topological insulators being confirmed experimentally more quickly and easily than TR-symmetry-broken topological insulators reliant on engineering particular magnetic orders~\cite{graph-TI,Min2006IntrinsicAR,QSHI-HgTe-Exp}. Following this idea in order to more rapidly observe finite-size topology in experiment, we study the time-reversal invariant finite-size topology of the QSHI and the strong TI (STI), by considering these systems in geometries as shown in Fig.~\ref{system-diagram}. We also note that, due to the vast experimental studies in TI ultra-thin films \cite{TI-thin-film-conduc,TI-crossover,TI-thin-phase-trans}, Van der Waals heterostructures \cite{vanwaals-hetero,Nature-hetero-antiferr,TI-hetero,robust-2dhetero}, and transition-metal dichalcogenides in particular given their large spin-orbit coupling \cite{TMD-1,TMD-topo}, there may already be signs of finite-size topology in previous experiments. Past work, for instance, indicates few-layer 1T'-MoTe\textsubscript{2} is semi-metallic~\cite{Song_2018}, while the monolayer is predicted to be a quantum spin Hall insulator~\cite{qian_2014}, suggesting the few-layer topology derives from the Weyl semimetal phase of the three-dimensional bulk, while the monolayer topological phase has a distinct origin due to a strictly two-dimensional bulk. \\

Since finite-size topological phases occur for the Kitaev chain and Chern insulator \cite{fst_one}, non-trivial finite-size topology is expected for TR-invariant systems of the QSHI and STI given concrete relationships between Hamiltonians for these topological phases: the Kitaev chain Hamiltonian may be used to construct the Chern insulator Hamiltonian, if many chains are coupled forming a 2D system \cite{Kitaev_2001}, and a Chern insulator Hamiltonian and its time-reversed partner are the basis of Hamiltonians for the QSHI\cite{kane2005quantum,kane2005z2}. We find that FST extends to these TRI topological phases. As Hamiltonians for TR-invariant topological phases are used to construct Hamiltonians for other topological phases, these results also reveal that a larger set of topological phases harbor FST: a Weyl semimetal phase\cite{TI-to-Weyl} Hamiltonian may be constructed from magnetically-doped STI and trivial insulator thin films stacked alternatingly, while a stack of QSHIs corresponds directly to the weak 3D TI \cite{Moorehomotopy2007,Fu-Kane-Mele}. Topological crystalline phases may furthermore be constructed, for instance, as Chern insulators within mirror subsectors or with the Chern insulator bulk confined to a mirror-invariant plane of a three-dimensional Brillouin zone \cite{Mirror-chern}. More generally, topological crystalline phases are characterized by considering symmetry-protection by crystalline point group symmetries \textit{in addition to} the internal symmetries of the ten-fold way. On a technical level this is accomplished by expressing the Hamiltonian in a block diagonal form using the additional symmetry, in each sub-sector internal symmetries are still present and thus can be analyzed by classification schemes obtained from the ten-fold way \cite{10-fold-way,Kitaev-K-theory-10fold}. 

In this manuscript, we first, section \ref{QSHI wire}, characterize finite-size topology in a QSHI wire. We start by considering a thin QSHI system with one open thin dimension and one infinite periodic dimension. The energy and Wilson loop spectra of this q(2-1)D system reveal that the non-trivial zones in phase space are a subset of the original 2D bulk topological regions. Furthermore opening boundary conditions again in the remaining periodic direction shows the presence of edge states localized on the q(2-2)D boundaries. We end this section by studying the response to on-site disorder and perturbations, where the robustness of the edge states indicates a link to the original 2D bulk gap. Afterwards in section \ref{WTe2-section}, we consider a more realistic and experimentally accessible system $1T'$ $\text{WTe}_2$ in the QSHI phase. We find similarly that this material realizes a finite-size topological phase with topologically-robust q(2-2)D edge states for a sawtooth ribbon geometry, the presence of this edge states is again verified to be predicted by a non-trivial Wilson loop spectrum. Extending our analysis to the 3D case we consider in section \ref{3D TI slab} the STI in a q(3-1)D slab geometry. In this case, interference between the STI Dirac cone surface states yields q(3-2)D edge states. We show the Wilson loop spectrum of the q(3-1)D bulk displays topologically non-trivial signatures in correspondence with these boundary states, indicating Wilson loop spectra are a robust bulk diagnostic of finite-size topology. Additionally we compute the magneto-electric polarizability, which should be trivially zero if the system is just a 2D QSHI, instead we encounter the response expected for the infinite 3D TI bulk. This central result allows us to contemplate the idea of detecting topological signatures of higher dimensional phases, say the 4D TI, in quasi lower dimensional systems.  Finally, we study the case of a STI in a q(3-2)D wire geometry, where we once again use the Wilson loop indicator to find a novel bulk-boundary correspondence restricted to a subset of the original 3D topological phase diagram. In this final case the number of edge states is seen to follow the number of $\pm \pi$ phases such that only even numbers of distinct edge states appear. In section \ref{Conclusions} we summarize our results and present some concluding remarks.

\section{QSHI wire} \label{QSHI wire}

\begin{figure}[ht]
  \centering
  \subfigure[]{\includegraphics[width=0.49\columnwidth]{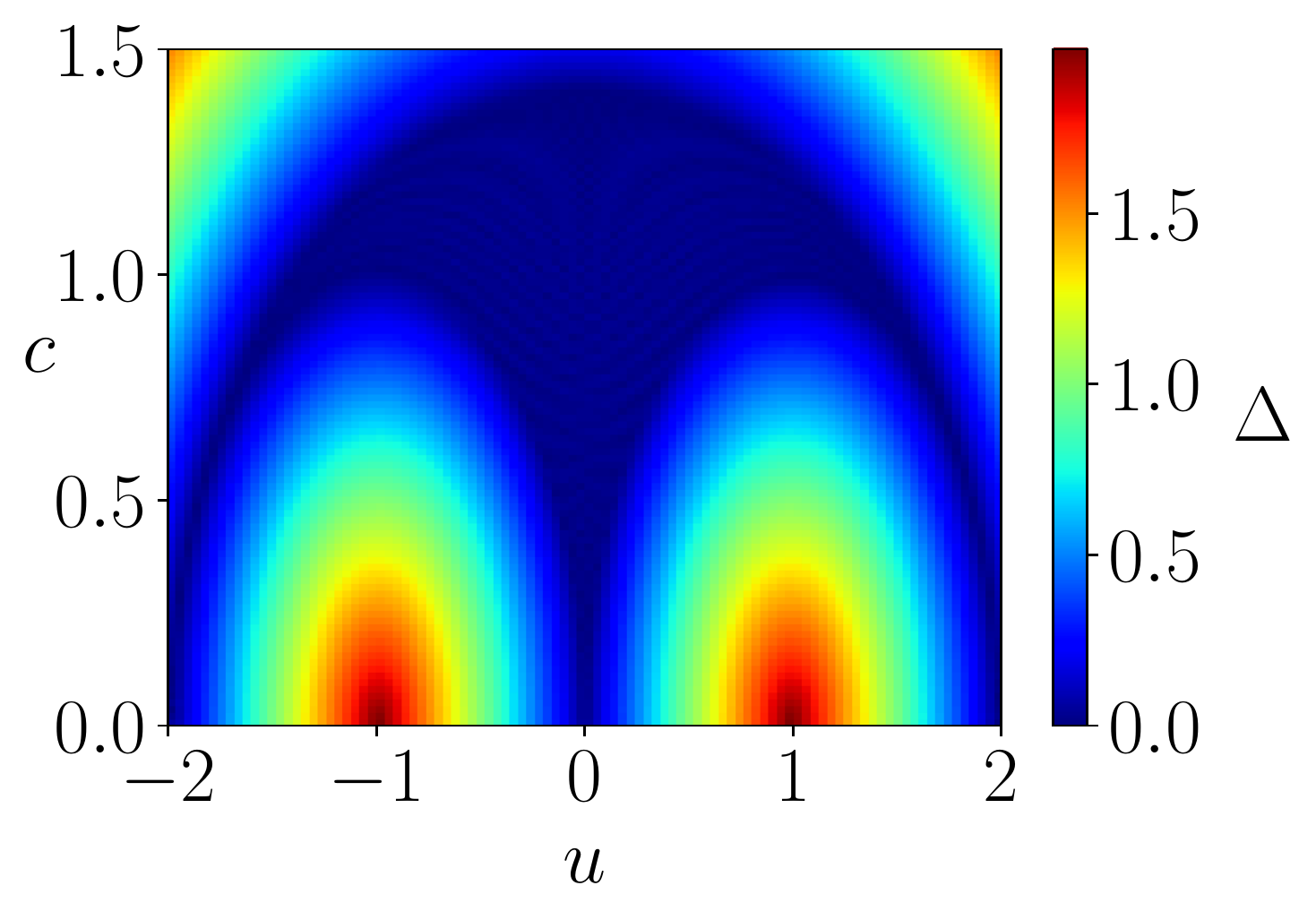}}
 \subfigure[]{\includegraphics[width=0.49\columnwidth]{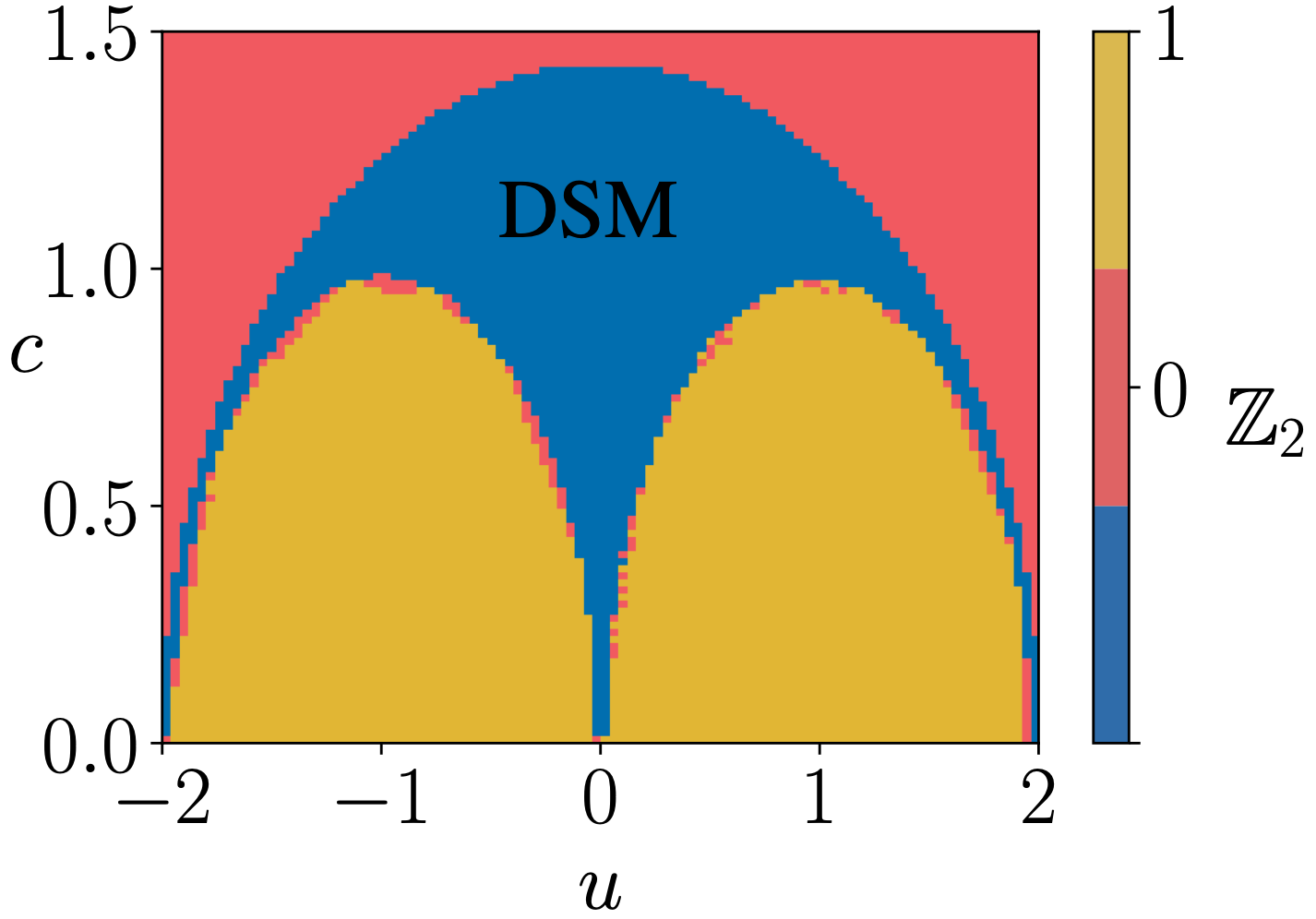}}
  \subfigure[]{\includegraphics[width=0.49\columnwidth]{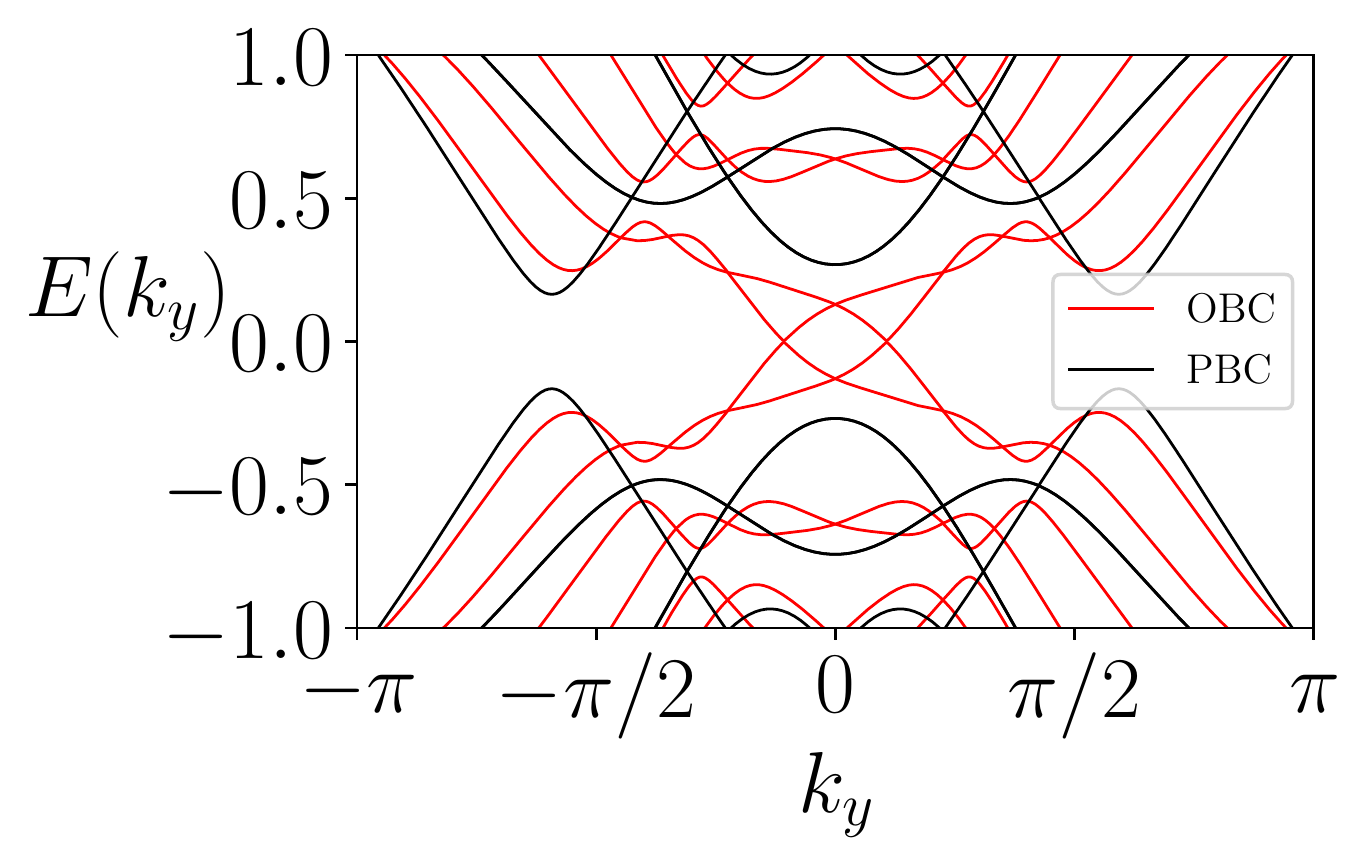}}
  \subfigure[]{\includegraphics[width=0.49\columnwidth]{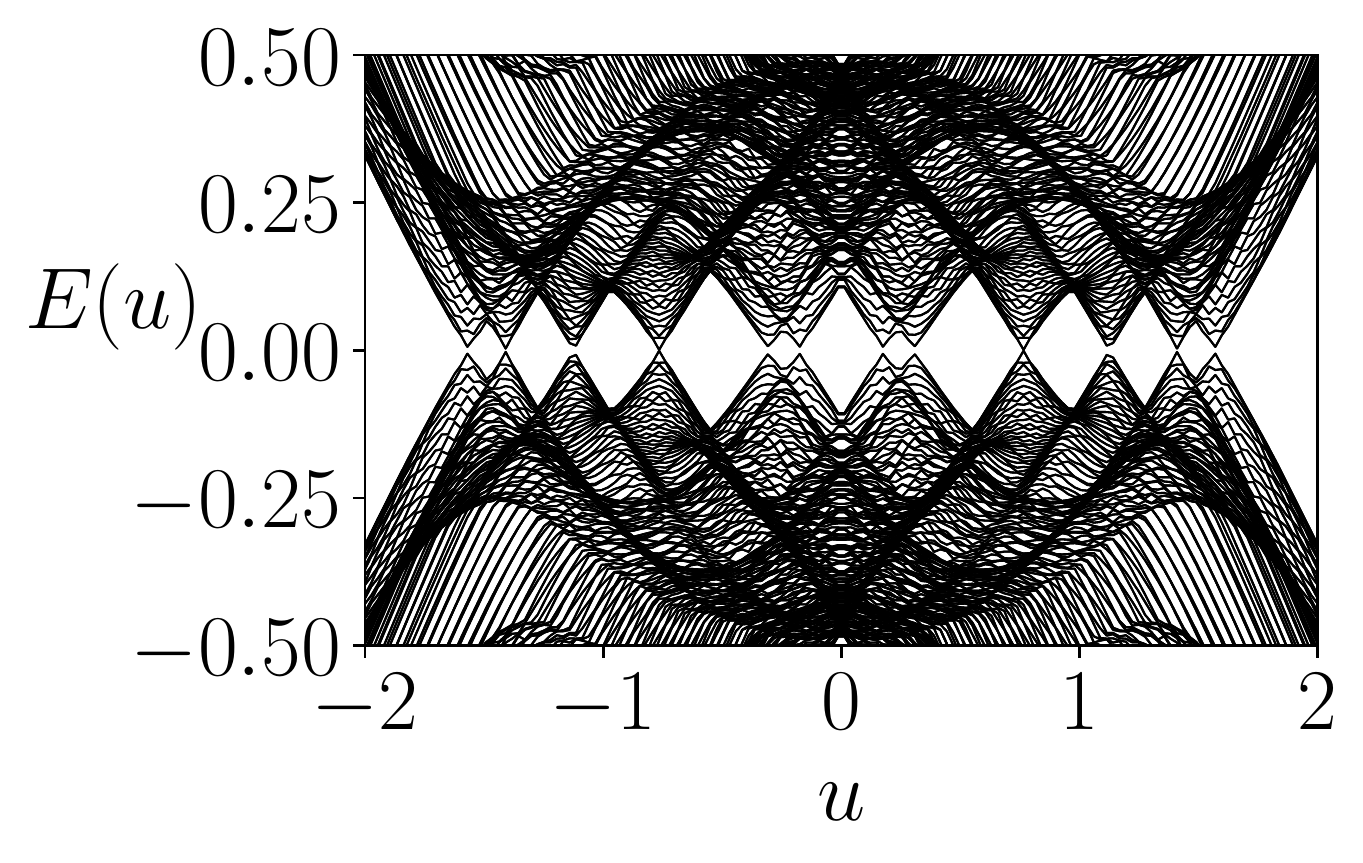}}
  \vspace{-0.3cm}
  \caption{ a) Direct gap heat plot of the 2D bulk hamiltonian eq. \eqref{QSHI} as a function of potential $u$ and spin-orbit coupling constant $c$, b) Topological phase diagram of the 2D bulk , showing the QSHI phase (yellow) and DSM gapless phase (blue) of the 2D bulk. c) Quasi-1D dispersion for PBC in $y$ and OBC (PBC) in $x$ with $N=6$ sites. The parameters for the gap closing with OBC in $x$ are  $u=0.76, c=0.8,t=1/2$
 d) Spectrum for PBC in $y$ as a function of the staggered potential $u$ and $c=0.8, t=1/2$.}
  \label{quasi1D-PBC-QSHI}
\end{figure}

 As a starting point of our analysis we consider a QSHI first considering the canonical Bernevig-Hughes-Zhang Hamiltonian for HgTe quantum wells \cite{QSHI-HgTe-Theory} where we also add a Rashba-type spin orbit coupling. Thus the Hamiltonian in momentum space has the form \cite{QSHI-ham}:
\begin{align}
h(k_x,k_y)=& (u+2t(\cos k_x+\cos k_y))\sigma_z+\sin{k_y}\ \sigma_y \label{QSHI} \\
+& \sin{k_x} \ s_z\sigma_x+c\ s_x\sigma_y, \notag
\end{align}
where $s_i,\sigma_i$ are Pauli matrices in spin and orbital space respectively. For simplicity we omit the identity in spin space and denote the tensor product by placing two matrices next to each other. The real number $u$ corresponds to a staggered potential, $t$ to a hopping parameter and $c$ is the spin orbit coupling that breaks $s_z$ spin symmetry. The phase diagram for this Hamiltonian includes both a region in which the QSHI phase is realized and a region in which the Dirac semimetal (DSM) phase is realized, as discussed in reference \cite{QSHI-ham} and plotted in Fig. \ref{quasi1D-PBC-QSHI} a) and b) . In the following analysis, we first consider the QSHI regime, and then that of the DSM. \\

Next we consider what happens if we open boundary conditions (OBC) in the $x$ direction for a small number of lattice sites $N$. Since the helical edge modes of the QSHI are not completely localized at the boundary, but instead decay exponentially into the bulk \cite{edgestates-QSHI}, these boundary states interfere in systems of finite-width. The lattice second quantized Hamiltonian with open boundary conditions in the $\hat{x}$-direction and periodic boundary conditions in the $\hat{y}$ direction is:
\begin{align}
    \hat{H}&= \sum_{k,n} \Psi_{k,n}^\dagger  \left( (u+2t\cos{k})\sigma_z+\sin{k}\ \sigma_y+c\ s_x\sigma_y \right)\Psi_{k,n}\notag\\
    &+\Psi_{k_y,n+1}^\dagger \left(t\ \sigma_z+\frac{i}{2}s_z\sigma_x \right)\Psi_{k,n}+\text{h.c.}\ ,
    \label{QSHI-q1D}
\end{align}

where $k\equiv k_y$ and $n$ runs over the $N$ sites of the open $x$ direction. Here, $\Psi_{k,n}$ are four component spinor fermion operators acting on the spin and orbit degrees of freedom. We first examine the spectrum of Eq.~\ref{QSHI-q1D} for small $N$ on the order of a few lattice constants. We first consider the spectrum for a particular point in phase space as shown in Fig.~\ref{quasi1D-PBC-QSHI} c) for a small system with $N=6$ with non-trivial 2D bulk invariant, as shown in Fig.~\ref{quasi1D-PBC-QSHI} b). Comparing the dispersion for the system with periodic boundary conditions in each direction (black lines) to that of the system with open boundary conditions only in the $\hat{x}$-direction (red lines), we see the periodic system is gapped, while the system with open boundary conditions is instead gapless. While gapless boundary states are expected due to the non-trivial bulk topology, the gaplessness in this case is not topologically-robust: the gapless boundary modes interfere in finite-size systems to open a hybridization gap in general. Under certain conditions, however, the boundary modes interfere destructively, corresponding to a fine-tuned gapless state when hybridization matrix elements pass through zero. Examining the spectrum for the q(2-1)D bulk as a function of $u$ as shown in Fig.~ \ref{quasi1D-PBC-QSHI} d) , we see this more general pattern of finite interference gaps, with a discrete set of $u$ corresponding to gap-closings and destructive interference between the helical boundary modes. We will show these gap-closings can correspond to topological phase transitions, and some of these gapped regions host finite-size topological phases.\\

\subsection{Periodic system}

\begin{figure}[ht]
    \subfigure[]{\includegraphics[width=0.49\columnwidth]{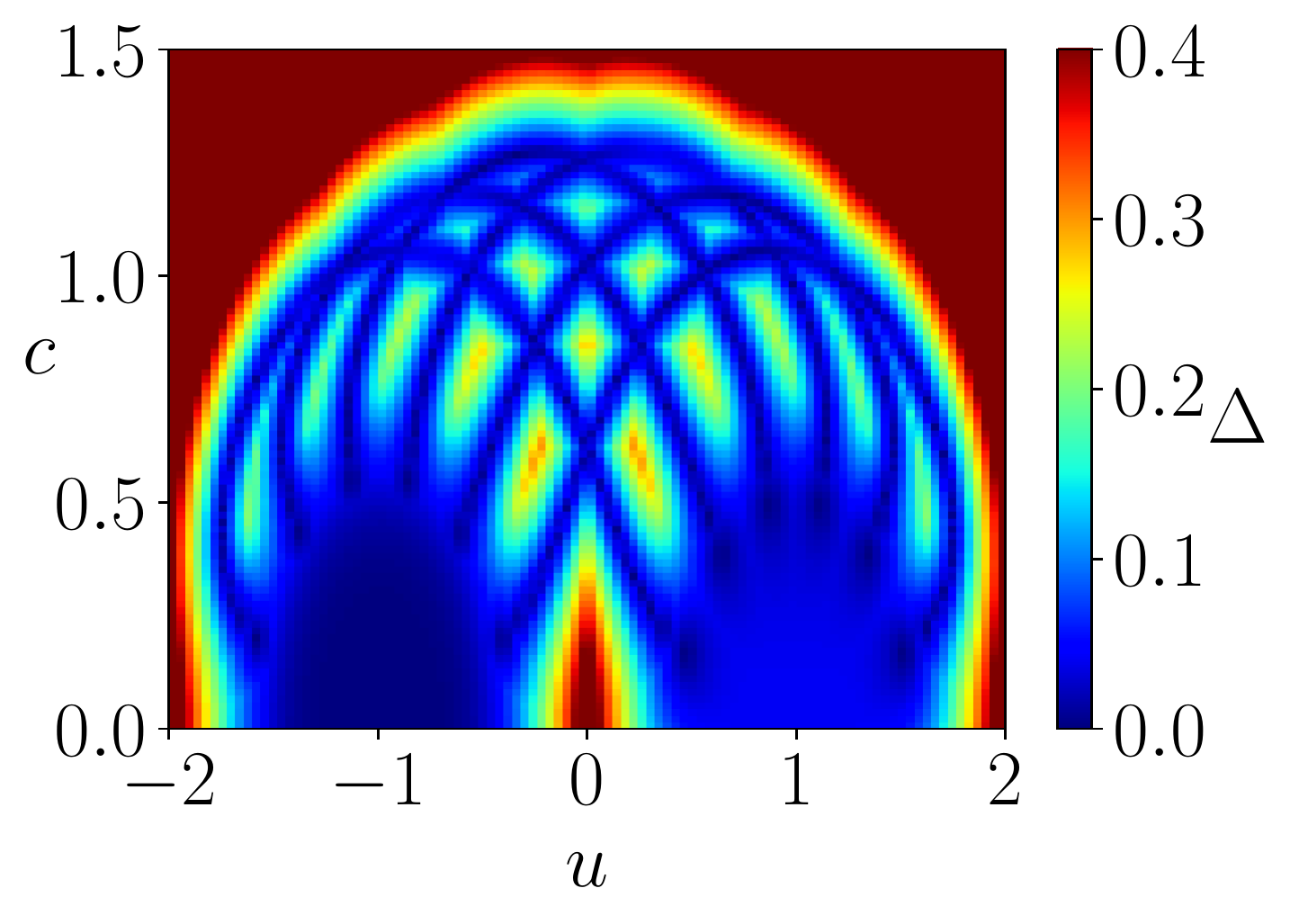}}
    \subfigure[]{\includegraphics[width=0.49\columnwidth]{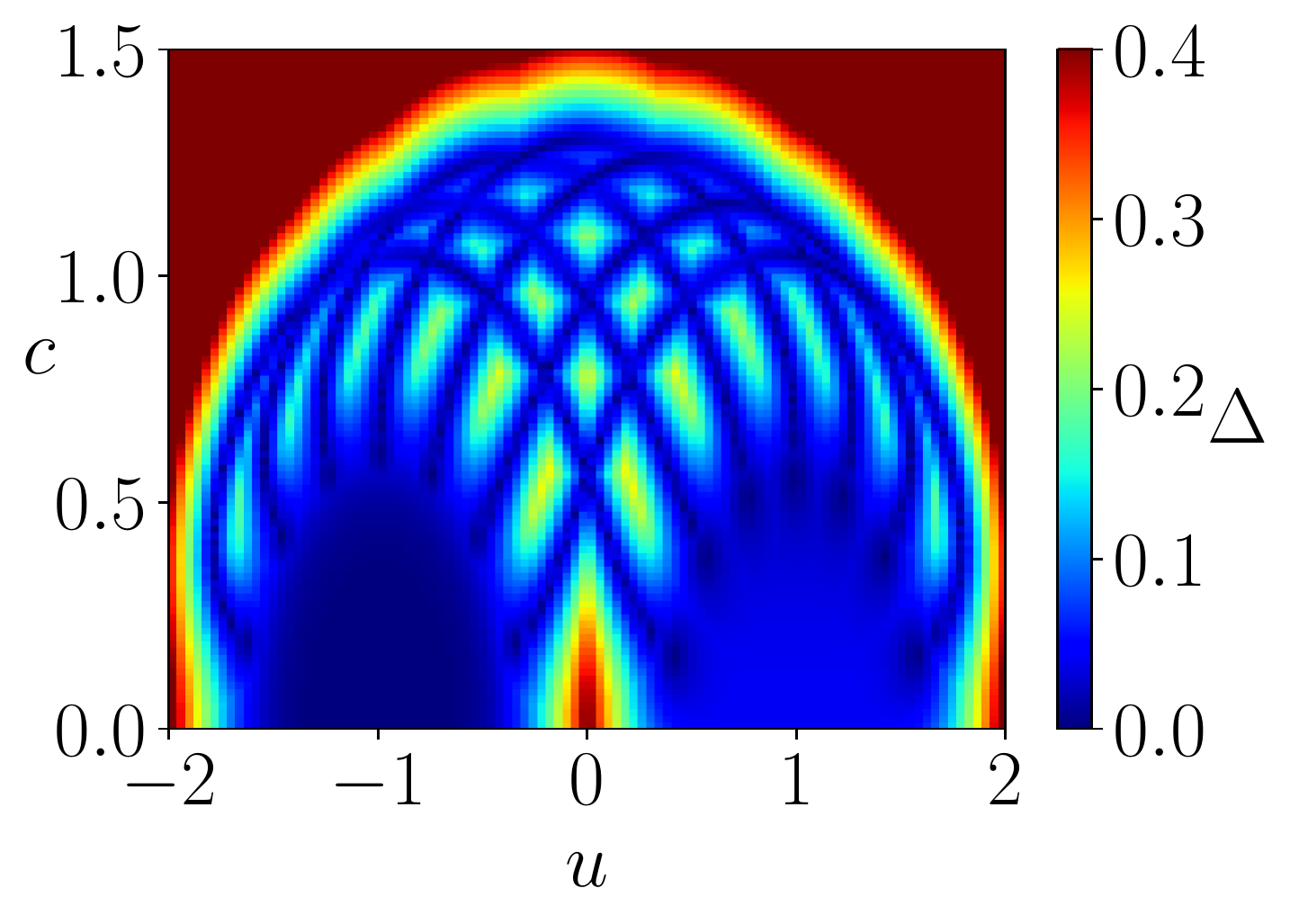}}
    \subfigure[]{\includegraphics[width=0.49\columnwidth]{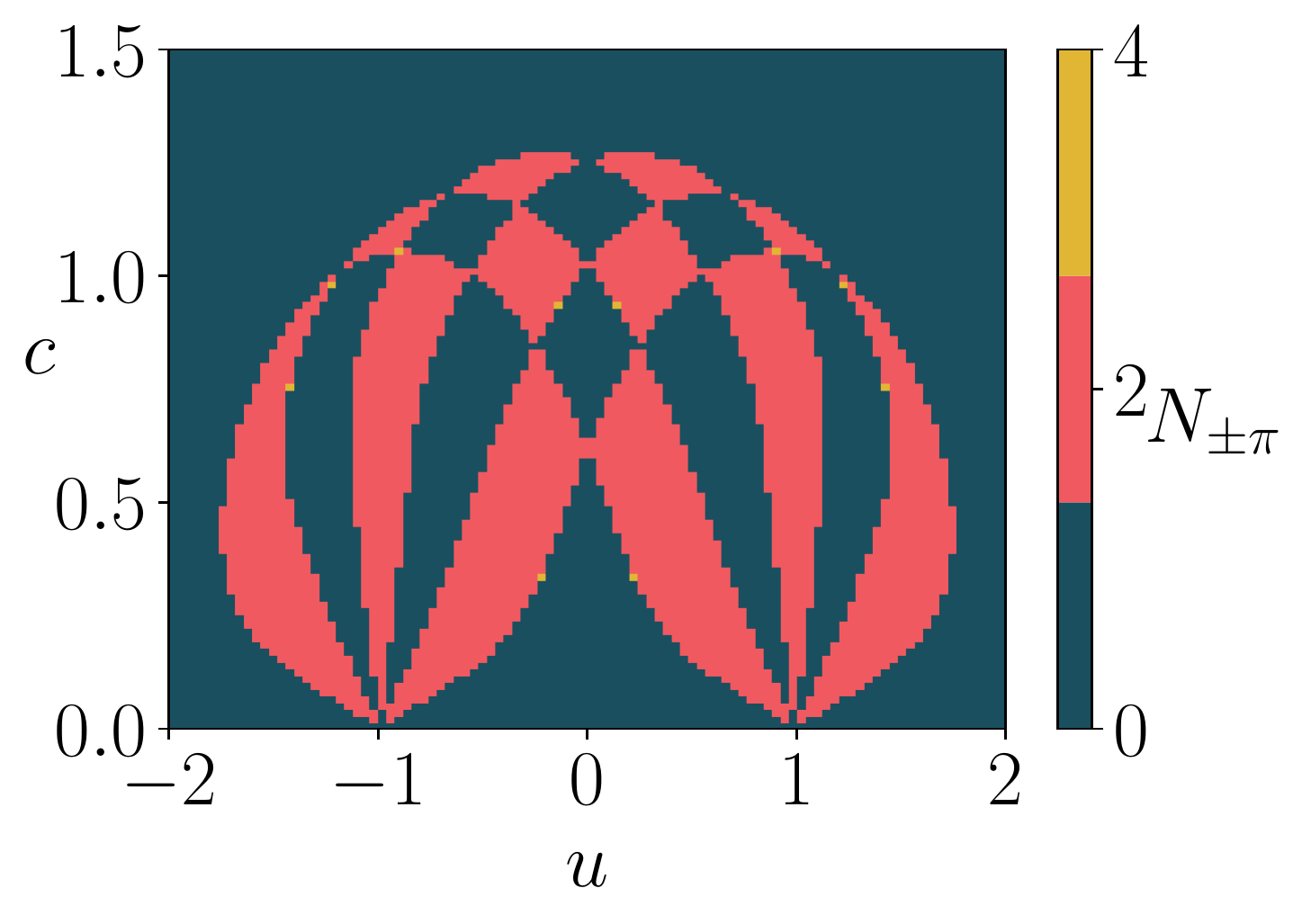}} 
    \subfigure[]{\includegraphics[width=0.49\columnwidth]{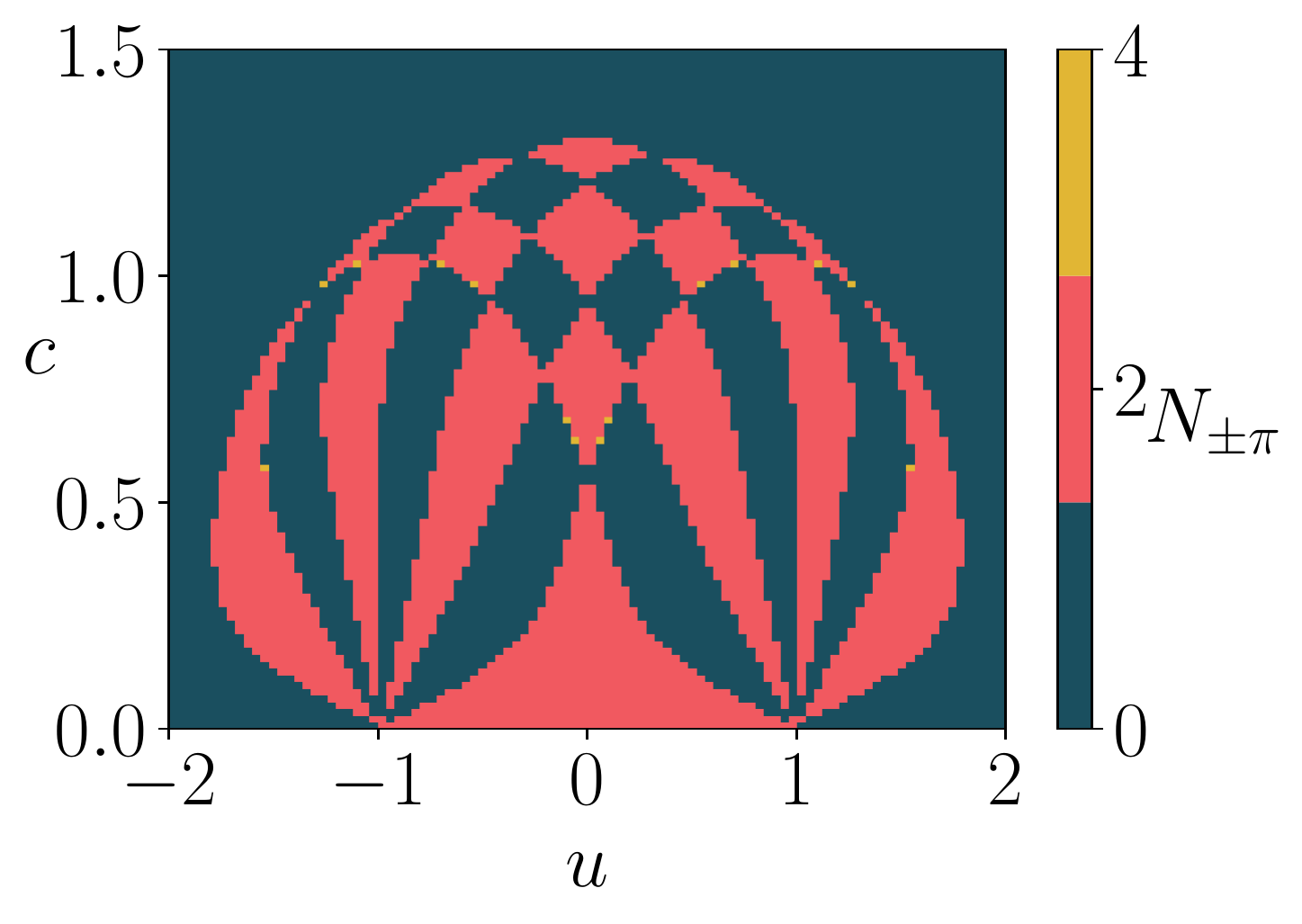}} 
  \vspace{-0.3cm}
  \caption{Quasi-(2-1)D minimum direct bulk gap for a) $N=6$, b) $N=7$ and $t=1/2$. Plot of the number of $\pm \pi$ phases $N_{\pm \pi}$ (red is 2, black 0) in the Wilson loop eigenvalues for c) $N=6$, d) $N=7$}
  \label{quasi1D-phasediagram}
\end{figure}

To characterize the finite-size topological phases of this time-reversal invariant system, we now re-interpret the original model with OBC in the $\hat{x}$-direction as a q(2-1)D bulk, and characterize topology of this q(2-1)D bulk system similarly to characterization of a $d$-dimensional bulk. We therefore first compute a phase diagram for the minimum direct gap over the Brillouin zone of the q(2-1)D bulk as a function of $u,c$ for fixed hopping $t=1/2$, shown in Fig.\ref{quasi1D-phasediagram} a),b) for $N=6$ and $N=7$ layers in the $\hat{x}$-direction, respectively. A dome forms in the phase diagram, consisting of a set of curved, stripe-like regions of finite minimum direct gap separated by lines along which the q(2-1)D minimum direct bulk gap is zero, with these lines  intersecting to form a checkerboard-like pattern at larger values of $c$. As the number of lattice sites in the $\hat{x}$-direction increases, the number of gap-closing lines increases while the regions of finite minimum direct gap decrease in size. This pattern is consistent with a picture of gap-closings due to interference between the helical boundary modes of the QSHI: the boundary modes in this q(2-1)D system possess a standing wave character, and the gap-closing lines correspond to hybridisation matrix elements passing through zero with tuning of system parameters. With increasing system size, this interference pattern becomes denser as the difference in wavelength between the oscillatory components of the helical boundary modes generically decreases.

It is particularly interesting to compare these phase diagrams for the q(2-1)D bulk with the counterpart phase diagram of the 2D bulk shown in Fig.~\ref{quasi1D-PBC-QSHI} a),b) , which reveals that different kinds of topological phases of the 2D bulk (and corresponding different gapless boundary states) yield different interference patterns as a function of $u$ and $c$. Notably, the checkerboard region of the phase diagram corresponds to the DSM phase region of the corresponding phase diagram for the 2D bulk, revealing that the DSM phase is generally gapped out in the q(2-1)D regime, and exhibits more complex interference pattern than does the QSHI.\\

As the 2D minimum direct bulk gap remains finite over the region of the phase diagram where we observe this interference pattern between helical boundary modes of the QSHI, and the 2D minimum direct bulk gap remains closed due to topologically-protected band-touchings of the DSM, topological invariants of the 2D bulk do not change within these regions. However, as subsets of each of these regions possess a finite minimum direct gap in the q(2-1)D spectrum, it is possible to further characterize the topology of the q(2-1)D system if suitable topological invariant(s) are identified. To further characterize finite-size topology of this quasi-1D TRI system with Wilson loop spectra, we compute the Wilson loop eigenvalues~\cite{Wilsonloop-inversion}, which distinguish between topologically-distinct phases of matter as they characterize holonomy in a system due to parallel transport through non-contractible loops in the $\text{BZ}$ \cite{Leone_2011,geo-compu,holo-atoms}. 

The Wilson loop spectra for the q(2-1)D system are computed by integrating \red{the} Berry connection over the remaining $k\equiv k_y$ momentum coordinate, using the following expression: \cite{Wilsonloop-inversion}

\begin{align}
 \displaystyle \mathcal{W}=\mathcal{P}e^{\textstyle -\int_{-\pi}^{\pi} d k A(k)}, \label{Wilson}
\end{align}

where $A(k)$ is the non-Abelian Berry connection over the occupied bands and $\mathcal{P}$ is the path ordering operator. Since we compute the Wilson matrix for a tight binding system we discretize Eq.~\eqref{Wilson}. The set of Wilson loop eigenvalue phases is the Wannier charge center spectrum characterizing polarization. In a topologically non-trivial phase, Wannier charge center(s) are fixed to value(s) of $\pm \pi$, so we compute the number of these non-trivial phases as $N_{\pm \pi}$. 
The phase diagrams characterizing $N_{\pm \pi}$ vs. $u$ and $c$ are shown in Fig.~\ref{quasi1D-phasediagram} c), d) for systems with $N=6$ or $N=7$ layers in the $\hat{x}$-direction, respectively.

These $N_{\pm \pi}$ vs. $c$ and $u$ phase diagrams shown in Fig.~\ref{quasi1D-phasediagram} c) and d) reveal alternating regions of $N_{\pm \pi} = 0$ and $N_{\pm \pi} = 2$, indicating the system undergoes a variety of topological phase transitions. We even observe stripe-like regions at smaller $c$, which intersect to form checkerboard patterns at larger $c$. These lines across which $N_{\pm \pi}$ changes in value are in direct correspondence with lines shown in Fig.~\ref{quasi1D-phasediagram} a) and b), respectively, along which the q(2-1)D minimum direct gap goes to zero. Taken together, these phase diagrams in Fig.~\ref{quasi1D-phasediagram} reveal a topological phase transition occurs every time the q(2-1)D minimum direct bulk gap goes to zero.

The phase diagrams for $N=6$ layers differ dramatically from those for $N=7$ layers, reflecting the dependence of this topology on finite-size effects. From the plots, one can see that, as the number of layers in the $\hat{x}$-direction increases, the number of topologically-distinct regions also increases in agreement with the number of lines along which the q(2-1)D minimum direct bulk gap is zero. The topological phase diagram of the 2D bulk Hamiltonian \eqref{QSHI} studied in Ref. \cite{QSHI-ham} is therefore being further divided into topologically-distinct regions in the q(2-1)D regime in a strongly $N$-dependent manner, revealing that topological phase transitions due to finite-size topology \cite{fst_one}, may occur without the minimum direct gap of the 2D bulk going to zero. \\


\subsection{Bulk-boundary correspondence and disorder}

\begin{figure}
    \subfigure[]{\includegraphics[width=0.49\columnwidth]{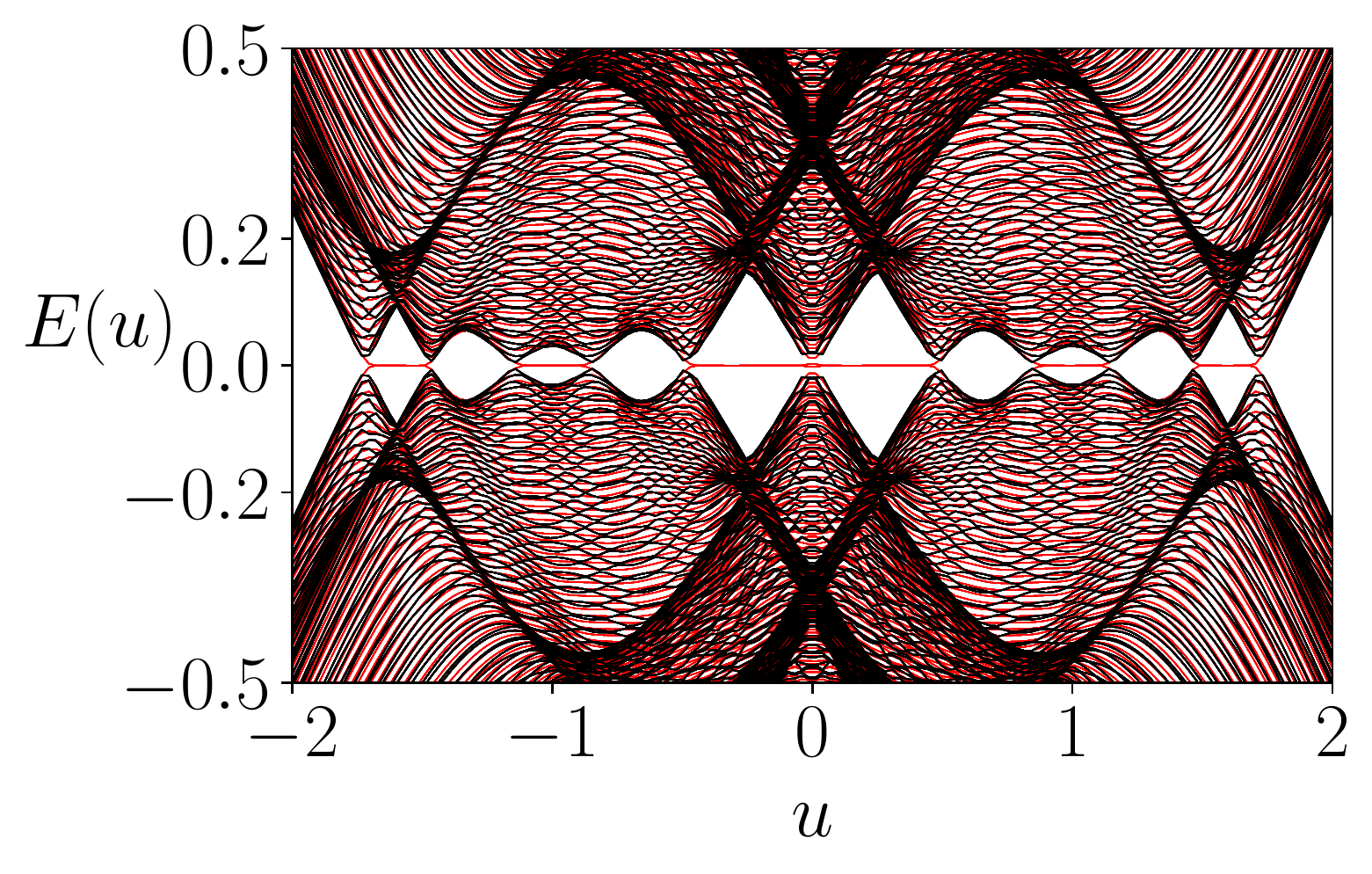}}
     \subfigure[]{\includegraphics[width=0.49\columnwidth]{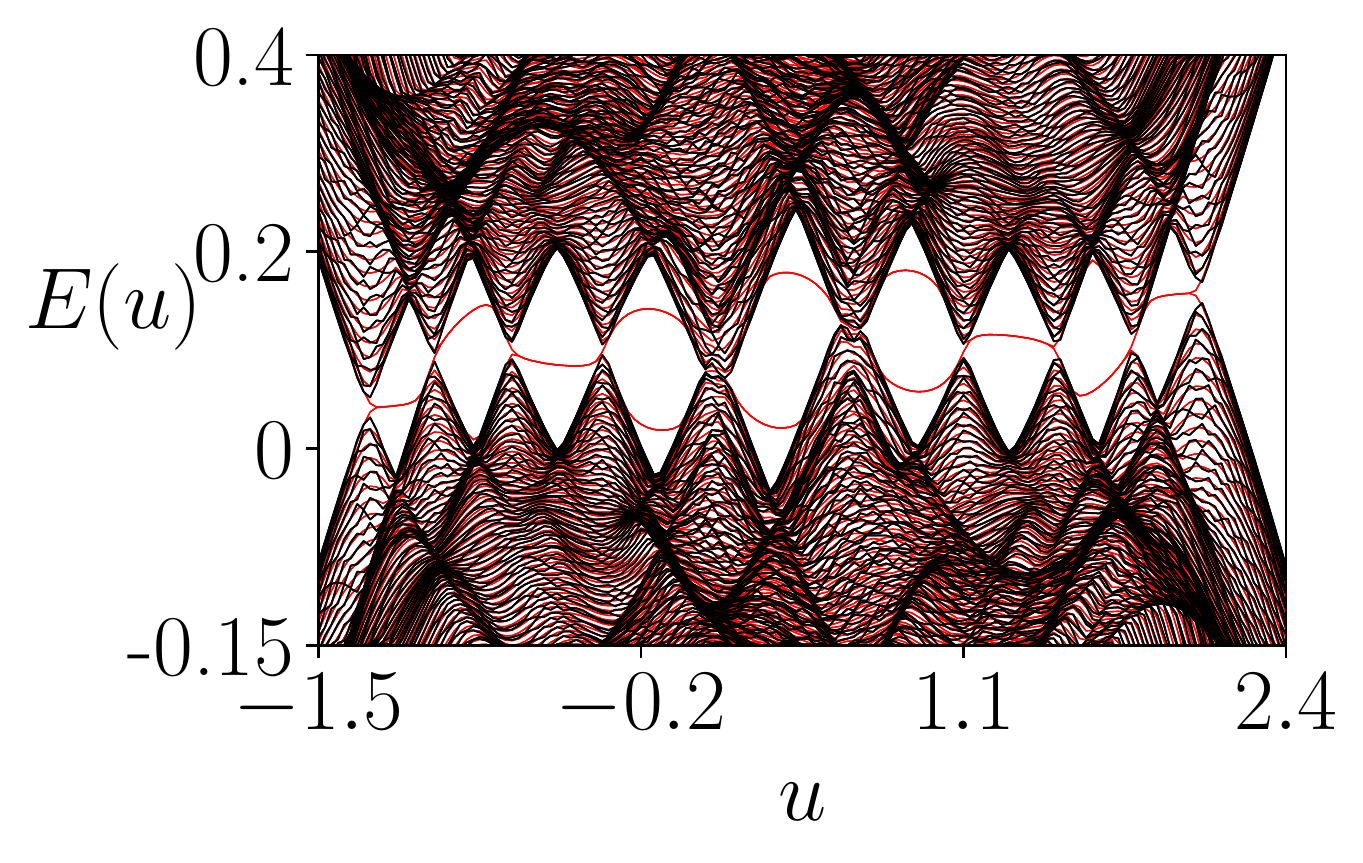}}
     \subfigure[]{\includegraphics[width=0.49\columnwidth]{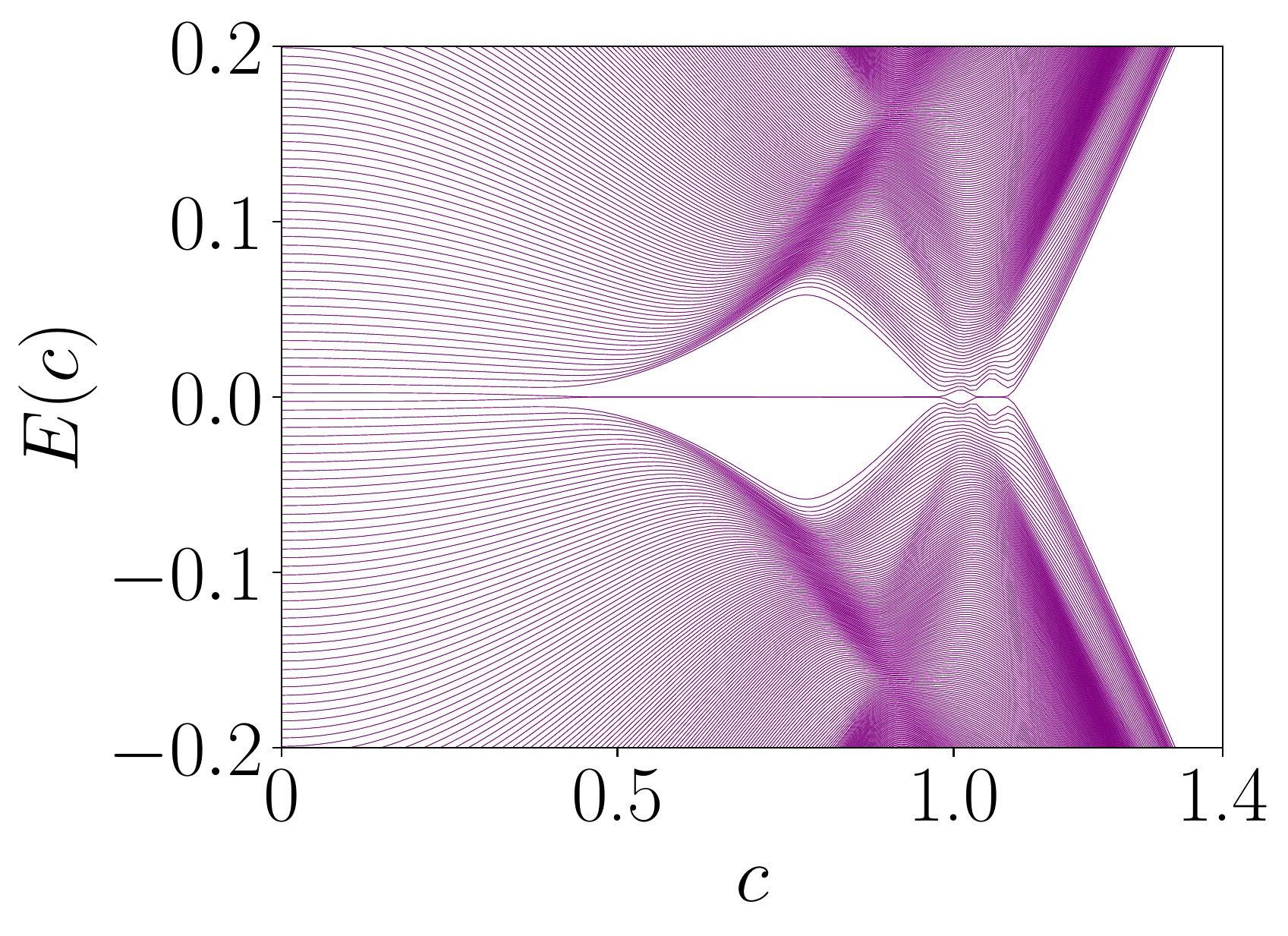}}
    \subfigure[]{\includegraphics[width=0.49\columnwidth]{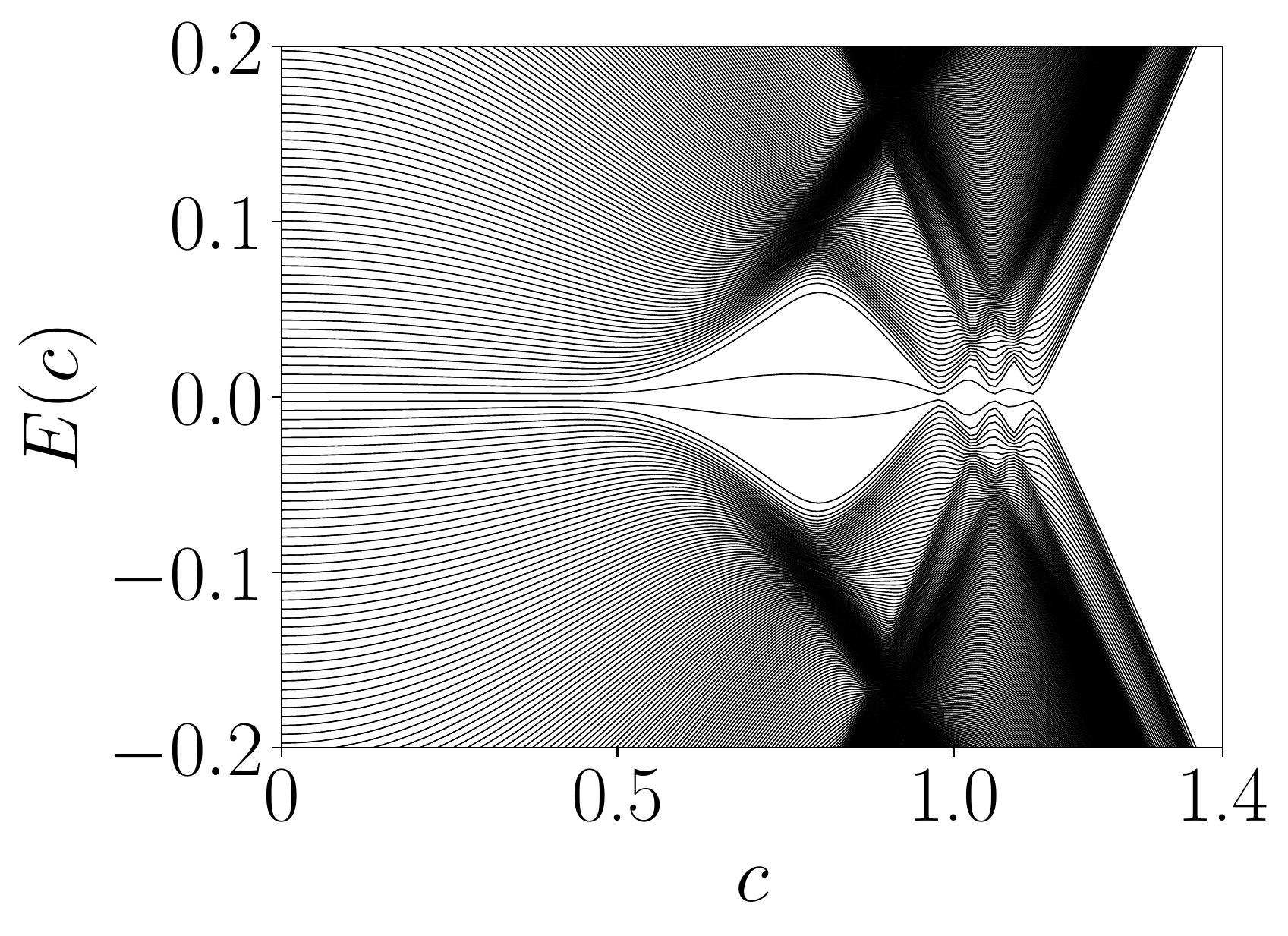}}
    \subfigure[]{\includegraphics[width=0.49\columnwidth]{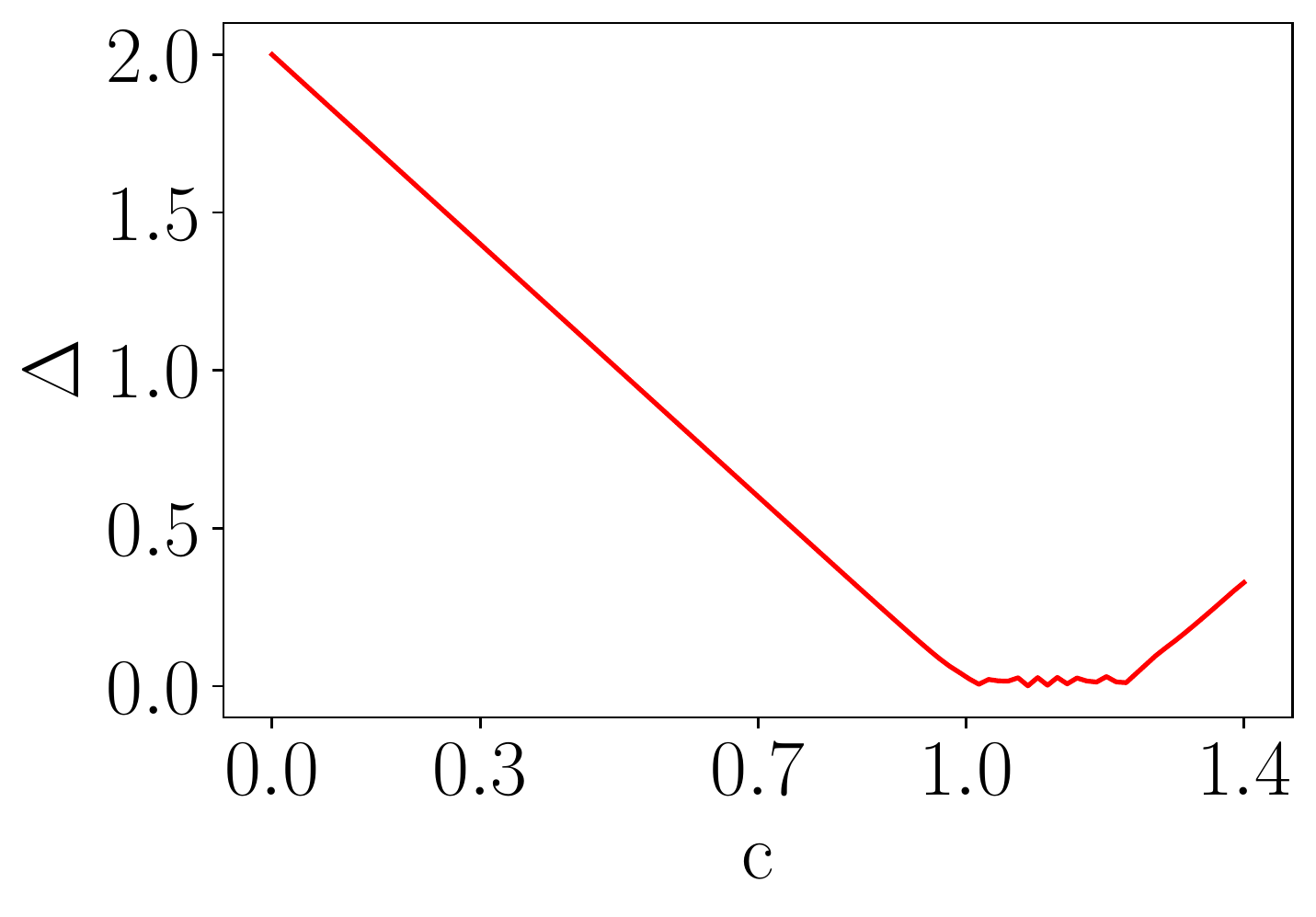}} 
    \subfigure[]{\includegraphics[width=0.49\columnwidth]{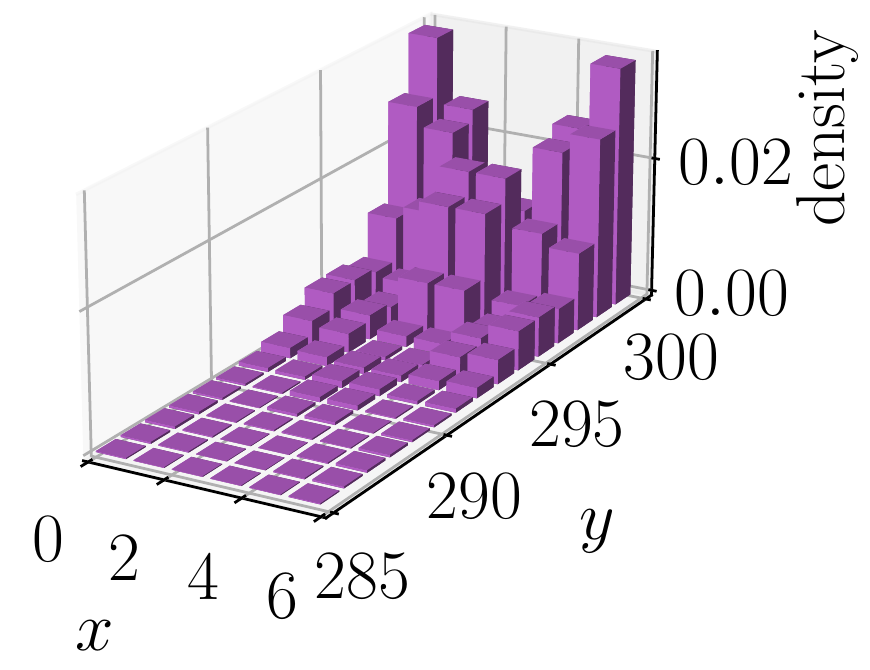}} 
    
  \vspace{-0.3cm}
  \caption{a) Spectrum for OBC (red)in both directions, PBC in $y$ (black)  as a function of the staggered potential $u$ and $c=0.6, t=1/2, N=6$.  b) Spectrum for OBC (red) in both directions, PBC in $y$ (black) as a function of the staggered potential $u$ and $c=0.8, t=1/2,N=6$ with a particle-hole symmetry and sublattice symmetry breaking on-site potential $0.1\cos(2\pi n/N) \sigma_x$. c) Disorder-averaged spectrum for $N=6$ sites and $u=1,t=1/2$ as a function of $c$ for 200 uniformly distributed random particle hole symmetric potentials of strength $\kappa=0.5u$. d) Disorder-averaged spectrum with the same parameters but for $200$ particle-hole symmetry-breaking disorder potentials of strength $\kappa=0.2u$. e) 2D minimum direct bulk gap as a function of $c$ for $u=1$. f)Density profile of q(2-2)D state for the same number of sites, $c=0.8,u=1.0$ and $L_y=300$.}
  \label{QSHI-disorder-states}
\end{figure}

 Having characterized finite-size topology of the q(2-1)D bulk of Hamiltonian Eq.~\eqref{QSHI} with open boundary conditions in the $\hat{x}$-direction and periodic in $\hat{y}$, we now explore the additional bulk-boundary correspondence of finite-size topological phases. We first study the spectral signatures of this bulk-boundary correspondence that appear for non-trivial Wilson loop spectra in accordance with the modern theory of polarization of Ref.~\cite{Pol-vanderbilt}. $N_{\pm \pi} \neq 0$ for the q1D bulk corresponds to topologically-protected, q0D bound states for open boundary conditions in the $\hat{y}$-direction in addition to open boundary conditions in the $\hat{x}$-direction. With system size in the $\hat{y}$-direction of $L_y$, such a bulk-boundary correspondence characterized in the q1D bulk by $N_{\pm \pi}$ is clear for $L_y \gg L_x$ as shown in Fig.~\ref{QSHI-disorder-states} a) b). In this case, one finds in-gap states close to zero energy within the q(2-1)D bulk gap of the energy spectrum. The separation in energy between these states decreases exponentially to zero as a function of $L_y$, realizing a four-fold degenerate manifold of zero-energy states. Such states are not present for periodic boundary conditions in the $\hat{y}$-direction, further indicating they appear as a consequence of bulk-boundary correspondence for the finite-size topological phase.\\

To further explore the extent to which these in-gap states are due to an additional bulk-boundary correspondence of finite-size topological phases, we compute the probability density for these in-gap states. We find these states are localized at the boundaries of the q1D system as shown in Fig.~\ref{QSHI-disorder-states} f). Probability density peaks at the corners of the system as seen in Fig.~\ref{QSHI-disorder-states} f) and decays over fifteen to twenty unit cells to zero for $x$ approaching the q(2-1)D bulk. As the system is time-reversal invariant, we also compute the spin polarization for these q(2-2)D boundary modes. We find the q(2-2)D boundary modes are spin-polarized in the $\pm \hat{z}$ direction, with a spin up/down pair localized at each end of the q(2-1)D system.

We also study the robustness of the in-gap q(2-2)D boundary states against symmetry-breaking due to disorder. We introduce disorder in the q(2-1)D system for OBC as a uniform random potential, with strength $\kappa$. The disorder average spectrum for $200$ disorder realizations of the form $\kappa_n s_0 \sigma_z$ is shown in Fig.~\ref{QSHI-disorder-states} c)  as a function of spin-orbit coupling $c$ for $\kappa_n \in (-0.2 u ,0.2 u)$. The q(2-2)D states are present over a wide range in $c$. Surprisingly, they survive even for disorder strengths $\kappa_n$ greater than the q(2-1)D minimum direct bulk gap for the QSHI phase. The edge states may move away from zero energy, if the perturbation breaks particle-hole symmetry, such effect is shown in \red{Fig.~\ref{QSHI-disorder-states} b)} as a function of $u$, where an onsite perturbation $0.1\cos(2\pi n/N) s_0\sigma_x$ is present. For disorder effects we considered a term of the form $V_i s_0 \sigma_0$, the spectrum is thus shown in Fig.~\ref{QSHI-disorder-states} d) where the edge modes deviate from zero energy, as in the case of an SSH chain with next nearest neighbours studied in reference \cite{SSH-NNN}. However, the q(2-2)D boundary modes persist and remain strongly localized, with their pair degeneracy preserved.\\

Interestingly, the topological invariant remains fixed at non-trivial values even when a particle-hole breaking term is present in the Hamiltonian. This reflects the dependence of this non-trivial topology on the presence of the topologically-protected boundary states of the higher-dimensional phase, which require only time-reversal symmetry to remain robust up to the minimum direct 2D bulk gap going to zero. We further find $N_{\pm\pi}$ still predicts the existence of edge states in the nontrivial region. The validity of the invariant and bulk-boundary correspondence in the absence of particle hole symmetry will be further verified for the case of the $1T'$-$\text{WTe}_2$ wire. The phase is therefore protected by time-reversal symmetry alone while particle-hole symmetry is required only to anchor the in-gap states to zero energy. This analysis parallels the one on ref.  \cite{SSH-NNN} where analogously the invariant and phase are protected by only inversion symmetry while both inversion and particle-hole symmetry secure the zero-energy value.  For particle-hole-symmetric disorder Fig.\ref{QSHI-disorder-states} c) the perturbation strength is protected not by the q(2-1)D gap but rather the 2D bulk gap of the system, which, for $c=0.8,u=1.0$, is $\Delta\approx 0.5$ as seen in Fig.~\ref{QSHI-disorder-states} e). If the perturbation breaks time reversal symmetry, however, the Kramers degeneracy of the q(2-2)D bound states is broken as expected.\\

We conclude from this analysis that the q(2-1)D wire with spinful time-reversal symmetry, realized for a system with a 2D bulk and open boundary conditions in one direction, exhibits finite-size topological phases for subsets of the topologically non-trivial regions of the 2D bulk topological phase diagram. In these subsets, bounded by lines along which the minimum direct q(2-1)D bulk goes to zero rather than the minimum direct 2D bulk gap, in general, the wire harbors topologically-protected q(2-2)D boundary modes for open boundary conditions in two directions appearing in Kramers pairs. In the quasi-1D bulk, these subsets correspond to Wilson loop spectra with some Wilson loop eigenvalue phases fixed to $\pm \pi$, protected by spinful time-reversal symmetry. These signatures of non-trivial topology are therefore associated with time-reversal invariant, q(2-1)D finite-size topological phases due to interference between topologically-protected gapless boundary modes resulting from 2D bulk topology, either of the quantum spin Hall insulator or the 2D Dirac semimetal. In contrast, the ten-fold way classification scheme of topological phases of matter determines a 1D system in class $\text{AII}$ \cite{ten-fold-way} has trivial topological classification. These results are therefore evidence of topologically non-trivial phases of matter outside of the ten-fold way classification scheme.

\section{$1T'$-$\text{WTe}_2$ wire} \label{WTe2-section}

Although the previous model is based on the celebrated  HgTe quantum wells, which allowed for the discovery of the first QSHI, it may not be most suitable for the experimental discovery of finite-size topology. The need for a small sample size in one direction may be easier to achieve for monolayers such as the $1T'$-$\text{TWe}_2$ QSHI \cite{WTe2-exp}. We thus consider the model studied in reference \cite{WTeT2-model} which combines density-functional theory calculations, symmetry considerations and fitting to experimental data. Since the finite-size topology comes from the hybridization of the edge states, we consider only the lattice termination which results in a Dirac crossing near the Fermi energy. That is we study the sawtooth $y$ ribbon with open boundary conditions in the $x$ direction. The $1T'$-$\text{WTe}_2$ Hamiltonian is presented in the Supplementary Material.\\

Under such circumstances we expect the Dirac crossing to gap out in general on larger and larger energy scales as the system size in the $x$ direction $N_x$ decreases. Such a gap opening does occur for the original material parameters for even $N_x$ values ranging from $4$ to $20$. It is worth noting that the gap still exists even for larger $N_x$ but its magnitude becomes very small compared to other material energy scales. Such a gapped system may then host topological q(2-2)D edge states if the Wilson loop spectrum is nontrivial. Specifically, we find that for $N_x=8,10,12$ the gap has a $\pm \pi$ phase in the Wilson loop spectrum. Even in the case where the spectrum is trivial, for example at $N_x=6$, we may still tune the system so that a gap closing occurs and a nontrivial Wilson loop phase appears. To do so, we may apply an electric field perpendicular to the monolayer corresponding to addition of a symmetry-allowed Rashba spin-orbit coupling term to the Hamiltonian. Since the original model already includes such couplings, we further include a change of the in-plane SOC parameters by a quantity $\Delta\lambda_{\text{SOC}}$.\\

\begin{figure}
    
    \subfigure[]{\includegraphics[width=0.49\columnwidth]{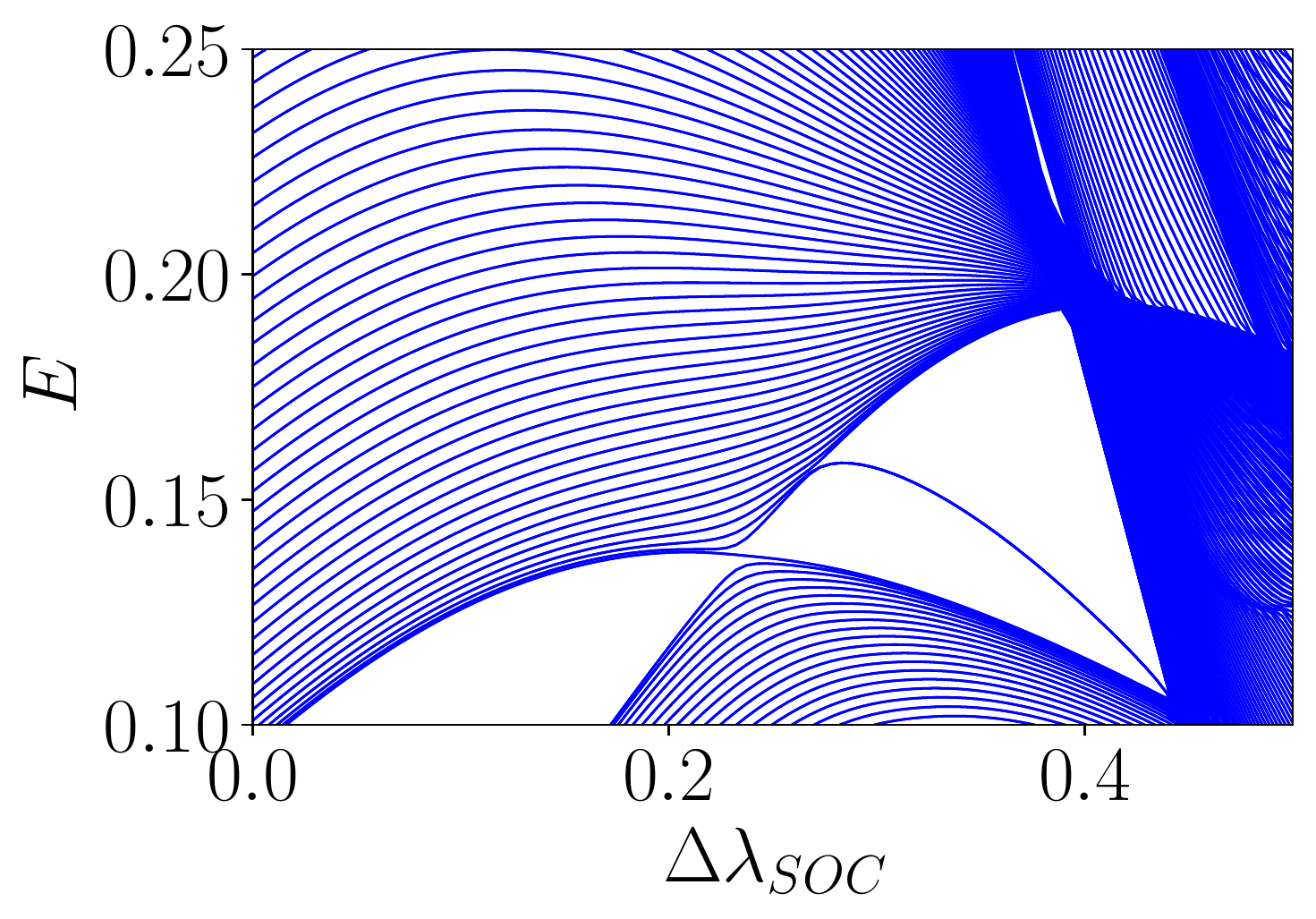}}
     \subfigure[]{\includegraphics[width=0.49\columnwidth]{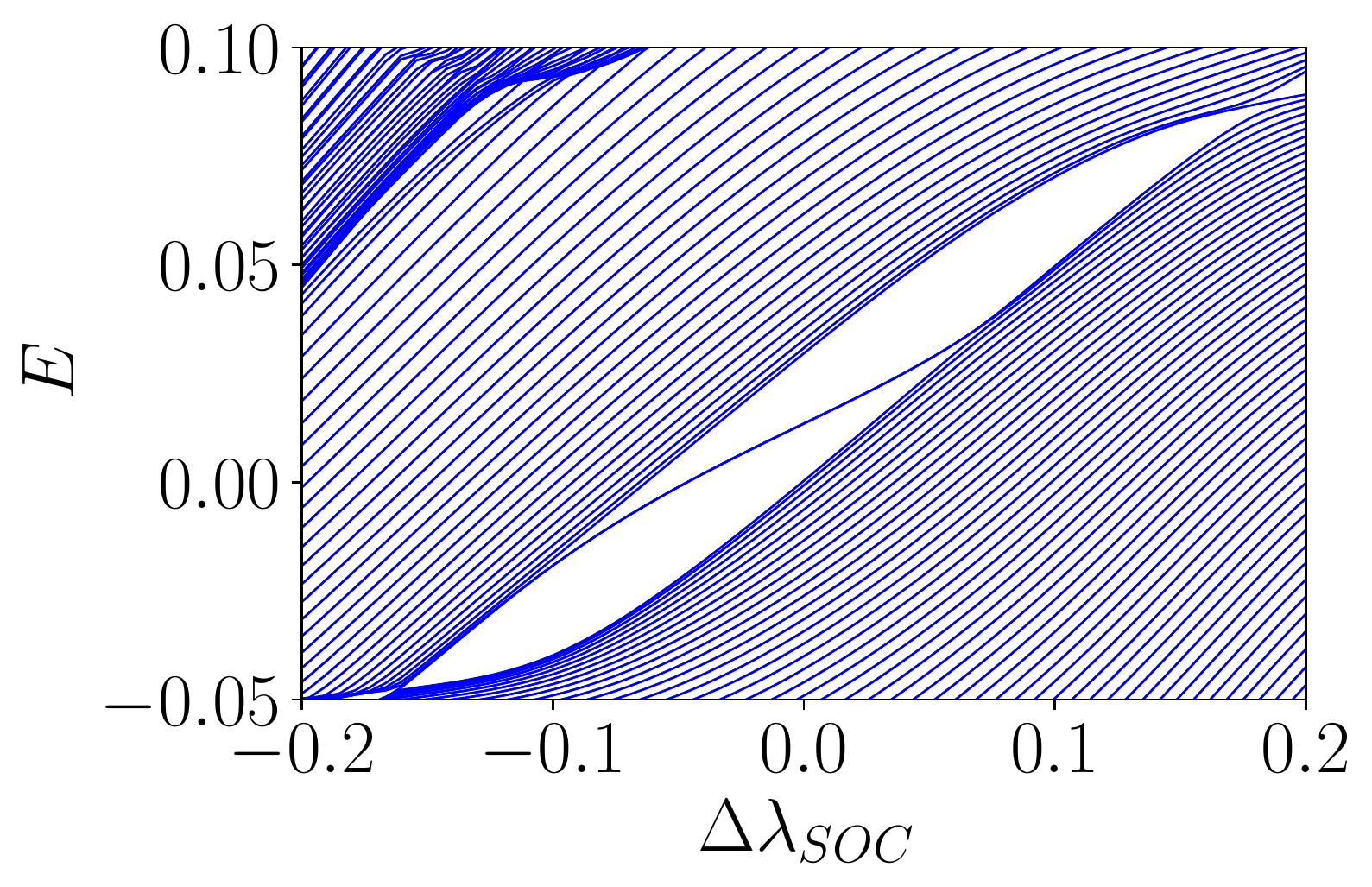} } 
    \subfigure[]{\includegraphics[width=0.49\columnwidth]{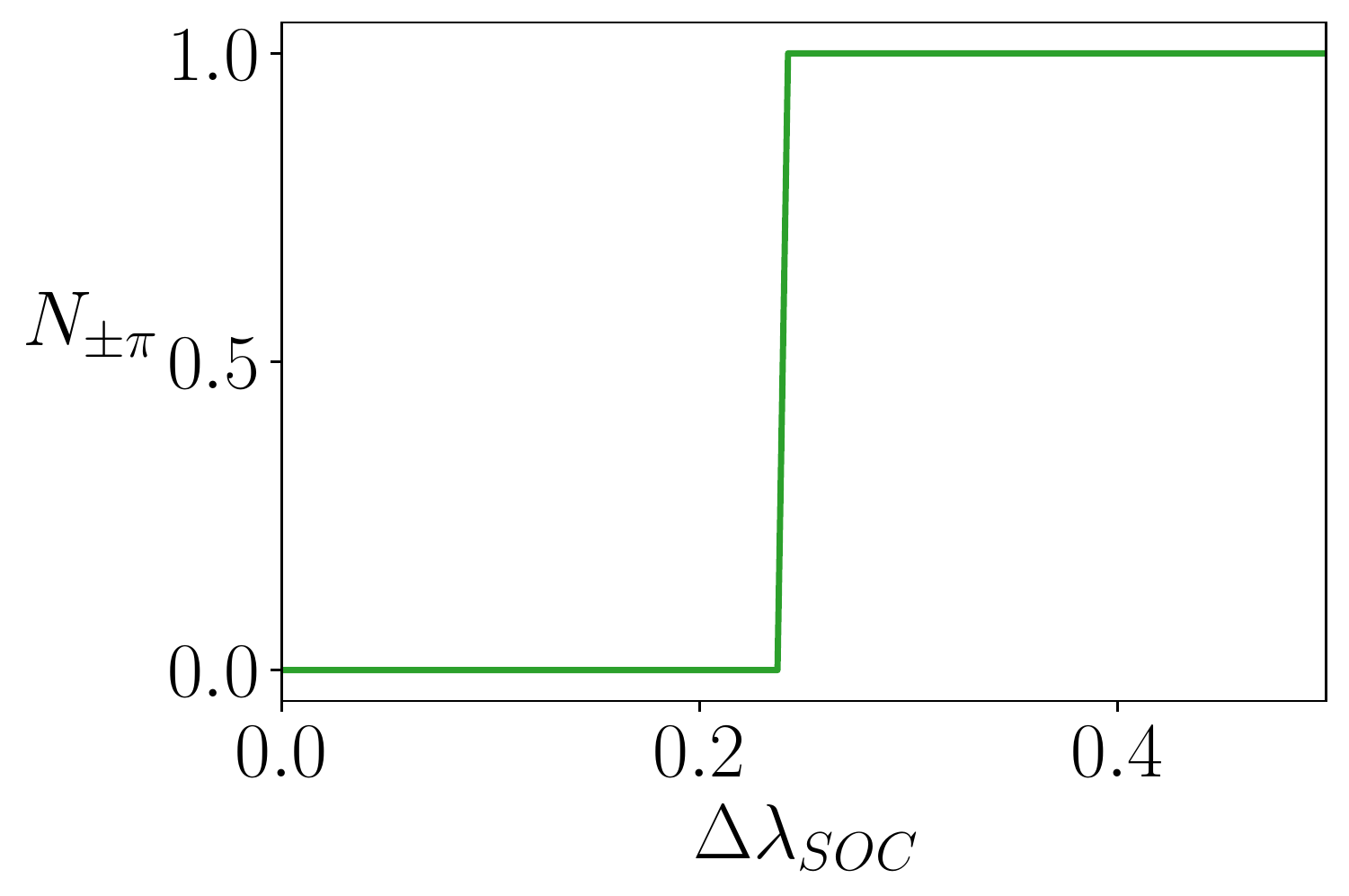}}
    \subfigure[]{\includegraphics[width=0.49\columnwidth]{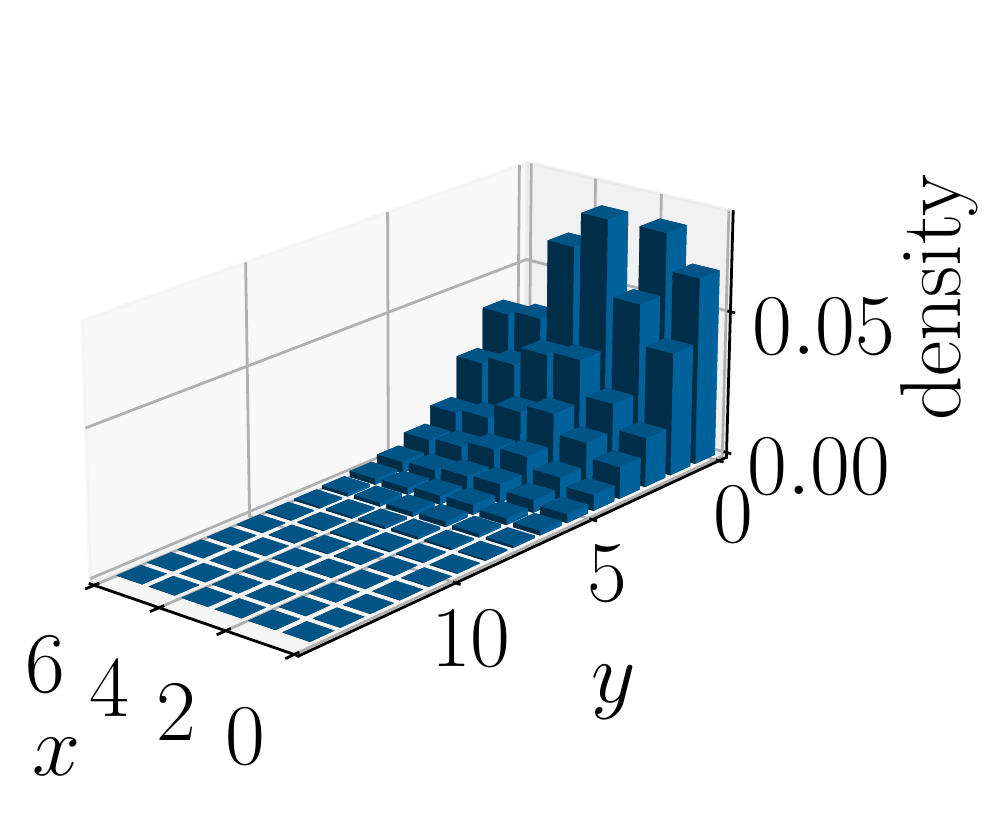}}

  \vspace{-0.3cm}
  \caption{  a) Energy spectrum (eV) as a function of the change in Rashba spin orbit coupling $\Delta\lambda_{\text{SOC}}$ for a $N_x=6$ saw-tooth terminated $1T'$-$\text{TWe}_2$ wire. b) Energy spectrum (eV) for the system with $N_x=12$ as a function of  $\Delta\lambda_{\text{SOC}}$ showing edge states at zero field.
  c) Number of Wilson loop spectrum $\pm \pi$ phases for $N_x=6$ as a function of $\Delta\lambda_{\text{SOC}}$ d) Wave function probability density for one edge state at $\Delta\lambda_{\text{SOC}}=0.35 eV$ and $N_x=6,N_y=200$ for the same saw-tooth terminated $1T'$-$\text{TWe}_2$ wire. }
  \label{WTe2}
\end{figure}

For the case of $N_x=6$, we obtain a phase diagram for the number of Wilson loop eigenvalues with phase $\pm \pi$, $N_{\pm \pi}$, vs. spin-orbit coupling anisotropy, $\Delta \lambda_{SOC}$, showing a change in $N_{\pm \pi}$ from $0$ to $1$ with increasing $\Delta \lambda_{SOC}$ as shown in Fig.~\ref{WTe2} a). Based on our previous results for a canonical toy model of the QSHI, we expect q(2-2)D edge states to appear if we open boundary conditions in the $\hat{y}$-direction as well. Such edge states due to an additional bulk-boundary correspondence characterized by $N_{\pm \pi}$ of the q(2-1)D bulk do exist, as shown in Fig.~\ref{WTe2} b)corresponding to a line of four-fold degenerate, in-gap states as a function of $\Delta \lambda_{SOC}$. The four-fold degeneracy corresponds to a two-fold Kramers degeneracy due to spinful time-reversal symmetry, and two-fold degeneracy corresponding to boundary modes localized at the left and right edge as in the previous simple model of eqref.~\eqref{QSHI}. For a fixed Rashba spin orbit coupling change of $\Delta\lambda_{\text{SOC}}=0.35 eV$, we find that the edge states localize on each end of the wire as shown in Fig. \ref{WTe2} c).

\section{3D TI slab} \label{3D TI slab}

We now study the finite size topology of the quintessential 3D TI in symmetry class $\text{AII}$ \cite{ten-fold-way,Kitaev-K-theory-10fold}. The classification in this case is dictated by four $\mathbb{Z}_2$ topological invariants $(\nu_0;\nu_1,\nu_2,\nu_3)$ \cite{Moorehomotopy2007,Fu-Kane-Mele} , where the last three invariants are related to the translational invariance of the system classifying the weak TI phase (WTI) while the $\nu_0$ parameter classifies the strong TI phase (STI). In the following, we study the STI phase with trivial lower-dimensional invariants corresponding to $(1;000)$, which is realized in the model Bloch Hamiltonian \cite{3DTI-ham}:

\begin{align}
        H(\bm{k})=-2\lambda\sum_\mu \sin{k_\mu} \sigma_z s_\mu + \sigma_x s_0(M-t\sum_\mu \cos{k_\mu}),\label{3DTI}
\end{align}

where the Pauli matrices $\{\sigma_i\}$,$\{s_j\}$ act on orbital and spin degrees of freedom respectively, $\lambda$ is a spin orbit coupling parameter that breaks spin conservation, $M$ is an onsite staggered potential, and $t$ is a nearest-neighbor hopping integral. In the following, we take all energies to be in units of $t$ by setting $t=1$, and consider the regime in which $1<M<3$ and $\lambda$ positive, for which the model realizes the desired strong TI phase. First, we specialize to the case of a thin slab in the $\hat{x}$-direction of $N$ layers. In that case, we expect in analogy to the QSHI that the Dirac cones from the upper and lower surfaces interfere due to the thin bulk and hybridize to open a gap even for open boundary conditions. \\

The hybridization gap, just as in the previous 2D case, is expected to sometimes protect a non-trivial finite size topological phase. To study this, we once again characterize topology using the Wilson loop spectrum, now as a function of $k_y$ or $k_z$ with OBC in $x$ to characterize topology first of a q(3-1)D bulk, in regions of the phase diagram where the 3D bulk corresponds to the strong TI. As the isotropy of the Hamiltonian Eq.~\eqref{3DTI} suggests, there is no difference between the Wilson loop eigenvalues of $W(k_y)$ and $W(k_z)$, thus we consider only $W(k_z)$, where each Wilson loop matrix is now defined as:
\begin{align}
 \displaystyle \mathcal{W}(k_y)=\mathcal{P}e^{\textstyle -\int_{-\pi}^{\pi} d k_z A_z(k_y,k_z)}, \label{Wilson-ky}\\
 \displaystyle \mathcal{W}(k_z)=\mathcal{P}e^{\textstyle -\int_{-\pi}^{\pi} d k_y A_y(k_y,k_z)}. \label{Wilson-kz}
\end{align}

\begin{figure}[ht]
  \centering
  \subfigure[]{\includegraphics[width=0.49\columnwidth]{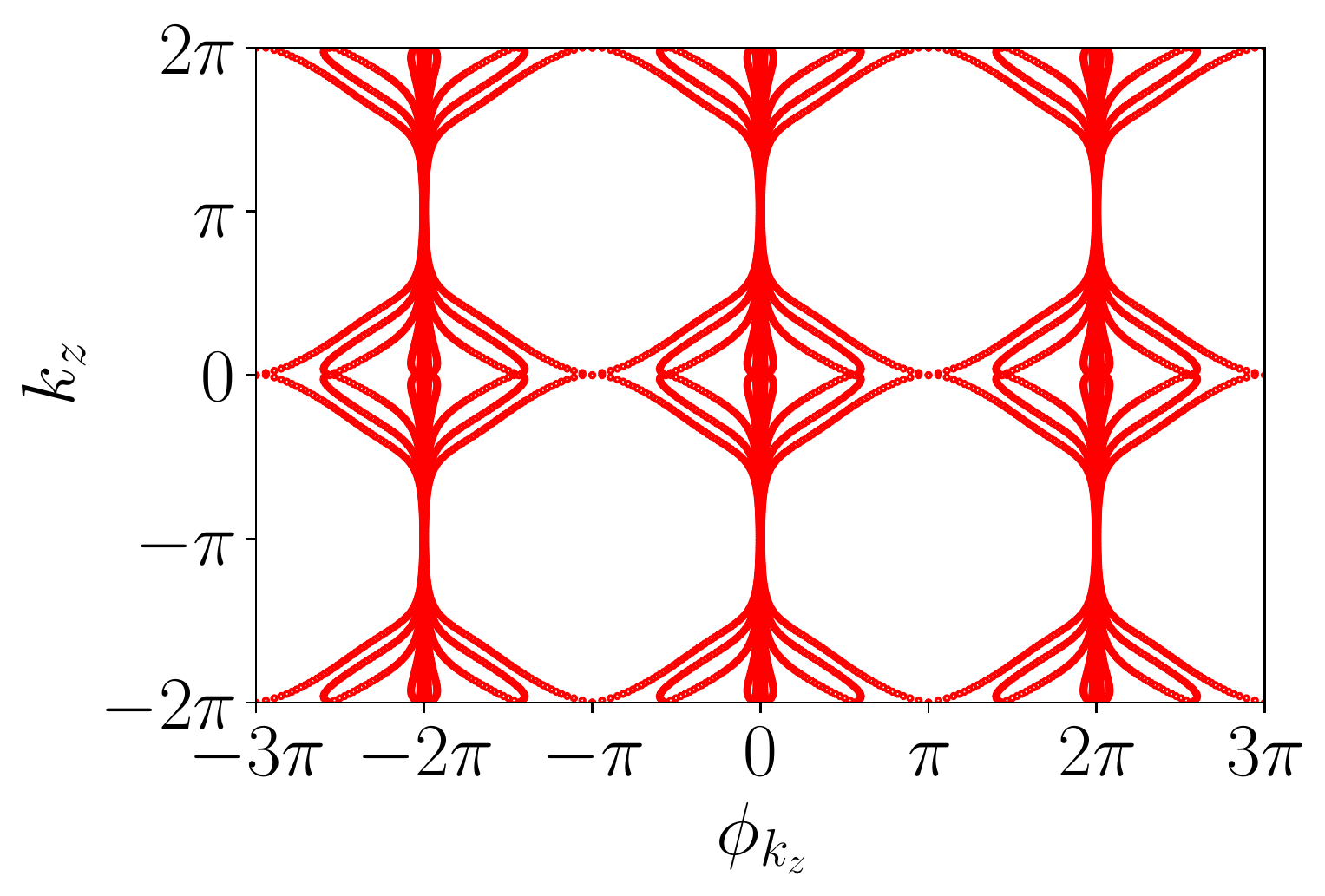}}
  \subfigure[]{\includegraphics[width=0.49\columnwidth]{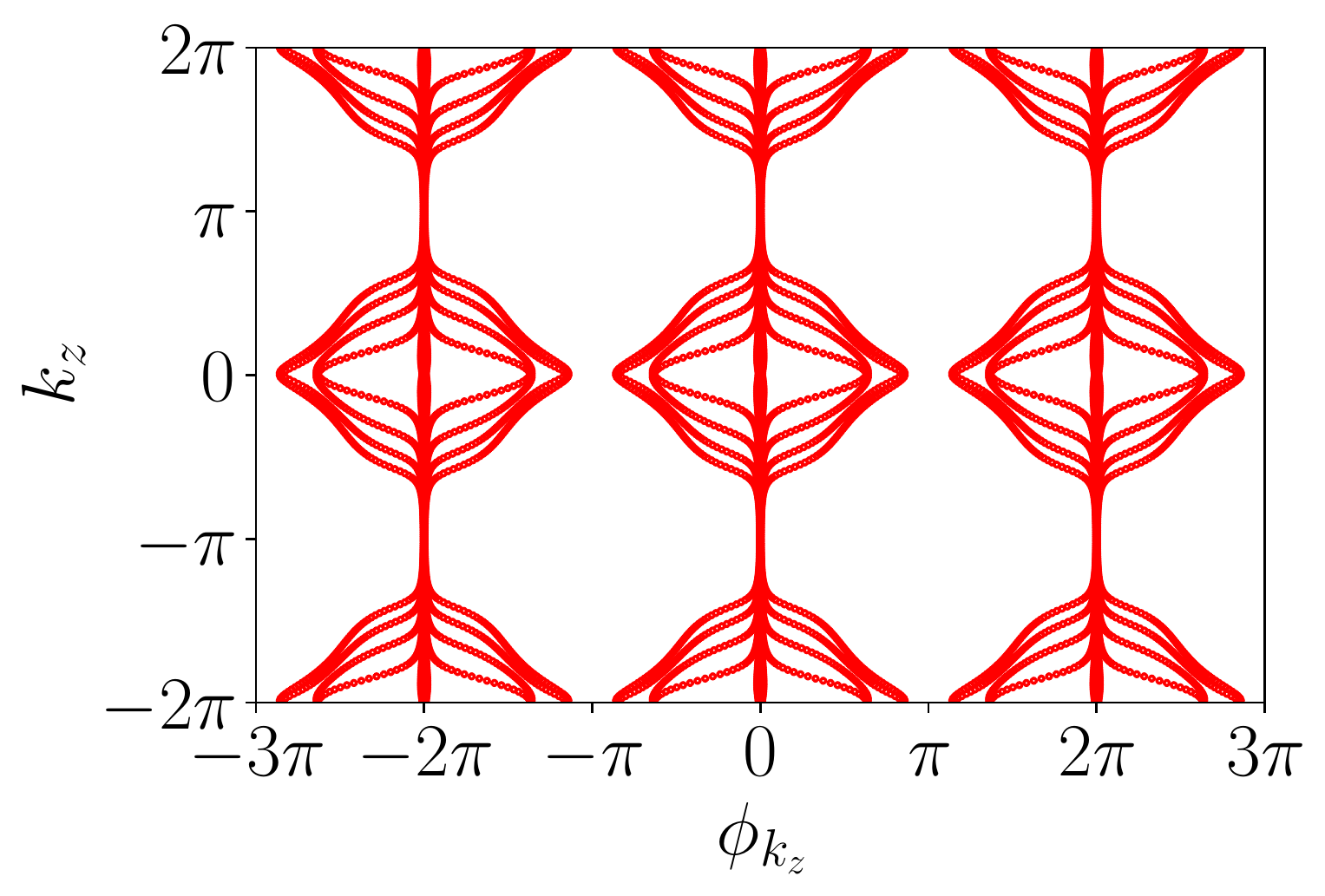}}
  \vspace{-0.3cm}
  \caption{a) Wilson loop nontrivial eigenvalue spectrum as a function of $k_z$ for OBC in $x$ and PBC in $y,z$. The parameters are $M=2.0,\lambda=0.2$ and $N=6$ layers.  
  b) Wilson loop trivial eigenvalue spectrum as a function of $k_z$ for OBC in $x$ and PBC in $y,z$. The parameters are $M=1.6,\lambda=0.1$ and $N=6$ layers. }
  
  \label{3DTI WL}
\end{figure}

A typical Wannier charge center spectrum vs. the remaining momentum component, $k_z$, is plotted in Fig.~\ref{3DTI WL}, which shows the spectral flow a) characteristic of a TR invariant topological insulator for some values and a trivial spectrum b) for others within the $(1;000)$ 3D bulk phase classification.\\

\begin{figure}[ht]
  \centering
  \subfigure[]{\includegraphics[width=0.49\columnwidth]{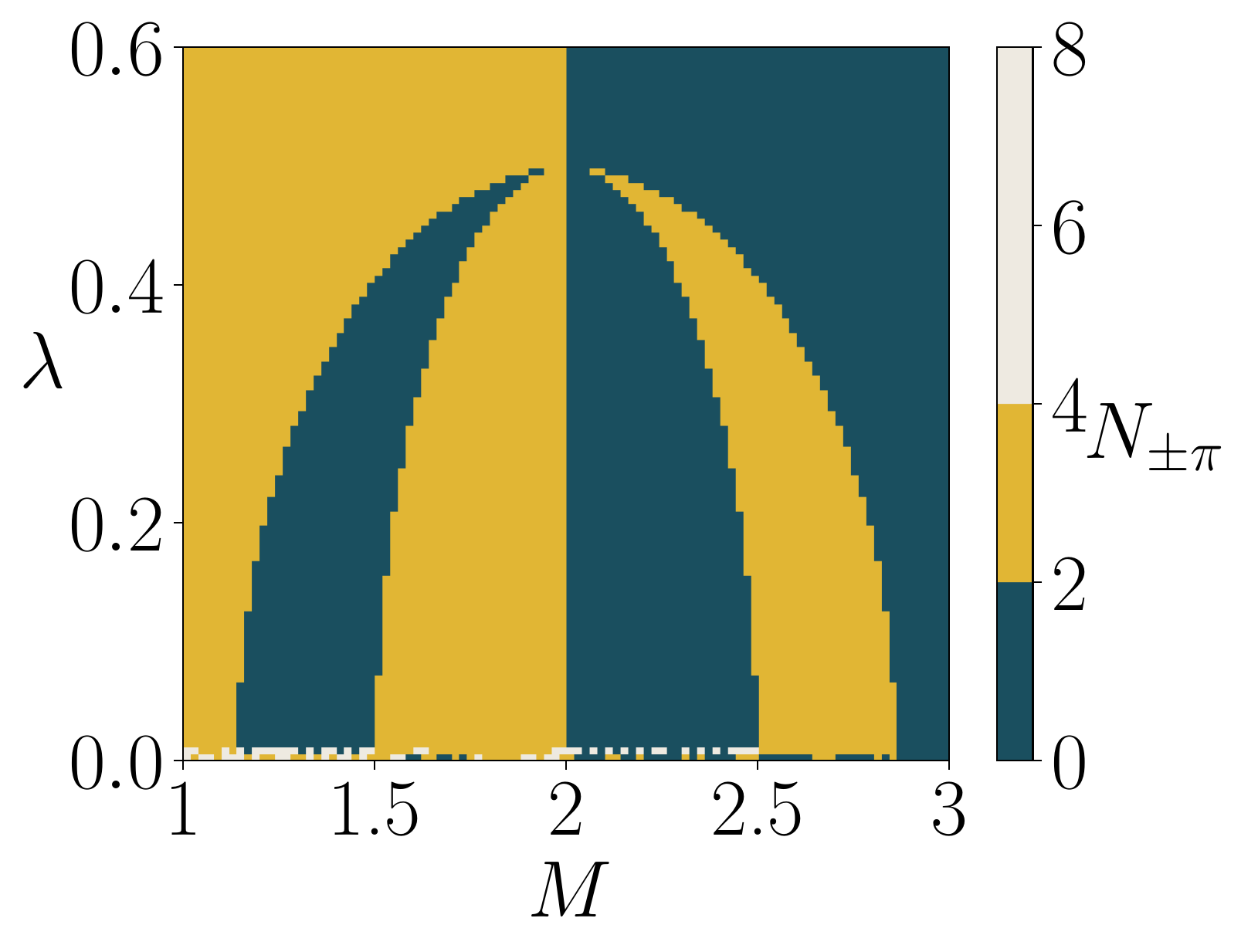}}
  \subfigure[]{\includegraphics[width=0.49\columnwidth]{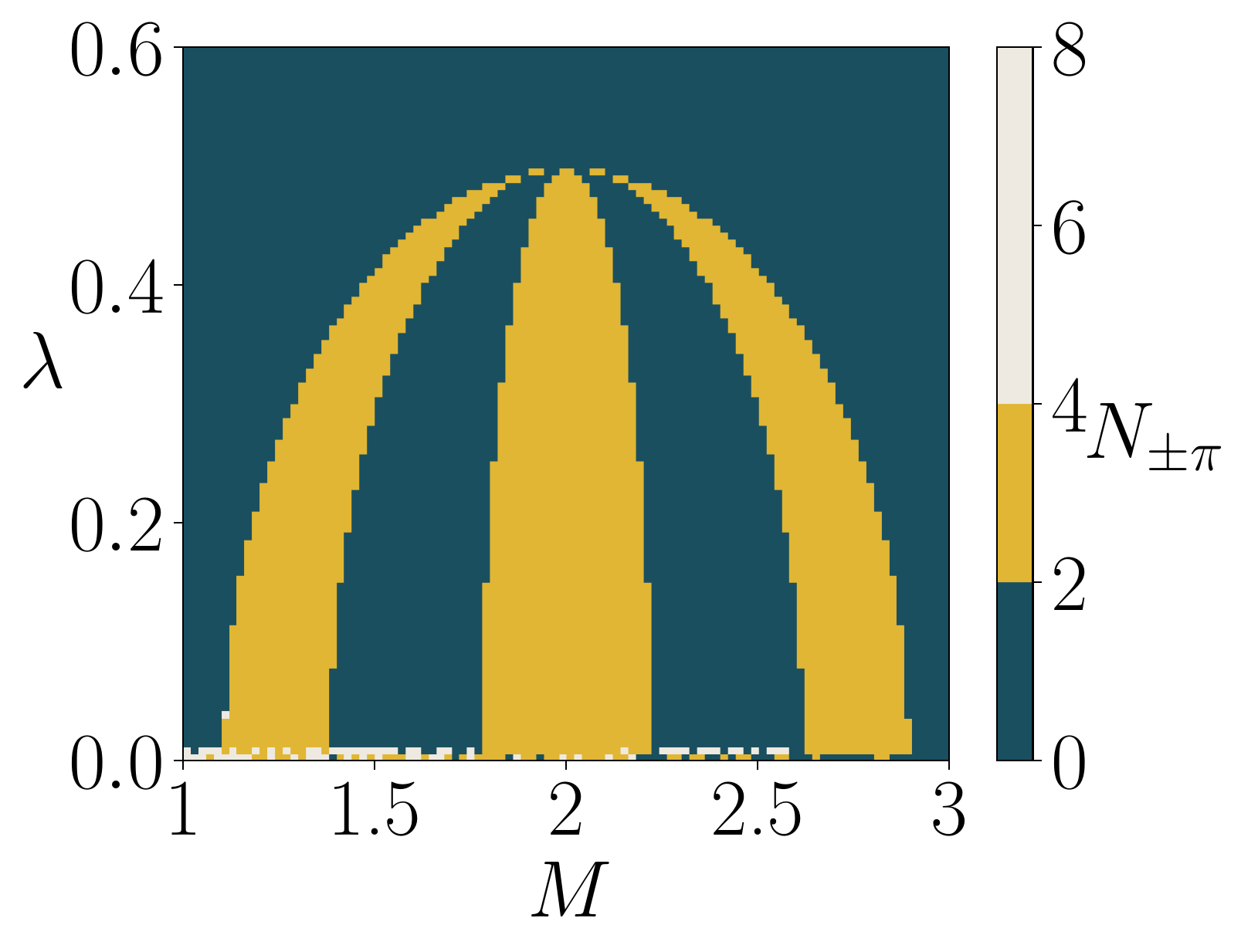}}
  \subfigure[]{\includegraphics[width=0.49\columnwidth]{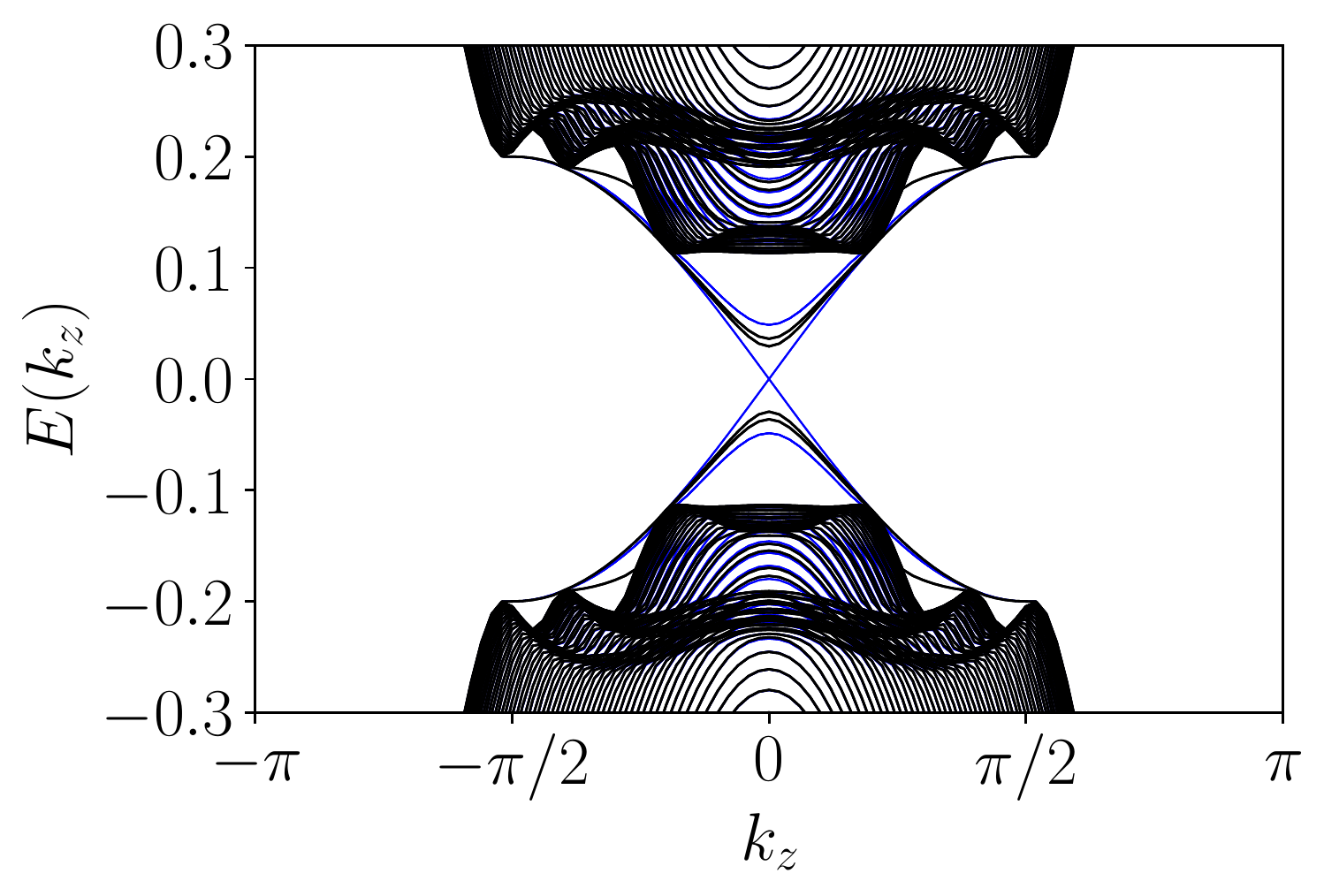} }
  \subfigure[]{\includegraphics[width=0.49\columnwidth]{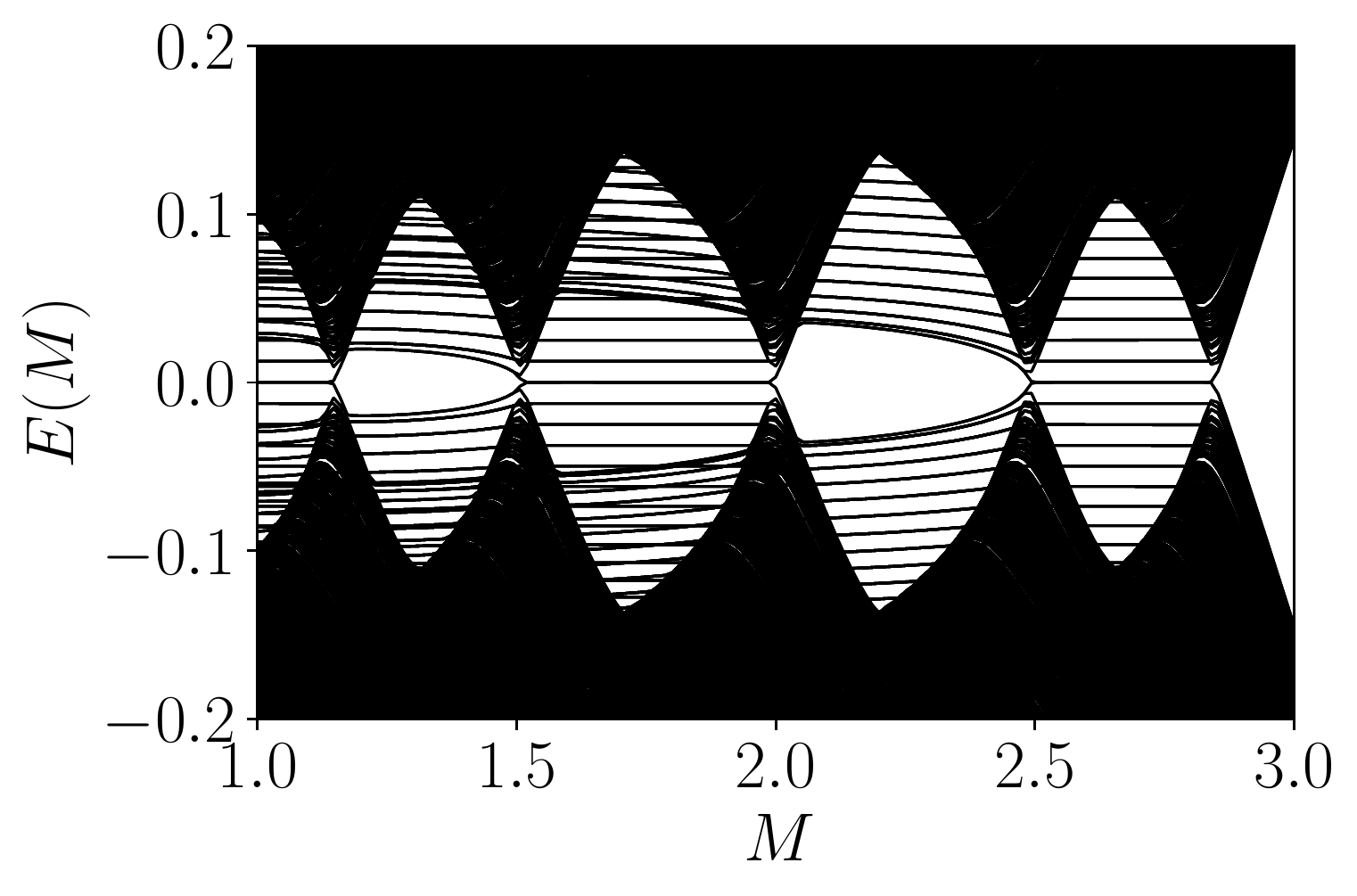}}
  
  \vspace{-0.3cm}
  \caption{
  Phase diagram from wilson loop spectrum for a slab with a) $N=5$, b) $N=6$, black-0, yellow-2 $\pm \pi$ phases. c)Quasi-1D dispersion for OBC in $x$ $N=6$, PBC in $z$ and OBC (PBC) in $y$ as a function of $k_z$. The parameters are  $M=1.8, \lambda=0.1$.  d) Spectrum for OBC in $x,y$ as a function of $M$, for discretized values of $k_z$ with $N=5$ and $100$ sites in the remaining directions, $\lambda=0.1$. }
  \label{3DTIslab}
\end{figure}

The non-trivial spectral flow only appears for certain parameter regimes: these regions can be distinguished in a systematic way by counting the number of fixed $\pm \pi$ phases in the Wilson loop spectrum, as in the previous case. This regions of parameter space which have an energy gap are again the "bubbles" observed in the QSHI case. We plot the phase diagram as a function of the model parameters in Fig.~\ref{3DTIslab} a),b) for $N=5,6$ layers, respectively. The phase diagram changes dramatically for each value of $N$, the number of layers, indicative of the finite-size topology  \cite{fst_one}. The pattern of trivial and nontrivial regions is entirely contained with the non-trivial region of the 3D bulk topological phase diagram for Hamiltonian \eqref{3DTI}, indicating the 3D minimum direct bulk gap remains finite during these topological phase transitions of the q(3-1)D bulk. We remark that the sudden changes in color near $\lambda=0$ seem to be just an artifact of the numerical precision when approaching the STI gap-closing in the 3D bulk.\\

One of the consequences of the system being in a topologically non-trivial regime, according to the Wilson loop spectra of the q(3-1)D bulk, is an additional bulk-boundary correspondence: opening boundary conditions in a second direction in these regions of phase space, we find topologically-protected q(3-2)D states that now are localized at the edges of the slab, as shown in Fig.~\ref{3DTIslab} c). These q(3-2)D states appear within the q(3-1)D bulk gap similarly to the case of q(2-2)D boundary states appearing in the q(2-1)D bulk gap of the QSHI. A topological phase diagram for the q(3-1)D system with open-boundary conditions in the $\hat{y}$-direction as a function of $M$ is also shown in Fig.~\ref{3DTIslab} d), demonstrating the direct correspondence of the topologically-protected boundary modes with the topologically non-trivial regions of the q(3-1)D bulk in Fig.~\ref{3DTIslab} a).  This indicates that $N_{\pm \pi}$, the number of $\pm \pi$ phases in the Wilson loop eigenvalue spectrum, characterizes finite-size topological phases resulting from interference of the topologically-protected Dirac cones of the 3D TI, in addition to characterizing finite-size topological phases due to interference between the helical boundary modes of the QSHI.

While the finite-size topological phase in the q(3-1)D system exhibits helical boundary modes analogous to those of the QSHI, this topological phase is not just a QSHI. Notably, the 3D minimum direct gap remains finite in this region of the phase diagram where non-trivial finite-size topological phases occur, so the 3D bulk is still in the topological phase $(1;000)$. The finite-size topological phase therefore exhibits signatures associated with a non-trivial intrinsically 3D topological invariant. This is indicated by adding perturbations to the system to probe the magneto-electric polarizability of the system, which depends on the intrinsically 3D topological invariant, a connection identified in previous work by Essin~\emph{et al.}~\cite{essin2009}.\\

\begin{figure}[ht]
  \centering
   \subfigure[]{\includegraphics[width=0.49\columnwidth]{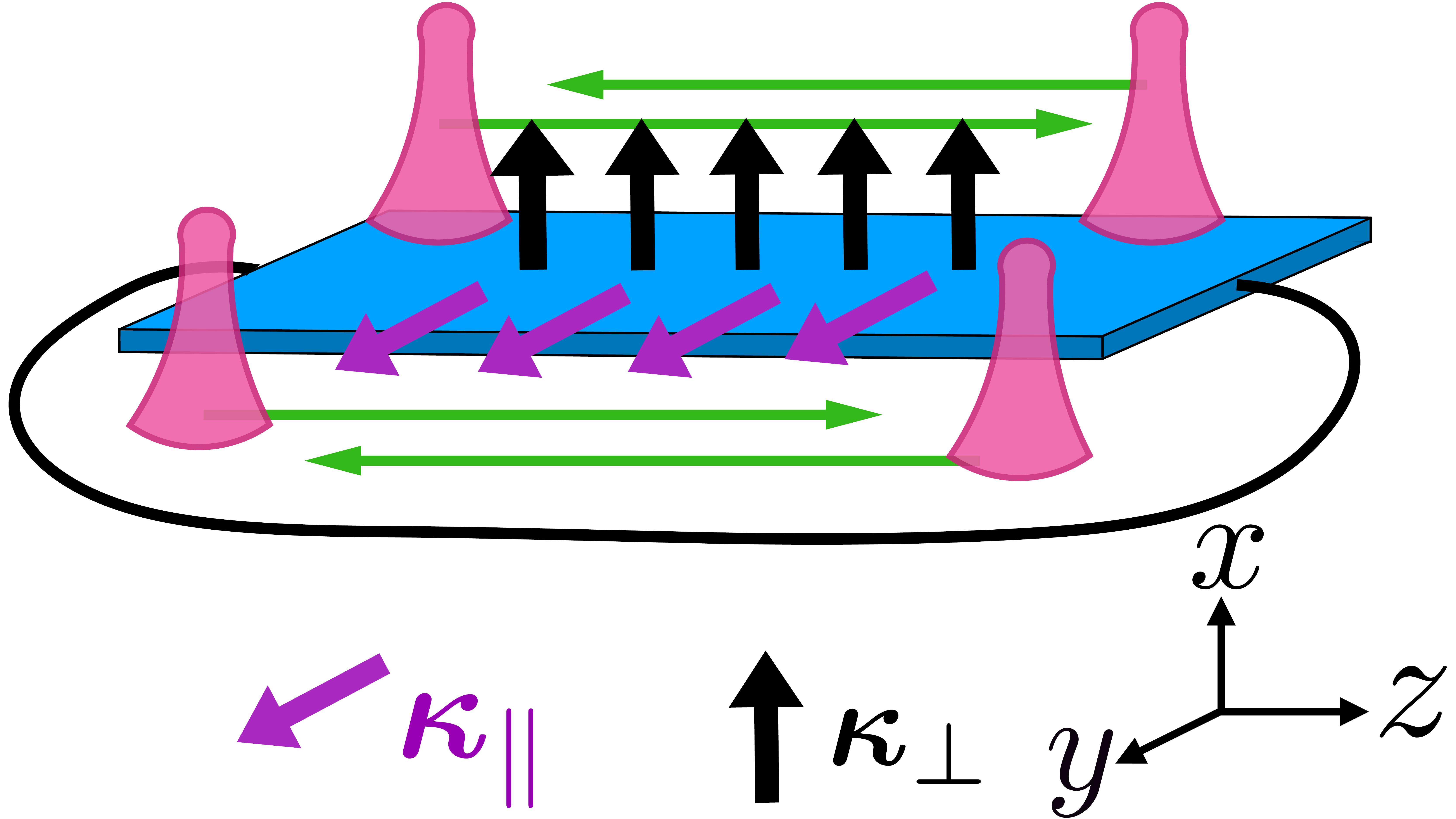}}
  \subfigure[]{\includegraphics[width=0.49\columnwidth]{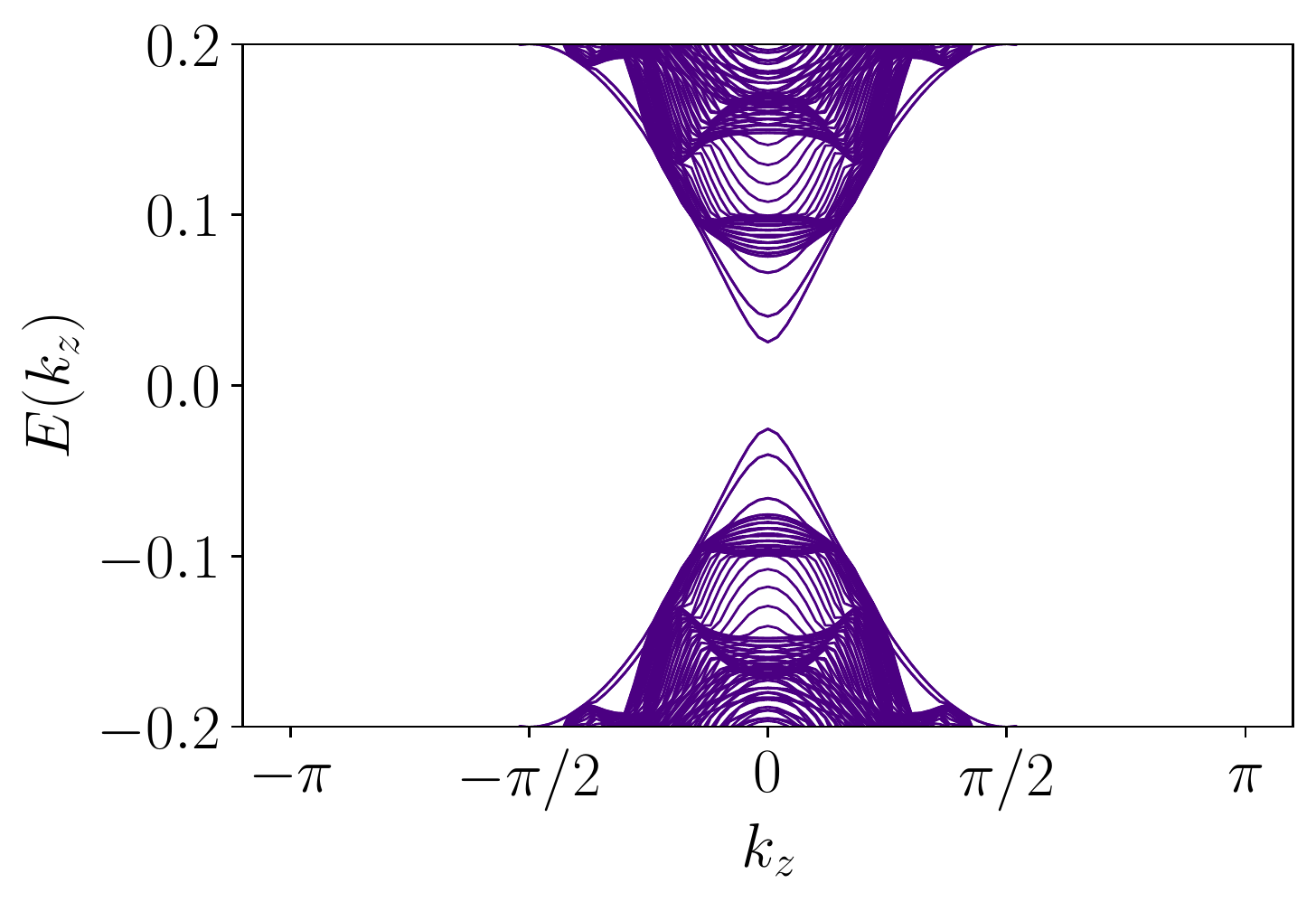}}
  \subfigure[]{\includegraphics[width=0.49\columnwidth]{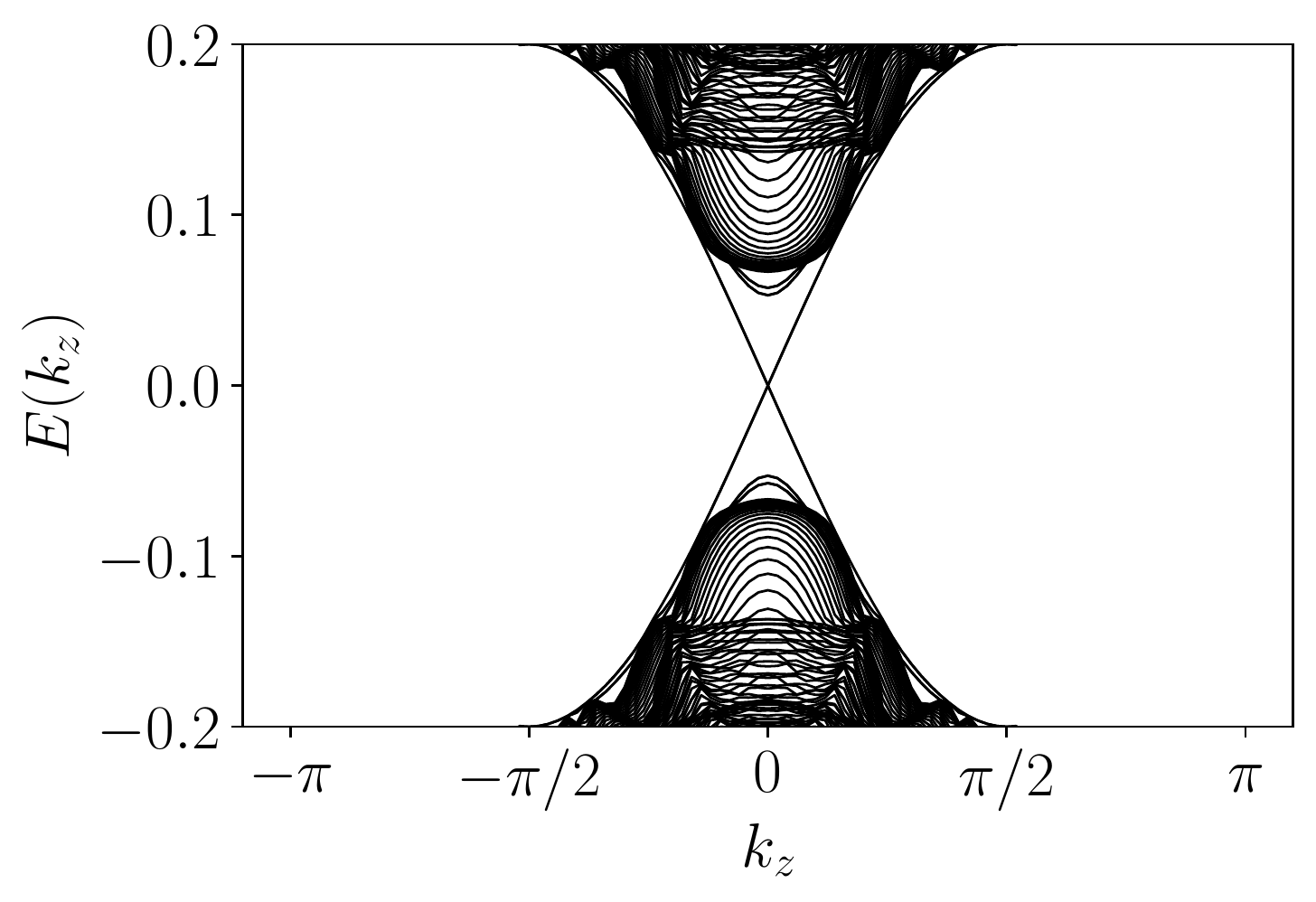}}
  \subfigure[]{\includegraphics[width=0.49\columnwidth]{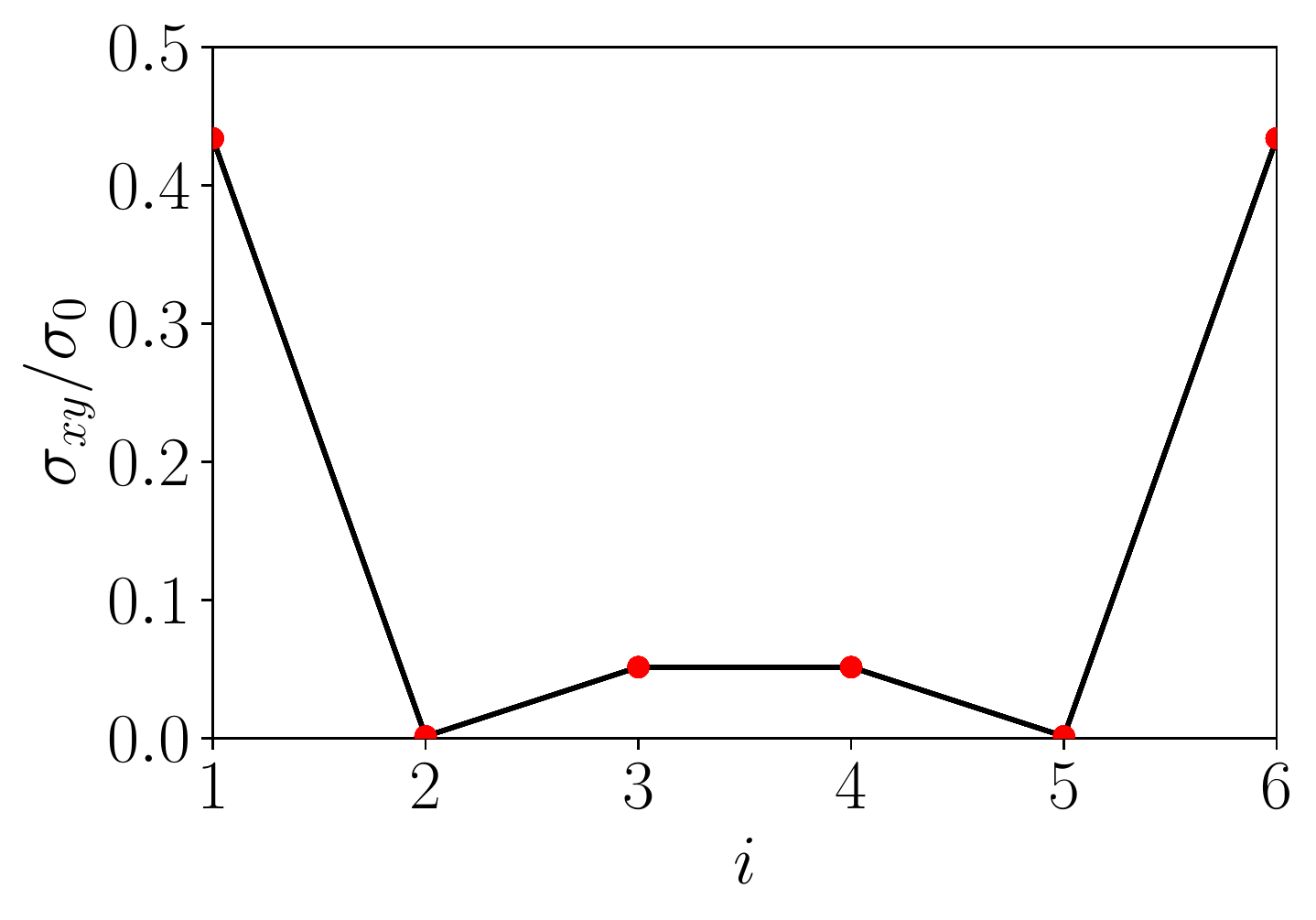}}
  
   \vspace{-0.3cm}
  \caption{ a) Diagram of perturbations and edge states (pink) on a q(3-1)D STI slab, parallel field (purple) and perpendicular field (black), we consider the perturbations for OBC in $x,y$ and PBC in $z$.
  b) Spectrum of STI slab with OBC in $x$ and $y$, $N_x=5$, $N_y=100$. A constant Zeeman field ordering $\kappa=0.1$ parallel to the slab top and bottom layers is present. The parameters are $M=1.8,\lambda=0.1$.  c) Same parameters as in c) but now with a perpendicular field the system exhibits a QHE.d) Layer conductivity (Chern number) for $N_x=6$, $M=2.0$,$\lambda=0.2$ . }
  \label{3DTIslab-pert}
\end{figure}

We then compare the result to a q(3-1)D stack of 2D QSHI in the $x$ direction with the same perturbation to demonstrate the finite-size topological phase of the q(3-1)D system is distinct from a QSHI. The type of perturbations we consider to determine the magnetoelectric polarizability in these two cases are TRS-breaking terms in the form of weak  Zeeman field in only the uppermost and lowest layers i.e. $V=\bm{\kappa}\cdot\bm{s} $ with $\mid \bm{\kappa}\mid \approx 0.1$. This could correspond to ferromagnetically-ordered magnetic dopants in just these layers. This situation is illustrated schematically in Fig. \ref{3DTIslab-pert} a). We first consider a Zeeman field oriented in the $yz$-plane and labeled  $\bm{\kappa}_{\parallel}$ as shown in Fig.~\ref{3DTIslab-pert} a) for both the $(1;000)$ system (q(3-1)D STI) and the $(0;001)$ system (q2D WTI, or stack of QSHIs). In this case, these systems react similarly, their spectra gapping out as shown in Fig.\ref{3DTIslab-pert} b). This is expected, as the perturbation breaks TR symmetry.\\

However, the responses of the two systems are strikingly distinct if we instead consider an applied Zeeman field oriented along the $\hat{x}$-axis, or field $\bm{\kappa}_{\perp}$ as shown in Fig.~\ref{3DTIslab-pert} a). In the case of the $(1;000)$ system (q(3-1)D STI), two of the q1D boundary states gap out, leaving two gapless boundary modes remaining. They constitute chiral boundary modes of a QHE on each surface as shown in Fig.~\ref{3DTIslab-pert} c), corresponding to a layer-dependent Hall conductivity $\sigma_{xy}$ as shown in Fig.~\ref{3DTIslab-pert} d). This is a known manifestation of the quantized magnetoelectric polarizability of the STI, resulting from the non-trivial value of the strong invariant, these results can be directly compared with past work for systems with a 3D bulk rather than q(3-1)D bulk~\cite{essin2009}. Notably, the layer-dependent Hall conductivity exhibits this non-trivial response only for topologically non-trivial finite-size topological phases as characterized by $N_{\pm \pi}$. This agrees with the response theory of previously-studied finite-size topological phases, and reflects the fact that only occupied states of the non-trivial bubbles carry Berry phase contributions to the underlying 3D topological invariant.  \\

In constrast, there is no topological magnetoelectric effect and no QHE in the surface layers of the QSHI stack. Instead, the states maintain their degeneracy and helicity, for small fields, which is consistent when we view the field as parallel to the edge surfaces hosting the helical boundary states of the QSHI layers. Therefore, although the spectrum of q(3-1)D STI slab appears to be similar to the spectrum of a QSHI stack, its response to TRS-breaking perturbations reveals that it is a finite-size topological phase arising from $(1;000)$ topology of the 3D bulk.

\section{3D TI wire}

We finally consider a q(3-2)D wire geometry for the  STI, demonstrating that finite-size topology arises from interference between topologically-protected boundary states of finite-size topological phases, as well as interference between the topologically-protected boundary states of topological phases in the ten-fold way. We therefore consider the same Hamiltonian Eq.~\eqref{3DTI} as considered in the previous section, but now with open boundary conditions in $\hat{x}$- and $\hat{y}$-directions and periodic boundary conditions in the $\hat{z}$-direction, with the number of lattice sites in the $\hat{x}$- and $\hat{y}$-directions,  $N_x$ and $N_y$, respectively, each much less than $N_z$, the number of lattice sites in the $\hat{z}$-direction (so $N_x,N_y \ll N_z$).

According to the ten-fold way classification scheme for topological phases of matter, this is an effectively 1D system in class $\text{AII}$, which is topologically-trivial~\cite{ten-fold-way,10-fold-way,Kitaev-K-theory-10fold}.  Here, we show finite-size topological phases are nonetheless possible in this system, first characterizing finite-size topology of the q1D bulk through analysis of Wilson loop spectra, and then by demonstrating an additional bulk-boundary correspondence yielding q(3-3)D topologically-protected boundary modes in the q1D wire, ultimately resulting from interference between the topologically-protected Dirac cone surface states of the STI.

\begin{figure}[ht]
  \centering
  \subfigure[]{\includegraphics[width=0.49\columnwidth]{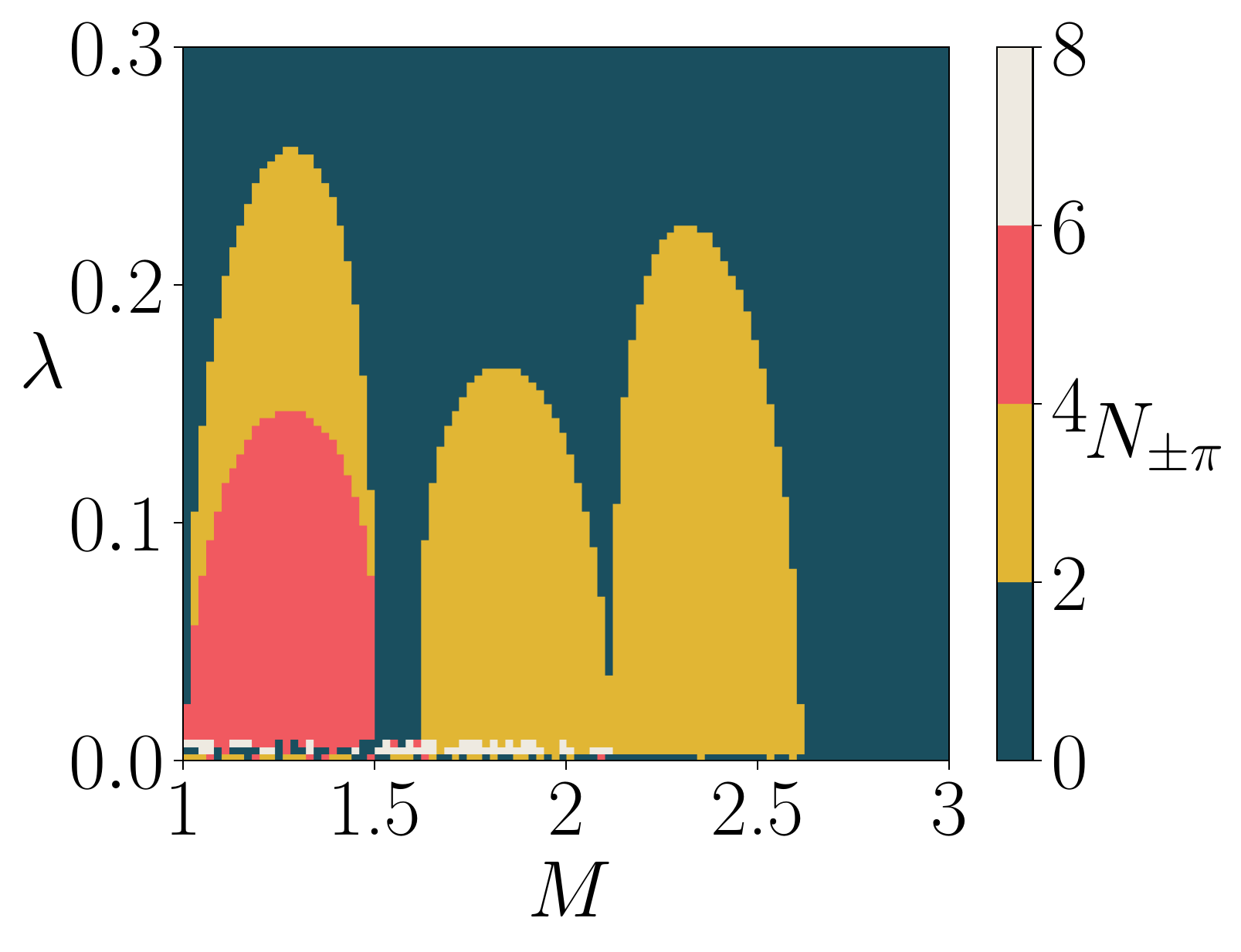}}
  \subfigure[]{\includegraphics[width=0.49\columnwidth]{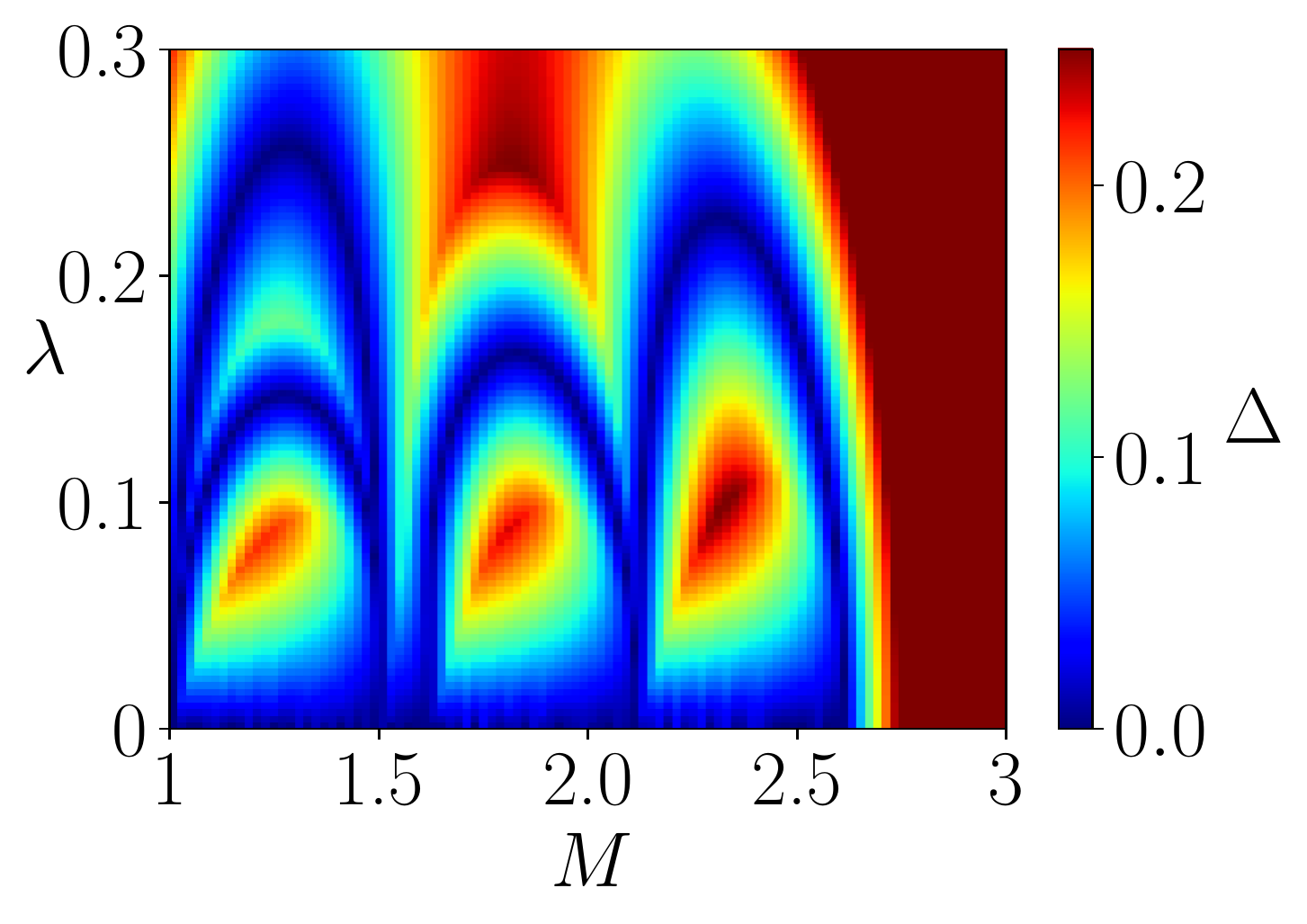}}
  \subfigure[]{\includegraphics[width=0.49\columnwidth]{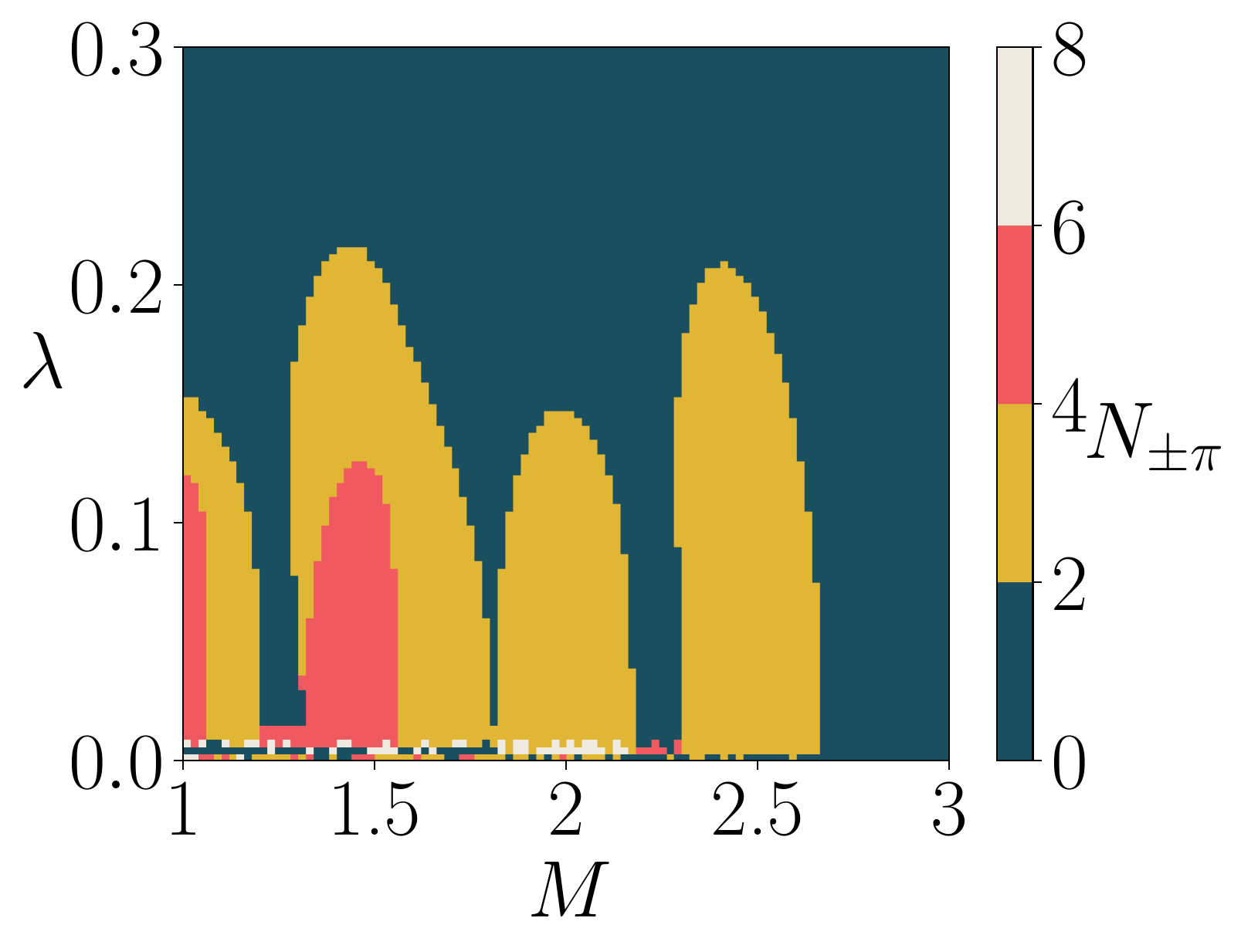}}
  \subfigure[]{\includegraphics[width=0.49\columnwidth]{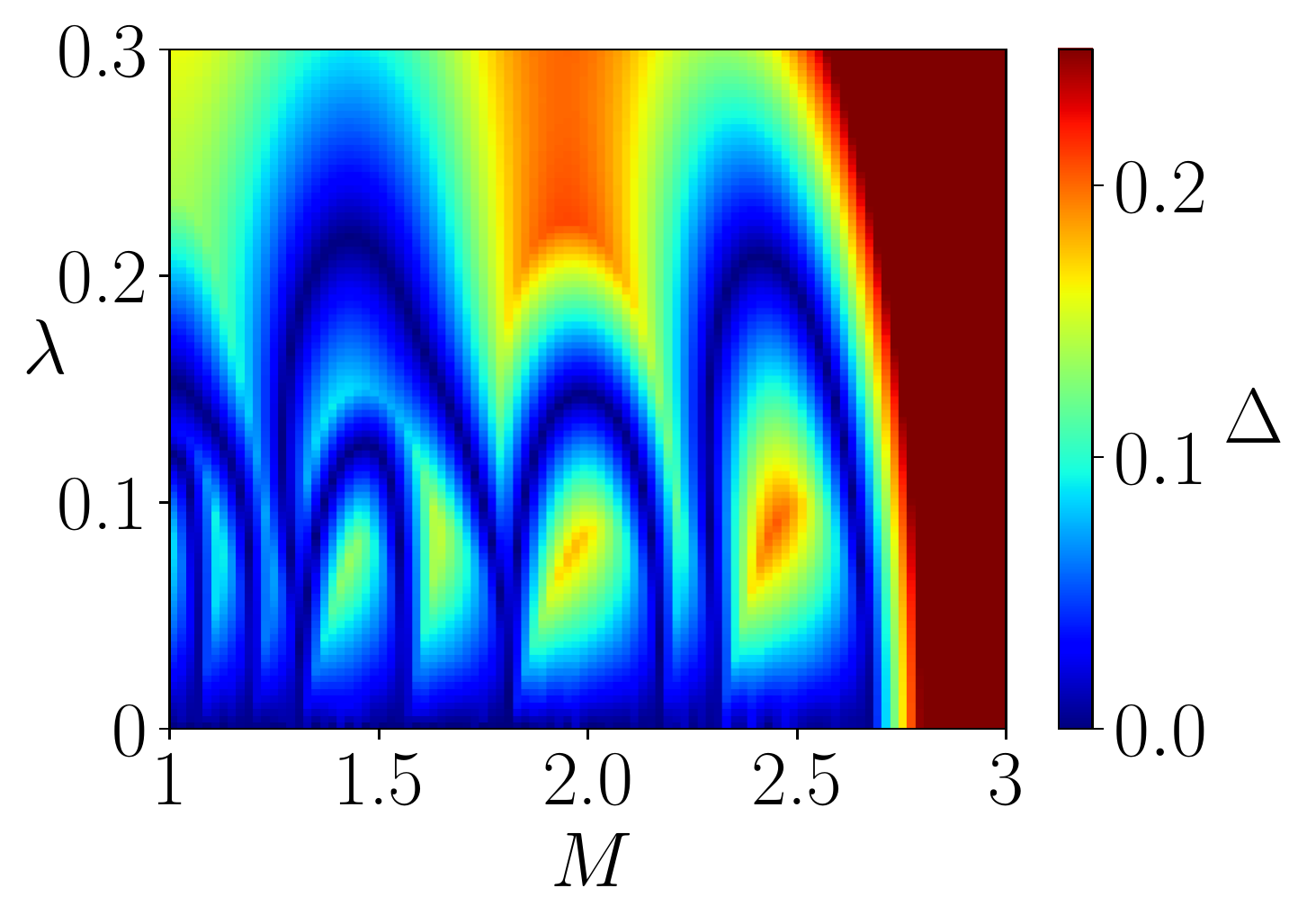}}
  \subfigure[]{\includegraphics[width=0.49\columnwidth]{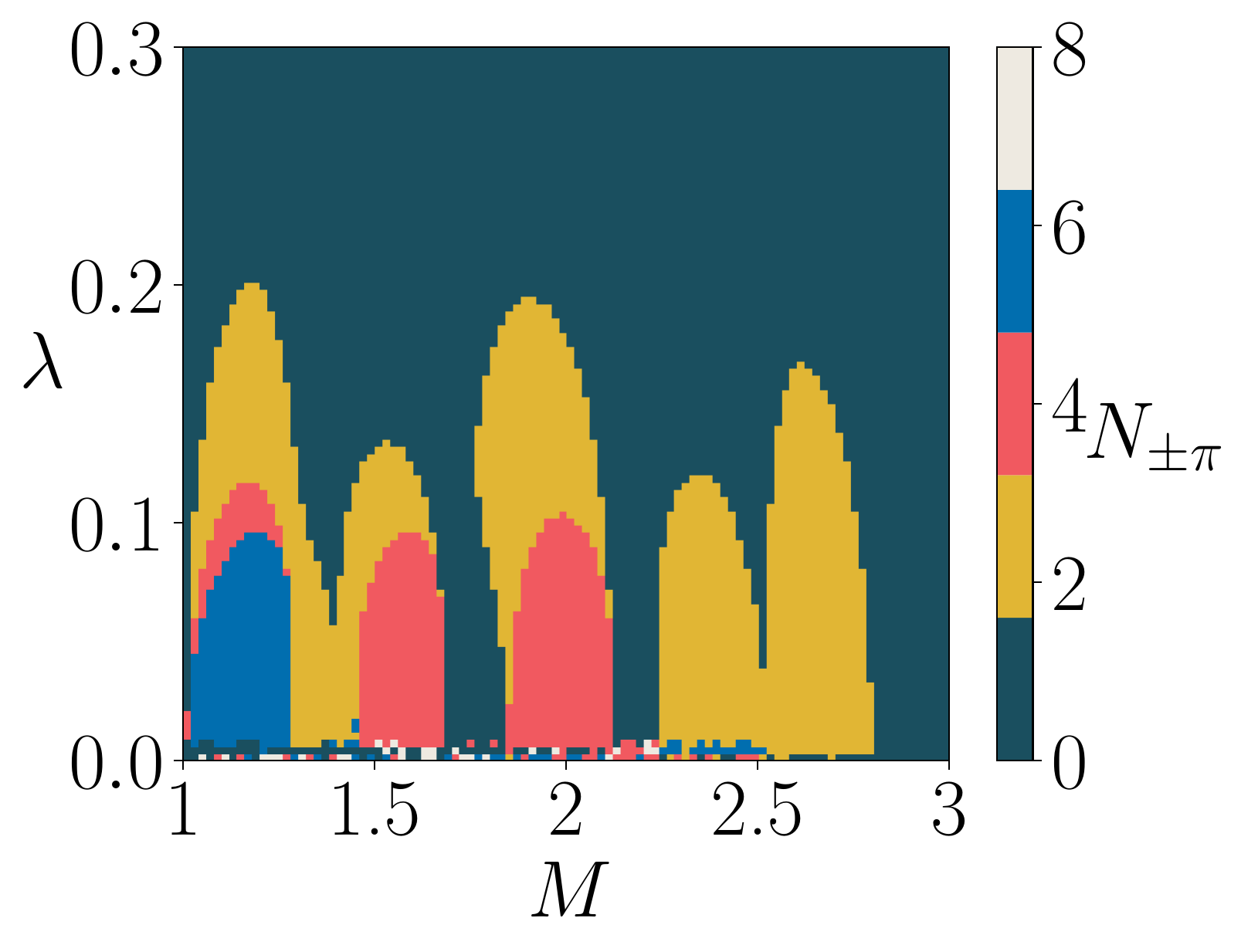}}
  \subfigure[]{\includegraphics[width=0.49\columnwidth]{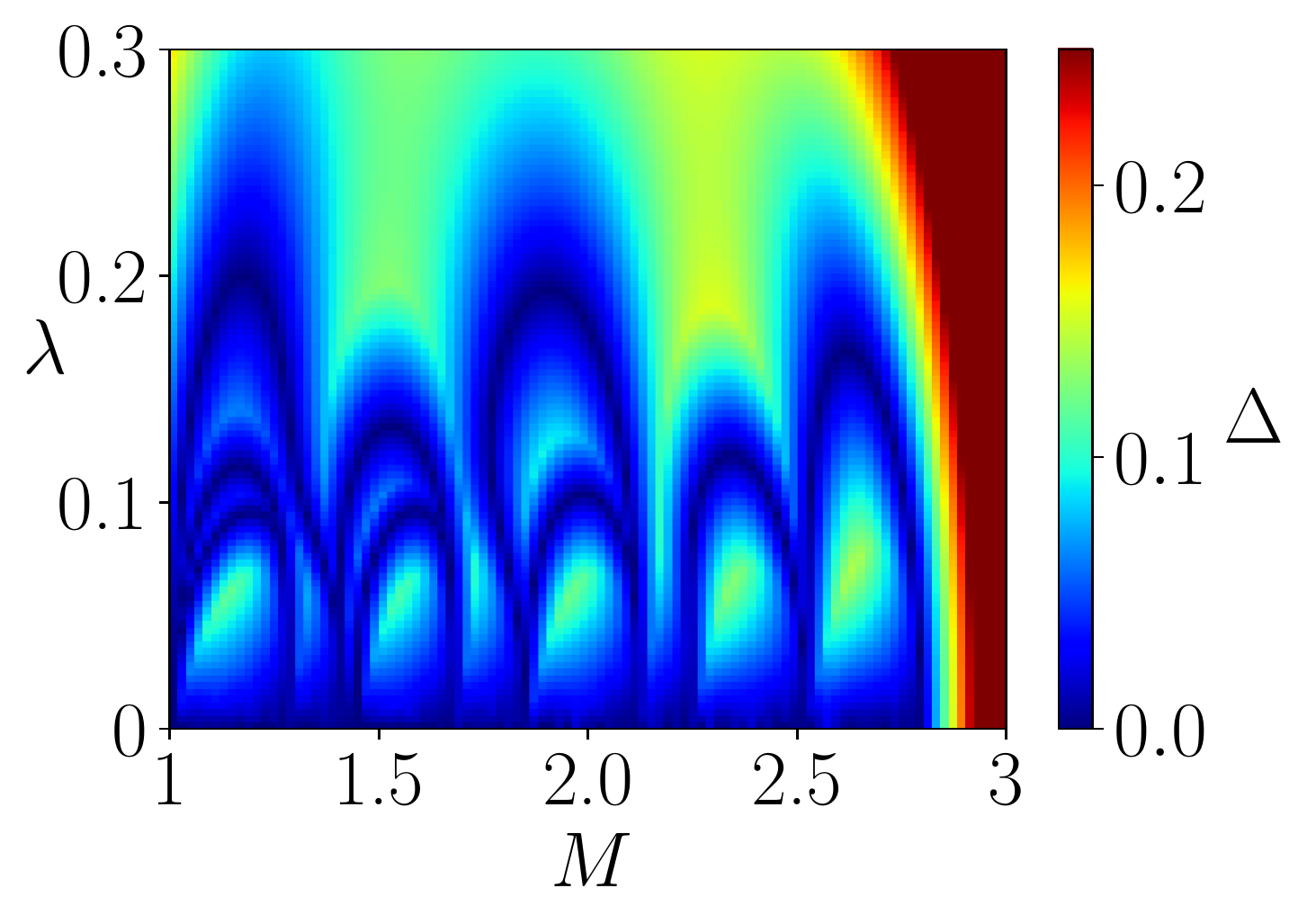}}
  
   \vspace{-0.3cm}
  \caption{
  a) Phase diagram for the same finite size parameters but PBC in z, red reflects 4 $\pm \pi$ phases, yellow 2, blue 0. with $N_x=N_y=4$ b) Heat plot of direct gap for the same system parameters. c),d) and e),f) same as before, but for $N_x=4, N_y=5$ and $N_x=N_y=6$ respectfully. }
  \label{3DTIwire-pd}
\end{figure}

For the q(3-2)D bulk of the STI with periodic boundary conditions only in the $\hat{z}$-direction, we compute Wilson loop spectra by integrating over this one good momentum component, similarly to the Wilson loop spectra calculations for the q(3-2)D bulk of the QSHI in Section II. The topological phase diagrams of the STI q(3-2)D bulk are then determined by computing the number of eigenvalues in the Wilson loop spectrum with $\pm \pi$ phases, or $N_{\pm \pi}$. Although in the previous cases the spectrum only had two $\pm \pi$ phases or none corresponding to nontrivial and trivial regions, we find for the 3D TI wire that some regions of the phase diagram have four $\pm \pi$ phase eigenvalues as shown in  red in Fig.~\ref{3DTIwire-pd} a) for $N_x=N_y=4$. In the case of  $N_x=N_y=6$, the phase diagram changes again and now there are not only two and four $\pm \pi$ phases but also six $\pm \pi$ phases as shown in blue in Fig.~\ref{3DTIwire-pd} e). The change of invariant is consistent with a gap closing of the q(3-2)D bulk as shown in Fig.\ref{3DTIwire-pd} b),f) where the gap closings coincide with the change of number of $\pm \pi$ phases in the Wilson loop spectrum.\\

\begin{figure}[ht]
  \centering
  \subfigure[]{\includegraphics[width=0.49\columnwidth,height=3cm]{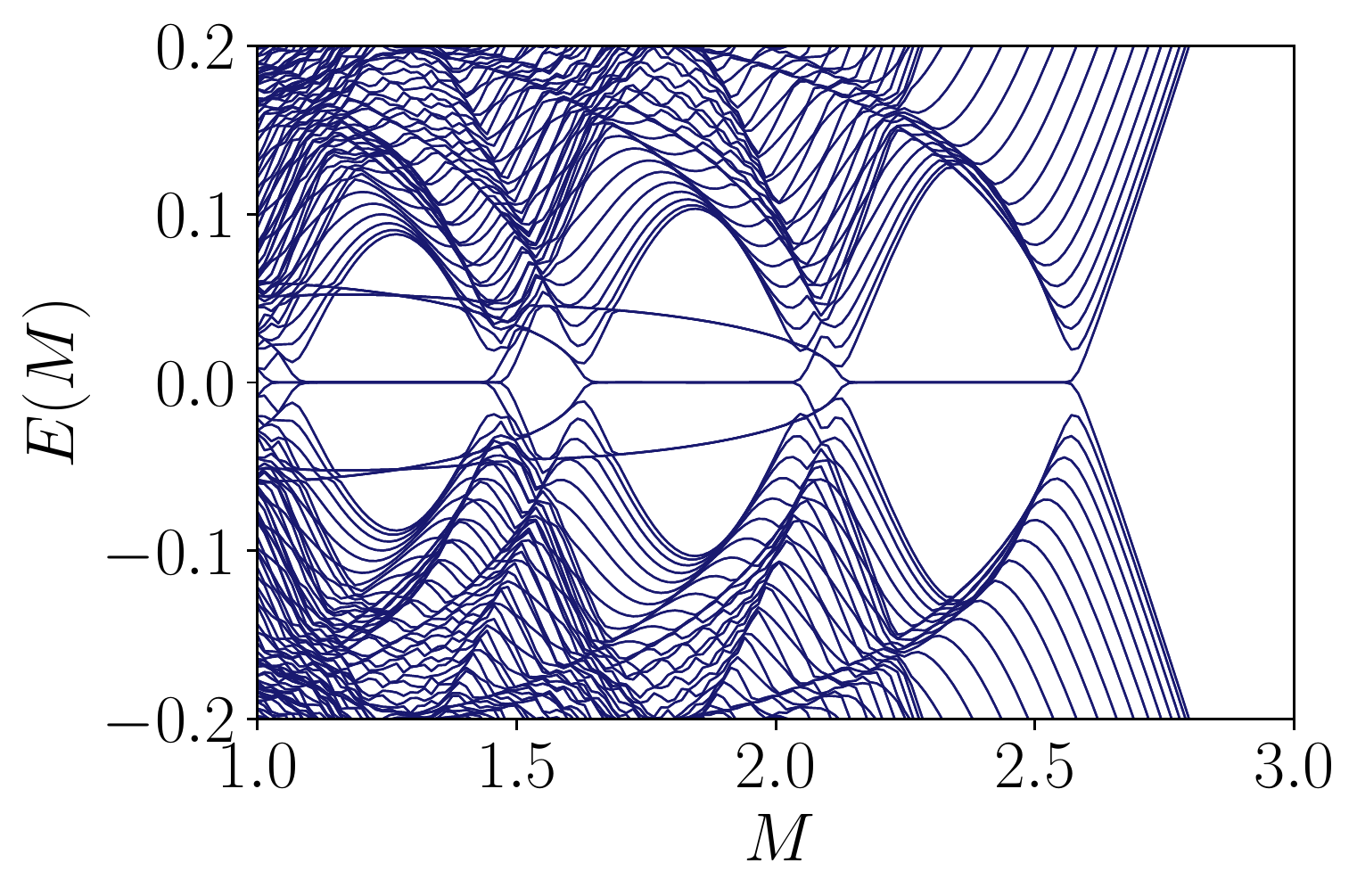}}
  \subfigure[]{\includegraphics[width=0.49\columnwidth,height=3cm]{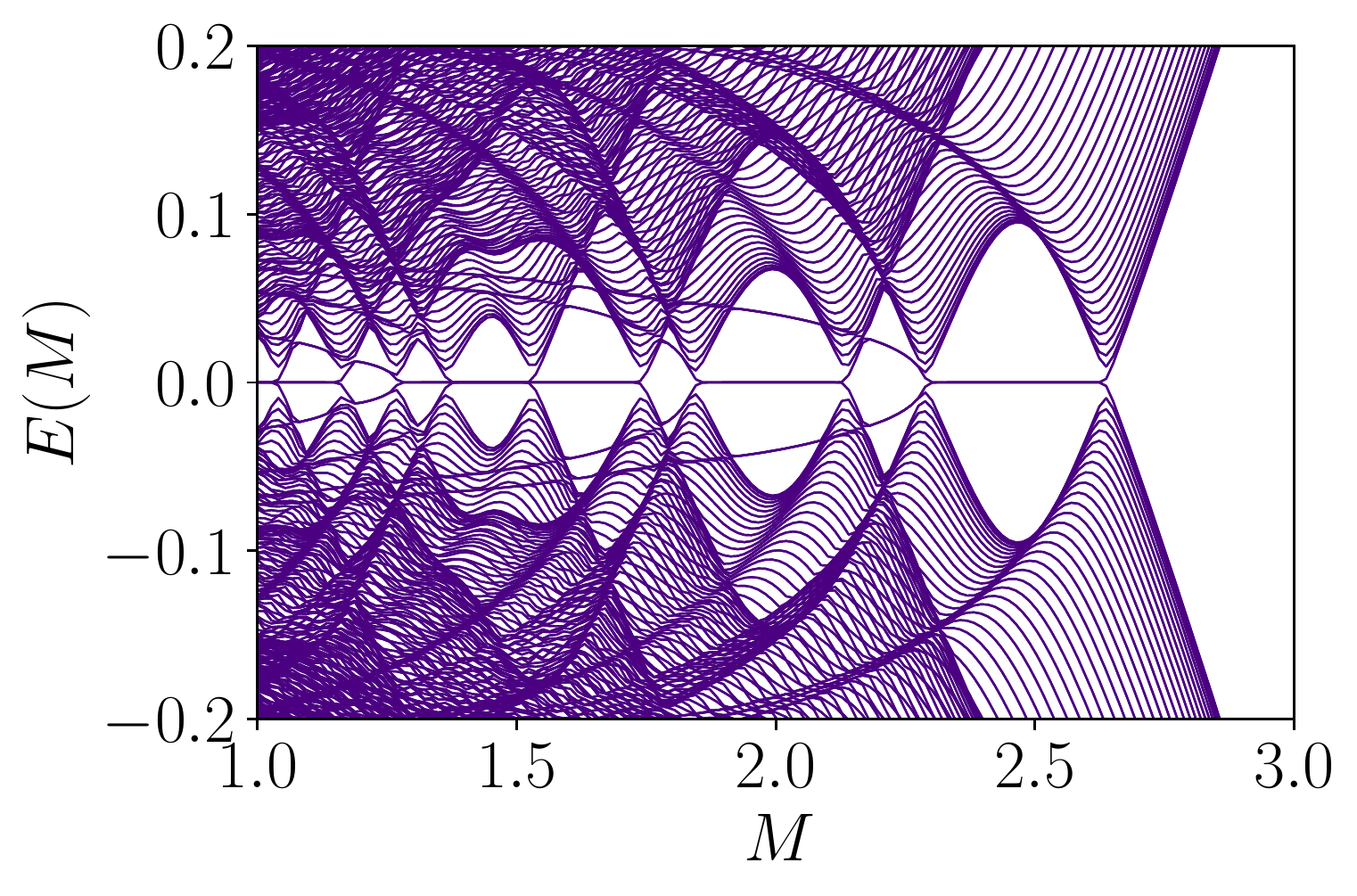}}
   \vspace{-0.3cm}
  \caption{
  a) Spectrum for OBC in x,y,z as a function of $M$, with $N_x,N_y=4$ and $N_z=50$ sites in the remaining directions with $\lambda=0.1$. b) Same plot but for $N_x=4,N_y=5$ and $N_z=100$. }
  \label{3DTIwire}
\end{figure}

As in previous cases, we find the topological phase diagrams for the q(3-2)D bulk of the STI depend strongly on system size. In the case of $N_x=4, N_y=5$, we find that the region with four $\pm \pi$ phases occurs for smaller parameter regions and is furthermore shifted in $M$ and skewed relative to the corresponding regions in the $N_x=N_y=4$, reflecting the difference between $x$ and $y$ directions in phase space. The skewness is also more prominent as we increase the number of sites as seen for $N_x=N_y=6$ in Fig.~\ref{3DTIwire-pd} e). In this case, even though the topologically non-trivial regions diminish in size, they are more strongly skewed. The deformation is such that, starting in a trivial region according to $N_{\pm \pi}$ near $M=1.75$, with almost zero $\lambda$, we can drive the system into a topologically non-trivial regime as determined by $N_{\pm \pi}$ by increasing the spin-orbit coupling to $\lambda\approx 0.1$. A comparison of  the phase diagrams in Fig.~\ref{3DTIwire-pd}, indicates that, as the number of sites increases, the regions present originally in $N_x=N_y=4$ remain for $N_x=N_y=6$ although now shifted to greater $M$ and reduced in size over phase space. We notice also that the new regions appear from the left.\\

The topological phase diagram and corresponding q(3-2)D minimum direct bulk gap phase diagram for $N_x=N_y=6$ are shown in Fig.~\ref{3DTIwire-pd} e), f), respectively. While there are similarities between results for this system size and the smaller ones, there is a topological region for which six Wilson loop eigenvalues have phases fixed to $\pm \pi$. The regions of greater $N_{\pm \pi}$ appear to be subsets of regions with lesser $N_{\pm \pi}$: the blue region is contained within a red region, and red regions are contained within yellow regions. The states again localize at the boundaries of the wire and states occur in Kramers pairs. These results indicate the number of $\pm \pi$ phases in the Wilson loop spectrum, $N_{\pm \pi}$, corresponds to half the number of zero energy edge states $N(E=0)$ in the q(3-2)D STI system with OBC in all directions. We can verify this relation appears to hold for the q(2-1)D QSHI wire and the q(3-1)D STI slab as well. As $N_{\pm \pi} > 2$ occurs for the q1D STI wire, this comes to suggest an integer classification $2 \mathbb{Z}$ for the q(3-2)D STI finite-size topological phases, to be explored in greater detail in future work.\\

\begin{figure}[ht]
  \centering
  \subfigure[]{\includegraphics[width=0.49\columnwidth,height=3cm]{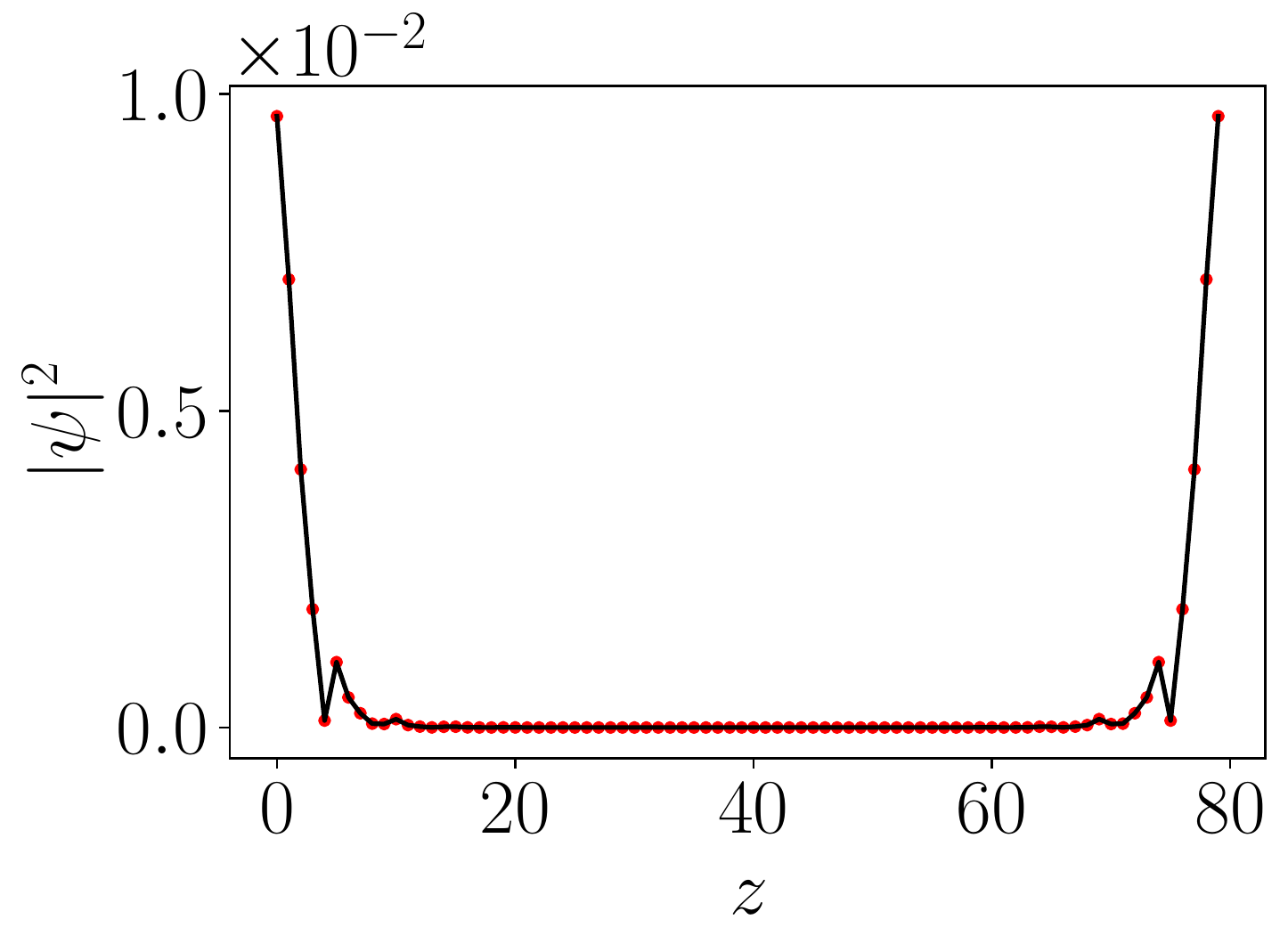} }
  \subfigure[]{\includegraphics[width=0.49\columnwidth,height=3cm]{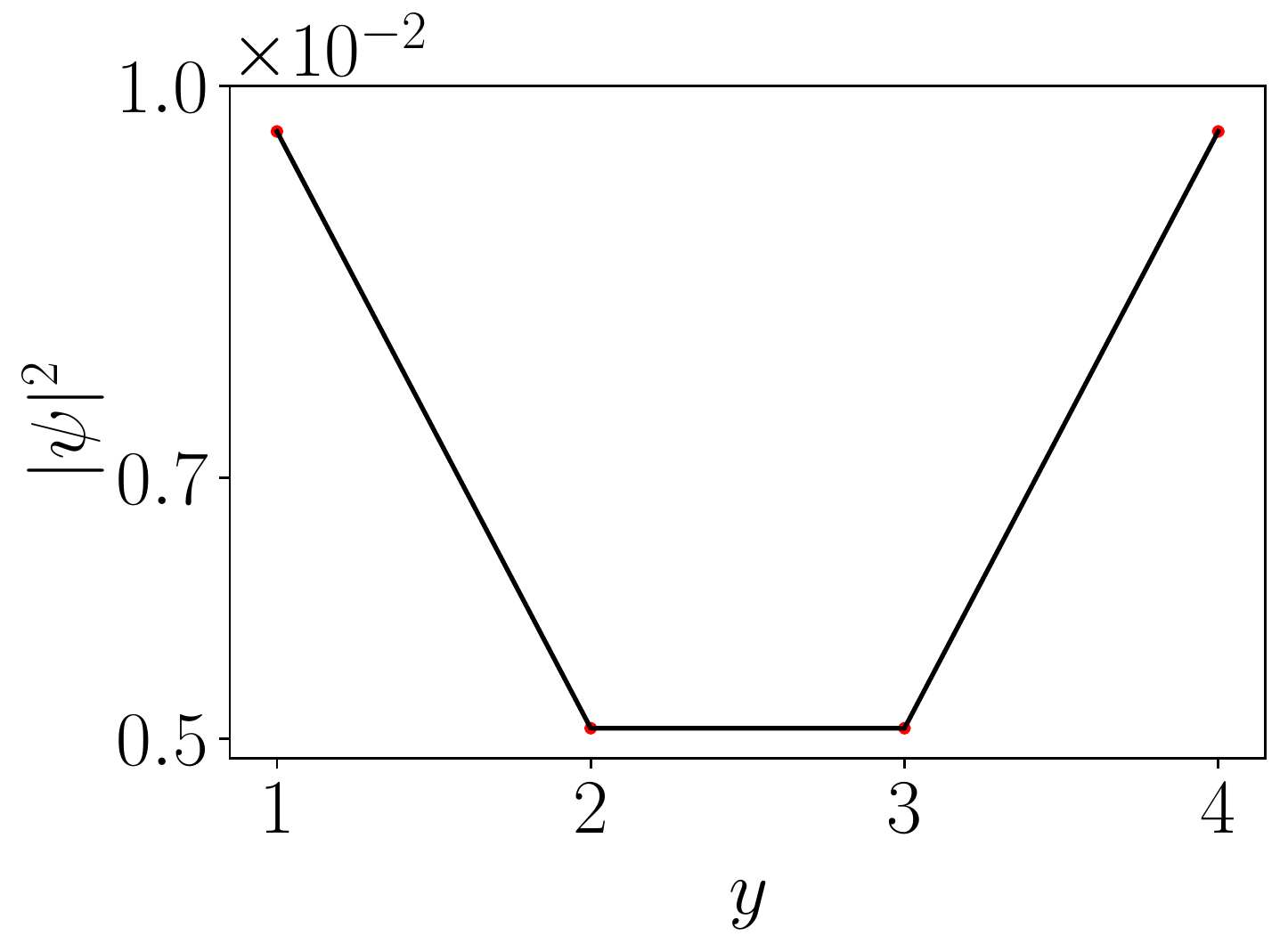}}
  \subfigure[]{\includegraphics[width=0.49\columnwidth,height=3cm]{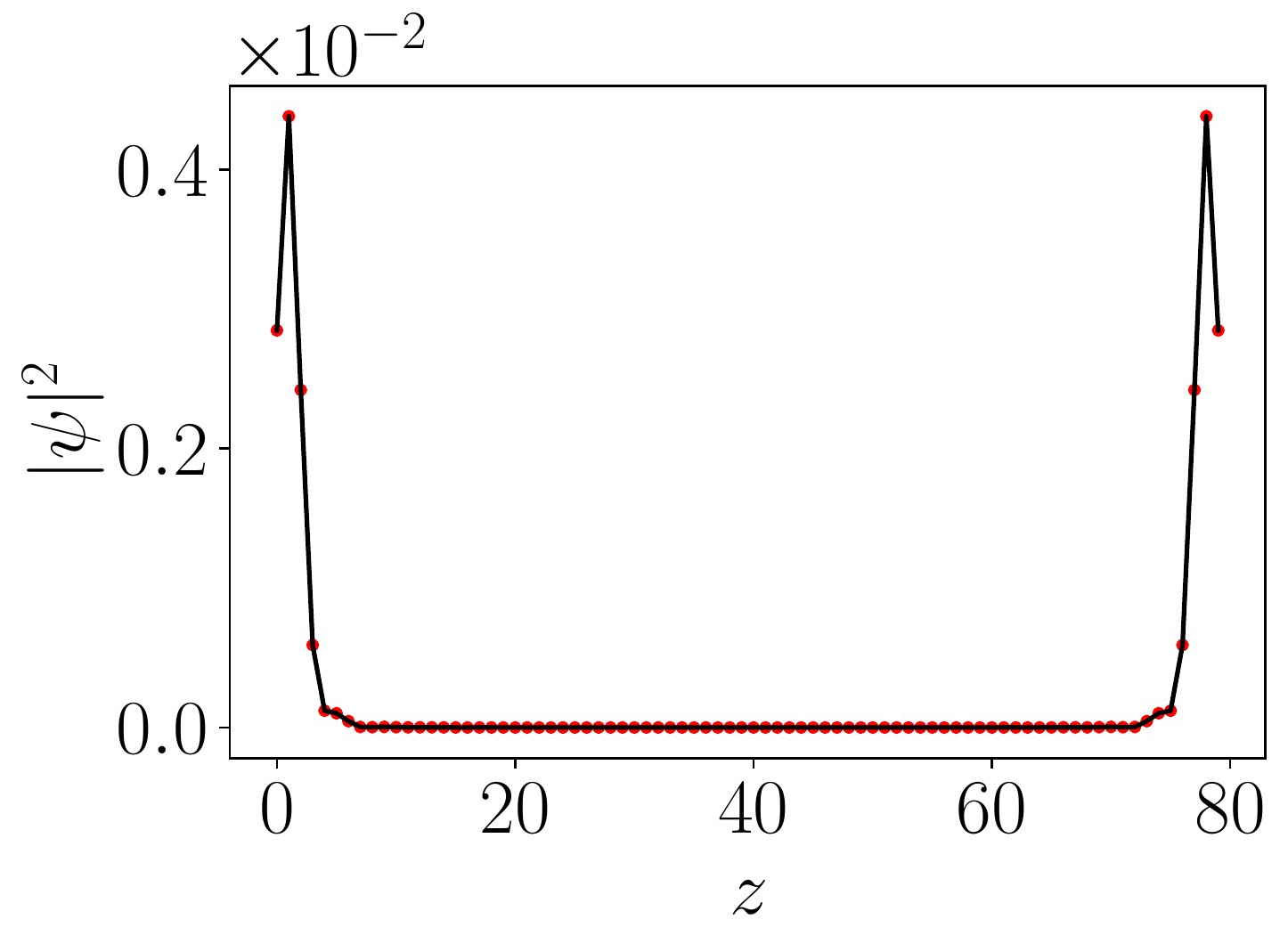}}
  \subfigure[]{\includegraphics[width=0.49\columnwidth,height=3cm]{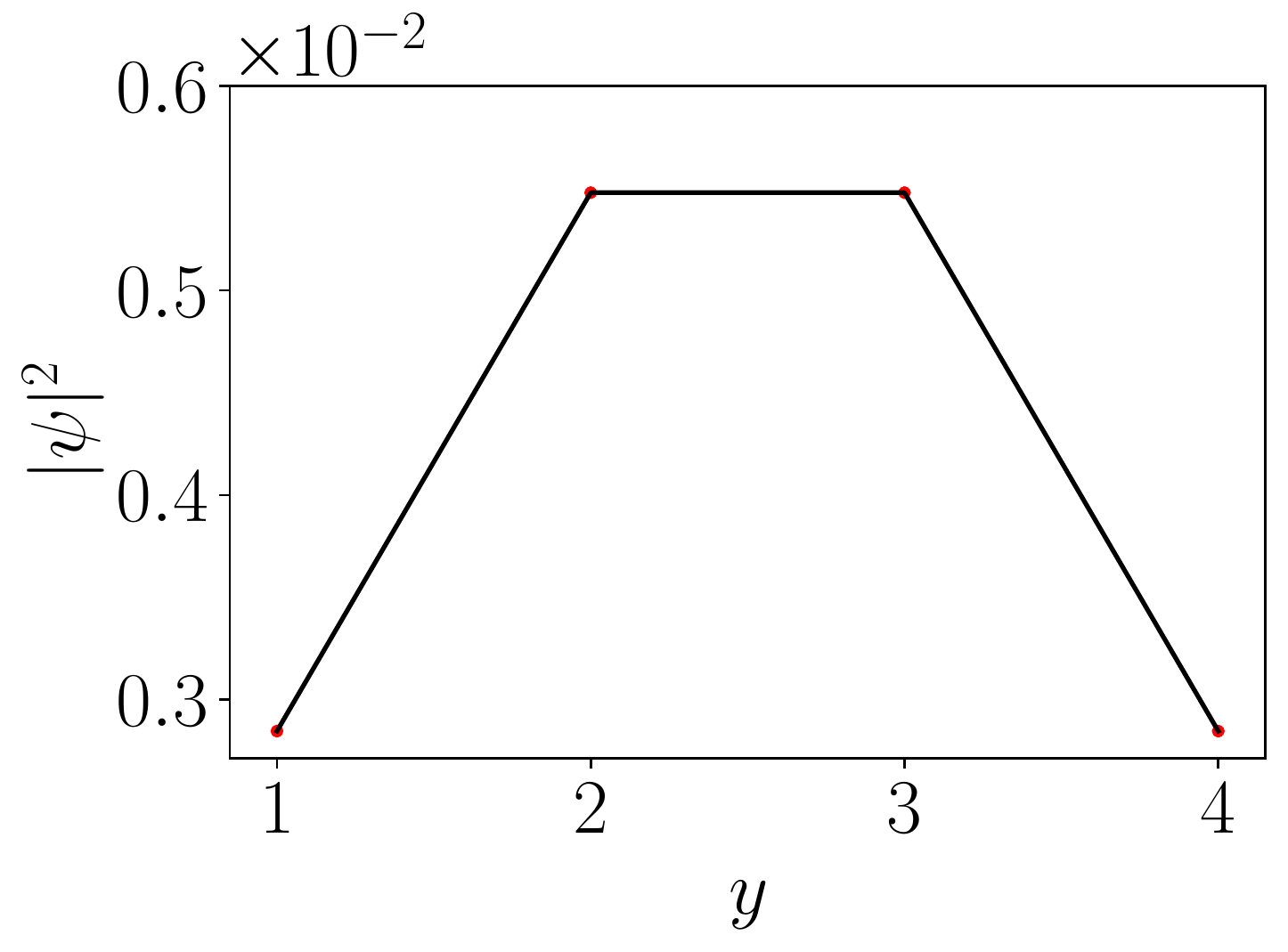}}
  \vspace{-0.3cm}
  \caption{
  a) Probability density plot for one q(3-3)D edge mode as a function  of wire length $z$ with $M=1.25,\lambda=0.1$,$N_x=N_y=4,N_z=80$ b) same edge state as a function of site index $y$ , c) Probability density for another quasi-0D edge mode within the same model parameters as a function of wire length d) now as a function of site index $y$. }
  \label{3DTIslab-states}
\end{figure}

Based on the topological phase diagrams for the q(3-2)D bulk, we now check for a finite-size topological bulk-boundary correspondence in this geometry by opening boundary conditions in the $\hat{z}$-direction, searching for topologically-protected boundary modes localized at the ends of the q(3-2)D wire. In analogy to the q(2-1)D QSHI wire, we study the non-trivial number of $\pm \pi$ eigenvalues in the Wilson loop spectrum, Fig.~\ref{3DTIwire} a),b). The system with $N_{\pm\pi} = 4$ phases has now eight edge states within the q(3-2)D bulk gap. These states occur in Kramers pairs, with each state in a given Kramers pair localized at the same edge There are, however, differences between these states observable in the probability density distributions. We show the probability densities as a function of layer index in each of the $\hat{z}$- and $\hat{y}$-directions, respectively,for four of the in-gap states, in Figs.~\ref{3DTIslab-states} a) and  b), respectively. The corresponding probability densities as a function of layer in the $\hat{z}$-direction and $\hat{y}$-direction for the other four in-gap states are shown in Figs.~\ref{3DTIslab-states} c) and  d), respectively. We see that the second set of four are distinguished from the first four by their localization: the second set of four are pushed inwards from the edge in both the $\hat{z}$- and $\hat{y}$-direction relative to the first four in-gap states. We find similar physics for $N_x=N_y=6$ in the $N_{\pm \pi} = 6$ phase: there are 12 q(3-3)D edge states at zero energy within the q(3-2)D bulk gap. This change in localization suggests that there is a distinction between edge states which may give way to distinct phases not distinguished by just the parity of the number of edge states.

\begin{figure*}
 \includegraphics[width=0.9\textwidth,height=10cm]{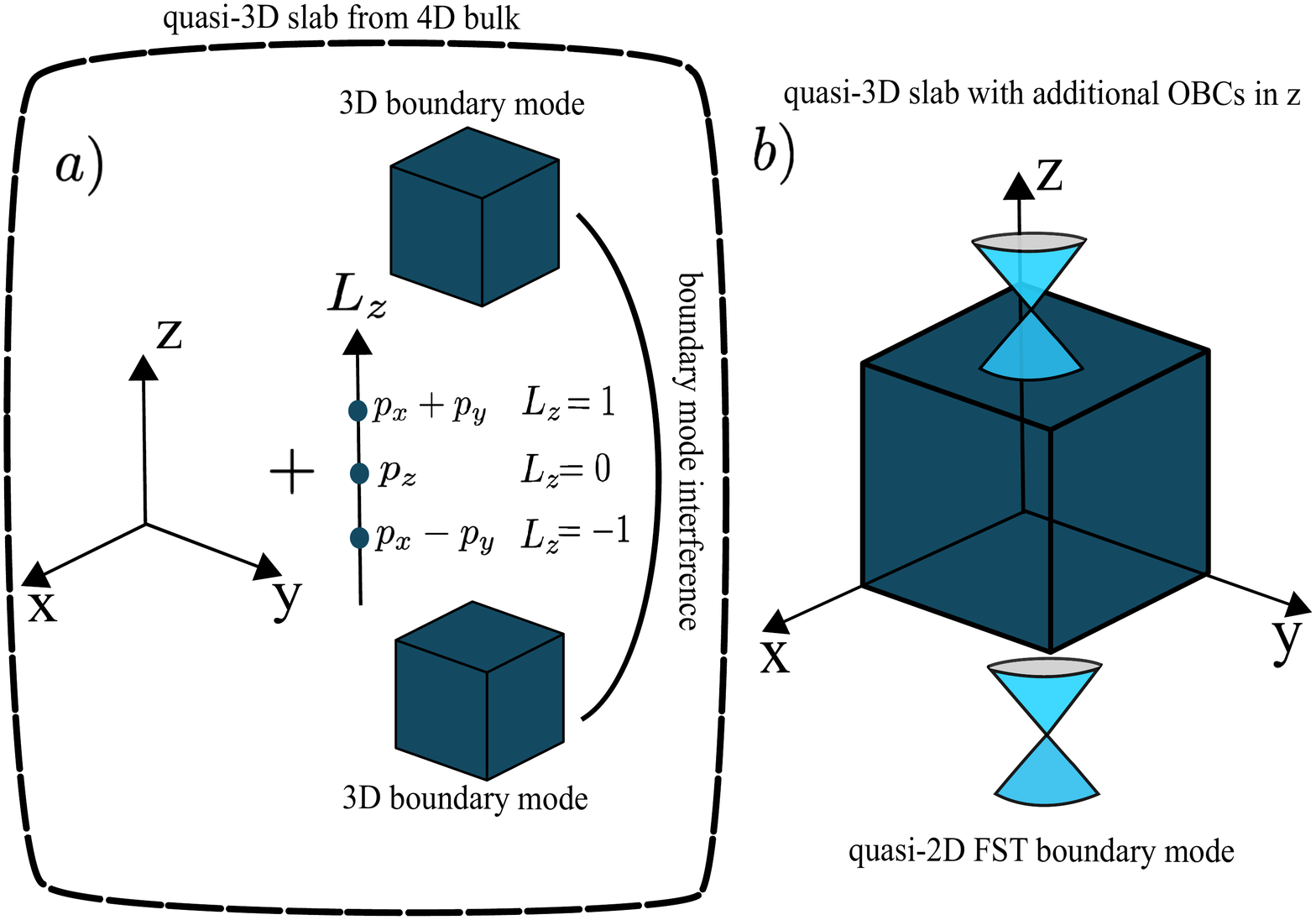}
  \caption{a) Schematic diagram of a 4D topological phase consisting of some $x,y,z$ real space directions and an $L_z$ orbital degree of freedom which gets thinned in the orbital direction. In this case the quasi-3D slab spectrum from the 4D bulk results from the hybridization of the original 3D boundary modes. b) The previous q(4-1)D slab possess an additional bulk-boundary correspondence when additional open boundary conditions in the $z$ direction are considered. This results in q(4-2)D boundary modes.  }
  \label{4d-fst}
\end{figure*}

\section{Concluding remarks} \label{Conclusions}

In this work, we have studied finite-size topology in time-reversal invariant systems, emerging from the hybridization of helical boundary modes in QSHIs and of Dirac cones in the strong TI. In the case of the QSHI, we find the helical boundary modes generically interfere to realize regions in phase space where the q(2-1)D bulk spectrum (periodic boundary conditions in one direction and open boundary conditions in the other) is gapped. These regions are separated from one another by critical points at which the q(2-1)D minimum direct bulk gap is zero. We characterize the topology of these gapped regions by computing Wilson loop spectra, finding topologically non-trivial gapped phases of the q(2-1)D bulk corresponding to a non-trivial number of Wilson loop eigenvalues with phase fixed to $\pm \pi$. \\

For open boundary conditions in each direction and a q(2-1)D wire geometry, these Wilson loop eigenvalues with phase $\pm \pi$ correspond to topologically-protected q(2-2)D boundary modes localized at the ends of the wire. These q(2-2)D boundary modes occur in Kramers pairs and are robust against disorder respecting spinful time-reversal symmetry, maintaining a fourfold degeneracy for particle-hole symmetric disorder, and splitting into doubly-degenerate Kramers pairs for disorder breaking particle-hole symmetry. In these cases, the in-gap, q(2-2)D modes are still topologically-robust in that they must correspond to time-reversal invariant charge transfer in an aperiodic Thouless pump from valence bands to conduction bands, and this connectivity between q(2-1)D bulk valence and conduction bands is observed in topological phase diagrams.

We first observe this finite-size topology of the QSHI for a canonical Hamiltonian describing HgTe quantum wells, but also find the finite-size topological phase occurs in a tight-binding model for 1T'-WTe\textsubscript{2} ribbons with sawtooth edges derived from density functional theory calculations, thus potentially relevant to experiment.
In the case of the strong topological insulator protected by time-reversal symmetry, we find finite-size topological phases both for q(3-1)D slab geometries and q(3-2)D wire geometries. Wilson loop spectra are used to characterize the topology of the q(3-1)D and q(3-2)D bulk: the winding of the Wilson loop eigenvalue phases characterizes the q(3-1)D topology, similarly to characterization of 2D topological phases in the bulk, while the q(3-2)D wire topology in this case is also characterized by the number of $\pm \pi$ Wilson loop eigenvalue phases as in the case of the q(2-1)D QSHI. For open boundary conditions in two directions, the q(3-1)D STI slab exhibits helical boundary modes in the finite-size topological phase, but also exhibits signatures of the magneto-electric polarizability of the STI, distinguishing this finite-size topological phase from the QSHI. In the case of the q(3-2)D wire, q(3-3)D boundary modes occur for open boundary conditions in all three directions, similarly to those of the q(2-1)D QSHI. However, results indicate that topological classification for the q(3-2)D STI is integer rather than $\mathbb{Z}_2$, with unusual localization of the quasi-0D topological boundary modes. Importantly, these results show that finite-size topology yields topologically-protected boundary modes of codimension greater than $1$.\\

We close by pointing out an intriguing possible extension of the work to a system with dimension $D>3$. For example, a four-dimensional topological phase is expected in a q(4-1)D setting, as pictured schematically in Fig.~\ref{4d-fst} for a small system size in some fourth dimension. These extra non-spatial dimensions could come from physical degrees of freedom, typically considered for three-dimensional systems, such as a $p$ orbital degree of freedom as considered in Fig.~\ref{4d-fst} a). One could now imagine an infinite 4D bulk defined by the $
\hat{x}$-, $
\hat{y}$-, $
\hat{z}$-, and $
\hat{L}_z$ (angular momentum)-axes. For non-trivial 4D bulk topological invariant, open boundary conditions in the $
\hat{L}_z$-direction and a large system size in the $
\hat{L}_z$-direction, three-dimensional topologically-protected boundary modes are realized. For the physical scenario of small system sizes in the $\hat{L}_z$-direction as shown in Fig.~\ref{4d-fst} a), these three-dimensional boundary modes could interfere to realize a topologically non-trivial FST phase, such that additional topologically-protected boundary modes are realized when boundary conditions are additionally opened in a second real-space direction, such as the $\hat{z}$-direction as shown in Fig.~\ref{4d-fst} b).
This raises the possibility of topologically-protected boundary states for systems in one to three-dimensions reflecting higher-dimensional, $D>3$ bulk topology. 

\bibliography{finite-TR}

\clearpage

\makeatletter
\renewcommand{\theequation}{S\arabic{equation}}
\renewcommand{\thefigure}{S\arabic{figure}}
\renewcommand{\thesection}{S\arabic{section}}
\setcounter{equation}{0}
\setcounter{section}{0}
\onecolumngrid
\begin{center}
  \textbf{\large Supplemental material for ``Time-reversal invariant finite-size topology''}\\[.2cm]
  R. Flores-Calderon,$^{1,2}$ Roderich Moessner,$^{1}$ and Ashley M. Cook$^{1,2,*}$\\[.1cm]
  {\itshape 
  ${}^1$Max Planck Institute for the Physics of Complex Systems, Nöthnitzer Strasse 38, 01187 Dresden, Germany\\
${}^{2} $ Max Planck Institute for Chemical Physics of Solids, Nöthnitzer Strasse 40, 01187 Dresden, Germany\\}
  ${}^*$Electronic address: cooka@pks.mpg.de\\
(Dated: \today)\\[1cm]
\end{center}

\section{Tight-binding model for $1T'$-$\text{WTe}_2$}

In the main text we discussed a realistic model for realizing finite-size topology in a $1T'-\text{WTe}_2$ monolayer. We adapted the finite-size calculation from the model explored and derived in reference \cite{WTeT2-model}. The model predicts that the four bands closest to the Fermi level are composed mainly of contributions from two $3d_{x^2-y^2}$ type orbitals centered at W and two $5p_x$ type orbitals centered at a subset of Te. With this information at hand and some experimental fitting the authors construct a minimal tight binding model that includes also a spin orbit interaction given by:

\begin{align}
H_{\text{WTe}_2}(\bm{k})=& s_0 \Big({\left[\frac{\mu_{p}}{2}+t_{p x} \cos \left(a k_{x}\right)+t_{p y} \cos \left(b k_{y}\right)\right] \Gamma_{1}^{-} } +\left[\frac{\mu_{d}}{2}+t_{d x} \cos \left(a k_{x}\right)\right] \Gamma_{1}^{+}+t_{d A B} e^{-i b k_{y}}\left(1+e^{i a k_{x}}\right) e^{i \mathbf{k} \cdot \Delta_{1}} \Gamma_{2}^{+} \notag\\
&+t_{p A B}\left(1+e^{i a k_{x}}\right) e^{i \mathbf{k} \cdot \Delta_{2}} \Gamma_{2}^{-} +t_{0 A B}\left(1-e^{i a k_{x}}\right) e^{i \mathbf{k} \cdot \Delta_{3}} \Gamma_{3}-2 i t_{0 x} \sin \left(a k_{x}\right)\left[e^{i \mathbf{k} \cdot \Delta_{4}} \Gamma_{4}^{+}+e^{-i \mathbf{k} \cdot \Delta_{4}} \Gamma_{4}^{-}\right] \notag\\
&+t_{0 A B x}\left(e^{-i a k_{x}}-e^{2 i a k_{x}}\right) e^{i \mathbf{k} \cdot \Delta_{3}} \Gamma_{3}+\text { H.c. }\Big)+  {\left[\left(\lambda_{d x}^{z} s_{z}+\lambda_{d x}^{y} s_{y}\right) \sin \left(a k_{x}\right)\right] \Gamma_{5}^{+} }
+\left[\left(\lambda_{p x}^{z} s_{z}+\lambda_{p x}^{y} s_{y}\right) \sin \left(a k_{x}\right)\right] \Gamma_{5}^{-} \notag\\
&-i \lambda_{0 A B}^{y} s_{y}\left(1+e^{i a k_{x}}\right) e^{i \mathbf{k} \cdot \Delta_{3}} \Gamma_{6} -i\left(\lambda_{0}^{z} s_{z}+\lambda_{0}^{y} s_{y}\right)\left(e^{i \mathbf{k} \cdot \Delta_{4}} \Gamma_{4}^{+}-e^{-i \mathbf{k} \cdot \Delta_{4}} \Gamma_{4}^{-}\right)-i\left(\lambda_{0}^{z} s_{z}+\lambda_{0}^{y} s_{y}\right) \notag \\
& \times\left(e^{-i b k_{y}} e^{i \mathbf{k} \cdot \Delta_{4}} \Gamma_{4}^{+}-e^{i b k_{y}} e^{-i \mathbf{k} \cdot \Delta_{4}} \Gamma_{4}^{-}\right) +\text {H.c., }
\end{align}
,where $s_i$ are Pauli matrices representing the spin degree of freedom and they defined the gamma matrices as:
\begin{align}
\Gamma_{0} &=\tau_{0} \sigma_{0} \\
\Gamma_{1}^{\pm} &=\frac{\tau_{0}}{2}\left(\sigma_{0} \pm \sigma_{3}\right) \\
\Gamma_{2}^{\pm} &=\frac{1}{4}\left(\tau_{1}+i \tau_{2}\right)\left(\sigma_{0} \pm \sigma_{3}\right) \\
\Gamma_{3} &=\frac{1}{2}\left(\tau_{1}+i \tau_{2}\right) i \sigma_{2} \\
\Gamma_{4}^{\pm} &=\frac{1}{4}\left(\tau_{0} \pm \tau_{3}\right)\left(\sigma_{1}+i \sigma_{2}\right) \\
\Gamma_{5}^{\pm} &=\frac{\tau_{3}}{2}\left(\sigma_{0} \pm \sigma_{3}\right) \\
\Gamma_{6} &=\frac{1}{2}\left(\tau_{1}+i \tau_{2}\right) \sigma_{1}
\end{align}
,where $\tau_i,\sigma_i$ are Pauli matrices acting in sublattice and orbital degrees of freedom respectively. Finally the constants from the previous Hamiltonian that reproduce the experimental results are:
\begin{align}
\begin{array}{lccc}
\mu_{p} & -1.75 \mathrm{eV} & \lambda_{0 A B}^{y} & 0.011 \mathrm{eV} \\
\mu_{d} & 0.74 \mathrm{eV} & \lambda_{0}^{y} & 0.051 \mathrm{eV} \\
t_{p x} & 1.13 \mathrm{eV} & \lambda_{0}^{z} & 0.012 \mathrm{eV} \\
t_{d x} & -0.41 \mathrm{eV} & \lambda_{0}^{\prime y} & 0.050 \mathrm{eV} \\
t_{p A B} & 0.40 \mathrm{eV} & \lambda_{0}^{\prime z} & 0.012 \mathrm{eV} \\
t_{d A B} & 0.51 \mathrm{eV} & \lambda_{p x}^{y} & -0.040 \mathrm{eV} \\
t_{0 A B} & 0.39 \mathrm{eV} & \lambda_{p x}^{z} & -0.010 \mathrm{eV} \\
t_{0 A B x} & 0.29 \mathrm{eV} & \lambda_{d x}^{y} & -0.031 \mathrm{eV} \\
t_{0 x} & 0.14 \mathrm{eV} & \lambda_{d x}^{z} & -0.008 \mathrm{eV} \\
t_{p y} & 0.13 \mathrm{eV} & & \\
a & 3.477 \mbox{\normalfont\AA}  
 & b & 6.249 \mbox{\normalfont\AA} \\
\mathbf{r}_{A d} & (-0.25 a, 0.32 b) & \mathbf{r}_{B p} & (0.25 a, 0.07 b) \\
\mathbf{r}_{A p} & (-0.25 a,-0.07 b) & \mathbf{r}_{B d} & (0.25 a,-0.32 b)
\end{array}
\end{align}
Finally we extended the model with an inclusion of a perpendicular electric field in the weak limit where the effect manifests itself as a change of Rashba spin orbit interaction, in the previous Hamiltonian this is modelled as a replacement $\lambda^{i}\rightarrow\lambda^{i}+\Delta \lambda $, with $i=x,y$. 

\end{document}